%% file: main.tex
\documentclass[nofootinbib,aps,prb,twocolumn,notitlepage,superscriptaddress,preprintnumbers]{revtex4-1}

\usepackage[colorlinks=true,linktoc=page,citecolor=red,linkcolor=blue]{hyperref}
\usepackage[usenames]{color}
\usepackage{amsmath,amssymb,amsfonts,times}	
\usepackage{bm,braket,array}
\usepackage{multirow}
\usepackage[all]{xy}
\usepackage{amsthm}
\usepackage{amscd}
\usepackage{slashed} 
\usepackage{graphicx}
\usepackage{tikz-cd}
\usepackage{bbding}
\usepackage[normalem]{ulem}
\usepackage{float}
 
\theoremstyle{definition}

\theoremstyle{remark}

\def\re{{\rm Re\,}}
\def\im{{\rm Im\,}}
\def\Re{{\rm Re\,}}
\def\Im{{\rm Im\,}}

\def\tr{{\rm tr\,}}

\def\ker{{\rm Ker\,}}

\newcommand{\s}{\sigma}
\newcommand{\g}{\gamma}
\newcommand{\la}{\lambda}
\newcommand{\G}{\Gamma}

\newcommand{\R}{\mathbb{R}}
\newcommand{\Z}{\mathbb{Z}}
\newcommand{\T}{\mathbb{T}}

\newcommand{\bk}{{\bm{k}}}

\newcommand{\ba}{{\bm{a}}}
\newcommand{\bt}{{\bm{t}}}

\newcommand{\bG}{{\bm{G}}}
\def\widebar{\accentset{{\cc@style\underline{\mskip10mu}}}} 
\def\wideubar{\underaccent{{\cc@style\underline{\mskip10mu}}}} 

\newcommand{\bb}{\bm{b}}

\newcommand{\XBS}{X_{\text{BS}}}

\newcommand{\mZ}{\mathbb{Z}}

\begin{document}

\preprint{YITP-21-28}

\title{Symmetry indicator in non-Hermitian systems}
\author{Ken Shiozaki}
\thanks{
K.S. and S.O. contributed equally to this work.\\
ken.shiozaki@yukawa.kyoto-u.ac.jp}
\affiliation{Yukawa Institute for Theoretical Physics, Kyoto University, Kyoto 606-8502, Japan}
\author{Seishiro Ono}
\thanks{
K.S. and S.O. contributed equally to this work.\\
ken.shiozaki@yukawa.kyoto-u.ac.jp}
\affiliation{Department of Applied Physics, University of Tokyo, Tokyo 113-8656, Japan}

\begin{abstract}
Recently, topological phases in non-Hermitian systems have attracted much attention because non-Hermiticity sometimes gives rise to unique phases with no Hermitian counterparts. Non-Hermitian Bloch Hamiltonians can always be mapped to doubled Hermitianized Hamiltonians with chiral symmetry, which enable us to utilize the existing framework for Hermitian systems to classify non-Hermitian topological phases. While this strategy succeeded in the topological classification of non-Hermitian Bloch Hamiltonians in the presence of internal symmetries, the generalization of symmetry indicators---a way to efficiently diagnose topological phases---to non-Hermitian systems is still elusive. 
In this work, we study a theory of symmetry indicators for non-Hermitian systems. We define space group symmetries of non-Hermitian Bloch Hamiltonians as ones of the doubled Hermitianized Hamiltonians. Consequently, symmetry indicator groups for chiral symmetric Hermitian systems are equivalent to those for non-Hermitian systems. Based on this equivalence, we list symmetry indicator groups for non-Hermitian systems in the presence of space group symmetries. We also discuss the physical implications of symmetry indicators for some symmetry classes. Furthermore, explicit formulas of symmetry indicators for spinful electronic systems are included in appendices.
\end{abstract}

\maketitle


\section{Introduction}
\label{sec1}
In the last two decades, our understanding of the topological phases of matter has considerably progressed. 
In particular, the understanding of the topological nature of band theory of crystalline systems and the bulk-boundary correspondence---connection between nontrivial topology in bulk and surface states---has played an essential role in developing the study of topological phases~\cite{Hasan_Kane:RMP,Qi_Zhang:RMP,Chiu_Teo_Schnyder_Ryu:RMP}.
Internal and crystalline symmetries such as time-reversal, particle-hole, and magnetic space group symmetries impose restrictions on a Bloch Hamiltonian and result in various topological invariants in momentum space~\cite{Ryu_Schnyder_Furusaki_Ludwig:TenFold,Kitaev:Periodic,Teo_Kane:defect_zero_modes,Fu:TCI,Freed_Moore:TwistedEquivariant,Thiang:K-theoretic_classification,Slager:2013th,Chiu_Yao_Ryu:Reflection_TCI,Morimoto_Furusaki:Clifford,Shiozaki_Sato:TCI,Fang_Fu:Glide_TCI,Shiozaki_Sato_Gomi:Glide_TCI,Shiozaki_Sato_Gomi:Nonsymmorphic,Shiozaki_Sato_Gomi:WallpaperGroup,Khalaf_Po_Vishwanath_Watanabe:SymmetryIndicators,Song_Zhang_Fang:Diagnosis,Song_Zhang_Fang:Diagnosis,Shiozaki_Sato_Gomi:AHSS_band_topology,Ono_Watanabe:Unified_understanding,Cornfeld_Chapman:point_group,elcoro2020magnetic,Peng_Jiang_Fang_Weng_Fang:diagnosis_MSG}.
On the other hand, the nontriviality of topological invariants in momentum  space does not necessarily imply nontrivial boundary states in real space. 
To extract only nontrivial band insulators with boundary states, trivial insulators, so-called atomic insulators consisting of localized orbitals that do not have any boundary state, should be suitably subtracted. 
Based on this idea, a theory of symmetry indicators has been developed in Ref.~\onlinecite{Po_Vishwanath_Watanabe:Symmetry-based_indicators}. 
This method is formulated by subtracting atomic insulators from band structures satisfying all compatibility relations~\cite{Kruthoff_deBoer_vanWezel_Kane_Slager:Compatibility_relation,Po_Vishwanath_Watanabe:Symmetry-based_indicators,Bradlyn_etal:Topological_quantum_chemistry}.
The theory of symmetry indicators is a practical tool because it needs only numbers of irreducible representations at high-symmetry momenta that are easily computed. There exist various applications to materials searches~\cite{Tang2019_NP,Tangeaau8725,Zhang_Jiang_Song_Huang_He_Fang_Weng_Fang:Catalogue,Vergniory_Elcoro_Felser_Regnault_Bernevig_Wang:Catalogue,Tang_Po_Vishwanath_Wan:Catalogue} and generalizations to superconductors~\cite{Ono_Yanase_Watanabe:SI_for_TSC,Ono_Po_Watababe:Refined_symmetry_indicators_SC,Skurativska_Neupert_Fischer:Atomic_limit_SC,Geier_Brouwer_Trifunovic:Symmetry_indicator_SC,Shiozaki:Variants_symmetry_indicator,Ono_Po_Shiozaki:Z2_symmetry_indicator_SC,Ono_Shiozaki:CR_nodes}.
The strategy of subtracting atomic insulators has also been applied to detection boundary states localized at spatial regions with codimension higher than one, called higher-order topological insulators and superconductors~\cite{Fang_Fu:Rotation_anomaly,Benalcazar_Bernevig_Hughes:Quantized_electric_multipole_insulators,Schindler_Cook_Vergniory_Wang_Parkin_Bernevig_Neupert:Higher-order_topological_insulators,Khalaf_Po_Vishwanath_Watanabe:SymmetryIndicators,Khalaf:Higher-order,Trifunovic_Brouwer:Higher-Order,Shiozaki:Surface_states_magnetic_point_group,Okuma_Sato_Shiozaki:Real-space_AHSS,Song_Huang_Qi_Fang_Hermele:Topological_crystals,Cornfeld_Carmeli:tenfold,elcoro2020magnetic,Peng_Jiang_Fang_Weng_Fang:diagnosis_MSG}.

Recently, there has been much interest in the topological nature of non-Hermitian systems~\cite{
Esaki_Sato_Hasebe_Kohmoto:Edge_states_non-Hermitian,Sato_Hasebe_Esaki_Kohmoto:Time-Reversal_Non-Hermitian,Hu_Hughes:Absence_non-Hermitian,Xu_Wang_Duan:Weyl_Exceptional_Rings,Gong_Ashida_Kawabata_Takasan_Higashikawa_Ueda:Topology_NonHermite,MartinezAlvarez_BarriosVargas_FoaTorres:Non-Hermitian_robust_edge_states,Yao_Wang:NonHermitian_skin_effect,Yao_Song_Wang:Non-Hermitian_Chern_Bands,Kunst_Edvardsson_Budich_Bergholtz:Biorthogonal,Kawabata_Shiozaki_Ueda_Sato:NonHermite,Zhou_Lee:Periodic_table_NonHermite,Yokomizo_Murakami:Non-Bloch_Bans_Theory,Lee_Ahn_Zhou_Vishwanath:Topological_Correspondence,Kawabata_Bessho_Sato:Classification_Exceptional_Points}.
By a formal equivalence between single-particle quantum mechanics and the classical wave equation, band theory in solid-state physics is  straightforwardly applied to classical systems with crystalline structures such as optics, photonics, electric circuits, mechanical systems, etc (see Ref.~\onlinecite{Ashida_Gong_Ueda:NH_review} for a review).
Although symmetry and the band-gap in band theory are well understood, we need to revisit the symmetry and gap conditions in non-Hermitian systems.
As for symmetry, internal symmetries, which do not change the spatial position, are particularly significant because they bring basic building blocks of topological phases to us.
While internal symmetry classes in Hermitian systems are classified into ten independent classes (Altland-Zirnbauer classes~\cite{Altland_Zirnbauer}), there are 38 classes (Bernard-LeClair classes~\cite{Bernard_LeClair:NonHermitian}) in non-Hermitian systems due to non-Hermiticity~\cite{Kawabata_Shiozaki_Ueda_Sato:NonHermite}.
As for the gap condition, while we usually impose a finite energy gap at $E=0$ on Hermitian systems, the line and point gap conditions are considered in non-Hermitian systems.
There are two types of line gaps, real and imaginary line gaps, which are defined as the eigenvalues $E$ of the Hamiltonian satisfying $|\re E|>0$ and $|\im E|>0$, respectively. 
When the Hamiltonian has a real (imaginary) line gap, it can be continuously deformable into an (anti-)Hermitian Hamiltonian, preserving internal symmetries and the real (imaginary) line gap~\cite{Kawabata_Shiozaki_Ueda_Sato:NonHermite,Ashida_Gong_Ueda:NH_review}.
In this sense, topological phases of non-Hermitian Hamiltonians with the real (imaginary) line gap condition are not essentially different from those of Hermitian systems.
On the other hand, the point gap condition, which is defined as the Hamiltonian having no zero eigenvalues $|E|>0$~\cite{Gong_Ashida_Kawabata_Takasan_Higashikawa_Ueda:Topology_NonHermite}, gives rise to a new topological phase that did not exist in the Hermitian system and is now being investigated for its properties. 
The non-Hermitian skin effect~\cite{MartinezAlvarez_BarriosVargas_FoaTorres:Non-Hermitian_robust_edge_states,Yao_Wang:NonHermitian_skin_effect} is a prototypical example of non-Hermite topological phases~\cite{Zhang_Yang_Fang:Winding_skin_effect,Okuma_Kawabata_Shiozaki_Sato:NonHermitian_skin_effect}.
A three-dimensional intrinsic non-Hermite topological phase is also known, which exhibits a chiral magnetic effect~\cite{Terrier_Kunst:dissipative_analog,Bessho_Sato:topological_duality,Denner_etal:exceptional_topological_insulators,Kawabata_Shiozaki_Ryu:TFT_nonHermitian_systems}.
Symmetry-protected non-Hermitian topological phases are also becoming understood.
A sort of time-reversal symmetry (reciprocity) leads to $\Z_2$ non-Hermitian  skin effects in one and two dimensions, where for two dimensions the conventional non-Hermitian skin effect is absent~\cite{Okuma_Kawabata_Shiozaki_Sato:NonHermitian_skin_effect}.
We note that intrinsic non-Hermitian topological phases, which are defined as topological phases with the point-gap condition that can not be deformed into a line-gap topological phase while keeping the point gap, were classified in Ref.~\onlinecite{Okuma_Kawabata_Shiozaki_Sato:NonHermitian_skin_effect}. 
Non-Hermitian analogs of higher-order topological insulators are also  discussed~\cite{Okugawa_Takahashi_Yokomizo:second_order_nonHermitian_skin_effects,Kawabata_Sato_Shiozaki:HigherOrder_nonHermitian_skin_effect}.

The purpose of this paper is to formulate a symmetry indicator for non-Hermitian systems.
Originally, the theory of symmetry indicators for Hermitian systems has been formulated by utilizing space group symmetries.
However, since non-Hermitian Bloch Hamiltonians do not necessarily describe crystalline electronic systems, we need to extend the definition of space group symmetries themselves. 
There is a canonical mapping between a non-Hermitian Hamiltonian and an Hermitian Hamiltonian with chiral symmetry, which is called a doubled Hermitianized Hamiltonian (see \eqref{eq:doubledHH} below)~\cite{Gong_Ashida_Kawabata_Takasan_Higashikawa_Ueda:Topology_NonHermite}.
In this work, we define space group symmetries for doubled Hermitianized  Hamiltonians as symmetries of non-Hermitian systems.
This procedure corresponds to viewing a Bloch Hamiltonian $H_\bk$ as a map from a crystalline system ${\cal H}_B$ to another one ${\cal H}_A$, and independent space group symmetries of respective systems ${\cal H}_B$ and ${\cal H}_A$ impose restrictions on the Hamiltonian. 
As a result, we have symmetry constraints for group element $g$ of a space group such that
\begin{align}
A^\bk_g H_\bk = 
\left\{\begin{array}{ll}
H_{p_g \bk} B^\bk_g &(c_g=1),\\ 
H_{p_g \bk}^\dag B^\bk_g &(c_g=-1), \\ 
\end{array}\right. 
\label{eq:NH_sym_intro}
\end{align}
depending on a sign representation $c_g \in \{\pm 1\}$ of space group (See Sec.~\ref{sec:NH} for details). 
The sign $c_g = \pm 1$ indicates the commutation or anticommutation relation between the chiral operator and the space group symmetry operator, which is an input data like the symmetry property of the gap function of a superconductor. 
Because of the mapping between non-Hermitian Hamiltonians in a given symmetry setting and doubled Hermitianized  Hamiltonians, symmetry indicator groups for chiral symmetric Hermitian systems are equivalent to those for non-Hermitian systems.
In this paper, we present symmetry indicator groups of chiral symmetric Hermitian systems in all space groups taking into account all possible sets of $c_g$, which have not been presented in the literature. The result is summarized in Appendix~\ref{app1}. We also discuss some examples of symmetry classes and discuss the relationship between nontrivial symmetry indicators and the corresponding topological boundary modes. 
As shown in Sec.~\ref{sec:ex}, we see that nontrivial symmetry indicators do not always detect intrinsic non-Hermitian topological phases such as non-Hermite skin effects. 
We also list explicit formulas of symmetry indicators for the factor system of spinful electrons in Appendix~\ref{KeySGs}.

In Hermitian systems, it is known that there exist bands that cannot be deformed into atomic insulators while keeping the gap in the entire Brillouin zone (BZ), but the same as atomic insulators at all high-symmetry momenta. When a topological band transition occurs at a generic momentum of which the little group is trivial, then the topological invariants, like the Chern number, can change nontrivially, resulting in boundary gapless states, even if the symmetry indicator index is unchanged. This is a common feature of the symmetry indicator, and this is also true in non-Hermitian systems formulated in this paper. Let us describe this point by using a simple one-dimensional non-Hermitian example. Suppose an inversion-like symmetry $H(k)^\dag=H(-k)$ is imposed on the non-Hermitian Hamiltonian $H(k)$. Under the point-gap condition $\det H(k) \neq 0$ for all $k \in [0,2\pi]$, one can define the winding number $W = \frac{1}{2 \pi i} \int_0^{2\pi} d \log \det H(k)$ which takes a value in integers. From the inversion-like symmetry, we have the $\mathbb{Z}_2$ indicator formula $(-1)^W = {\rm sign} \det H(\pi)/ {\rm sign} \det H(0)$. It should be noted that the $\mathbb{Z}_2$ indicator detects only the even and odd parts of the winding number $W$. In fact, when topological band transitions occur at generic two inversion symmetric $k$-points $k_* \in (0,\pi)$ and $k_*+\pi \in (\pi,2\pi)$, the winding number $W$ changes by an even integer. 

Before moving on to the body of this paper, several comments are in order. 
Symmetries treated in this work are limited such that transposition and complex conjugation are not involved, as written in Eq.~\eqref{eq:NH_sym_intro}.
This restriction is technical, not physical. Our classification of the symmetry indicator groups applies only to the trivial Bernard-LeClair class with additional crystalline symmetry written in Eq.~\eqref{eq:NH_sym_intro} without time-reversal/transposition symmetry.
Although there is no reason that a factor system of a non-Hermitian system is the same as that in electric systems, we show symmetry indicator groups for factor systems of spinful and spinless electron systems as examples.
When the sign representation $c_g$ is trivial, the chiral symmetry is preserved inside sectors of irreducible representations, meaning that no symmetry indicators exist. 
However, in the presence of chiral symmetry, various topological invariants are defined on a one-dimensional subspace of momentum space~\cite{Shiozaki_Sato_Gomi:AHSS_band_topology}, which are invariants beyond counting numbers of irreducible representations.

Upon preparing this manuscript, two related papers appeared~\cite{Okugawa_Takahashi_Yokomizo:generalized_inversion_symmetry, Vecsei_Denner_Neupert_Schindler:Symmetry_indicator_nonHermitian}, where symmetry indicators for non-Hermitian systems with a certain inversion symmetry were discussed. 
Their discussions rely on concrete expressions of topological invariants. One of the most significant merits of symmetry indicators is that we can diagnose topological phases even when we do not know topological invariants. In fact, to our knowledge, we have not discovered all topological invariants that characterize possible topological phases in all symmetry settings.
Furthermore, we note that even for inversion symmetry, the definition of symmetry itself in this paper is more general than that in Refs.~\onlinecite{Okugawa_Takahashi_Yokomizo:generalized_inversion_symmetry, Vecsei_Denner_Neupert_Schindler:Symmetry_indicator_nonHermitian}

This paper is organized as follows.
In Sec.~\ref{sec2}, we formulate the symmetry indicator for the chiral symmetric Hermitian systems.
In Sec.~\ref{sec:NH}, we introduce the space group symmetries for non-Hermitian systems adopted in this paper.
In Sec.~\ref{sec:ex}, we discuss some examples of symmetry indicators for non-Hermitian systems.
In Sec.~\ref{sec:con} we close this paper with conclusion and outlook.
Appendix~\ref{app1} lists the symmetry indicator groups for spinful and spinless fermions in space groups. 
The formulas of the symmetry indicators for factor systems of spinful fermions are also shown in Appendix~\ref{KeySGs}.

\section{Symmetry indicator in Hermitian chiral symmetric systems}
\label{sec2}
In this section, we formulate the theory of symmetry indicators for Hermitian systems with chiral symmetry, which will be utilized into symmetry indicators for non-Hermitian systems in Sec.~\ref{sec:NH}. 

\subsection{Symmetry and $\Z$ invariant}

Let $\tilde H_\bk$ be a Hermitian Bloch Hamiltonian on the BZ. 
In the following, we use the tilde symbols like $\tilde X$ for Hermitian systems to distinguish them from symbols in non-Hermitian systems.
In this paper, we employ the convention that a Hamiltonian $\tilde H_\bk$ is periodic $\tilde H_{\bk+\bG} = \tilde H_\bk$ by a reciprocal lattice vector $\bG$.
This corresponds to the Fourier transformation from real space to momentum space by $\ket{k,i} = \sum_x \ket{x,i} e^{ikx}$, where $x$ runs over the Bravais lattice, and $i$ stands for degrees of freedom inside unit cell. 
This choice of Fourier transformation does not respect the local position of degrees of freedom, which may affect the physical quantities such as the Berry phase but does not affect topological boundary states. 

Let $G$ be a space group, $\T\subset G$ be the subgroup of lattice translations (Bravais lattice), and $G/\T$ be the point group of $G$. 
Let us denote the point group operation and the (fractional) lattice translation of $g \in G$ by $p_g$ and $\bt_g$, respectively. 
The group law implies that $\bt_g+p_g\bt_h=\bt_{gh}$.
We also use the Seitz notation $\{p_g|\bt_g\}$ to represent elements of the space group.
Each element $g\in G$ is represented by a unitary matrix $\tilde U^\bk_g$, which satisfies 
\begin{align}
&\tilde U^\bk_g \tilde H_\bk = \tilde H_{p_g\bk} \tilde U^\bk_g, \label{eq:H_sym}\\
&\tilde U^{p_h\bk}_g \tilde U^\bk_h = z_{g,h} \tilde U^\bk_{gh}, \label{eq:proj_rep}
\end{align}
for $g,h \in G$.
Here, $z_{g,h} \in \text{U}(1)$ is a factor system (two-cocycle) that is determined by the physical system in which we are interested.
For instance, we can always choose $z_{g,h}=+1\ (g,h\in G)$ for spin integer systems. 
In our convention, $\tilde U^\bk_g$ is also periodic $\tilde U^{\bk+\bG}_g=\tilde U^\bk$ by a reciprocal lattice vector $\bG$. 
Note that, for translations $\bt \in \T$, $\tilde U^\bk_{\bt} = e^{-i \bt \cdot \bk}$. 
We assume any translation $\bt\in \T$ commutes with other symmetry operators. 
In other words, the phase factor $z_{g,h}$ depends only on the point group $G/\T$. 

The chiral symmetry is represented as 
\begin{align}
&\tilde \G \tilde H_\bk = - \tilde H_\bk \tilde \G, \label{eq:CS}\\
&\tilde \G^2=1.
\end{align}
Here, $\tilde \G$ is a unitary matrix that acts on internal degrees of freedom, which implies $\tilde \G$ is $\bk$-independent. We can always choose the basis such that $\tilde \G = {\rm diag}({\bf 1}_N, -{\bf 1}_M)$. 
One can assume $N=M$, otherwise the chiral symmetry implies the existence of $|N-M|$ zero-energy flat bands. 

In the presence of chiral symmetry, there is another input data to fix a symmetry class. 
Usually, a symmetry representation is sometimes projective in the one-particle Hilbert space. 
Thus, we have projective factors $c_g$ for $g \in G$ defined by
\begin{align}
\tilde U^\bk_g \tilde \G = c_g \tilde \G \tilde U^\bk_g 
\label{eq:cg_deg}
\end{align}
with $c_g\in \text{U}(1)$.
We note that the group multiplication law Eq.~\eqref{eq:proj_rep} ensures that $c_g$ is a group homomorphism, namely $c_gc_h=c_{gh}$. 
In addition, $\tilde \G^2 = 1$ implies that $c_g$ is a sign $c_g \in \{\pm 1\}$.  
In this paper, we simply assume $c_g=1$ for $g\in \T$ to keep the first BZ unchanged~\footnote{
One may have a nontrivial sign representation for a lattice translation subgroup~\cite{Shiozaki_Sato_Gomi:Glide_TCI}.
For example, let us consider the case $G=\Z$, the lattice translation symmetry in one dimension. 
For the generator $t \in \Z$ one can assign $c_t=-1$. 
In such a case, the lattice translation exchanges the chirality as $U_g^k = \begin{pmatrix}
&e^{-ik}\\
1\\
\end{pmatrix}$.
As a result, the genuine lattice translation is doubled to $2\Z = \{t^{2n}| n \in \Z\}$.
}.
In other words, the projective factor $c_g$ depends only on the point group $G/\T$.  
As a result, a symmetry class is specified by a space group $G$ and a sign representation of the point group $G/\T$. 
It is useful to extend the factor system for $G$ to the full symmetry group $G \times \{e,\G\}$. 
Here, we denoted chiral symmetry as an element of the symmetry group by $\Gamma$.
It is equivalent to fix symmetry representations for element $g\G$ with $g\in G$. 
Here, we define the representation of $g\G$ by $\tilde U_{g\G}^\bk := \tilde U^\bk_g \tilde \G$, that is we choose 
\begin{align}
    &z_{g,\G} = 1, \\
    &z_{\G,g} = c_g, 
\end{align}
for $g\in G$. 

Let us focus on a point $\bk$ in the BZ. 
The little group $G_\bk$ at $\bk$ is defined by $G_\bk = \{g \in G | p_g\bk = \bk+{}^\exists \bG\}$. 
The Hamiltonian is subject to the symmetry constraint 
\begin{align}
[\tilde U^\bk_g,\tilde H_\bk]=0, \quad g\in G_\bk,  
\end{align}
in addition to chiral symmetry Eq.~\eqref{eq:CS}. 
We fix a representative $\underline{g} \in \underline{g}\T$ for each coset of  $G/\T$. 
For trivial coset $e\T=\T$, we set the trivial group element $e$ as a representative. 
From the equality $\{ p_g|\bt_{\underline{g}} \} \{p_h|\bt_{\underline{h}}\}=\{ e|(\bt_{\underline{gh}}-p_h\bt_{\underline{h}}-\bt_{\underline{h}})\} \{ p_{gh}|\bt_{\underline{gh}}\}$ in $G$, we have 
\begin{align}
    &\tilde U^\bk_{\underline{g}} \tilde U^\bk_{\underline{h}}
    =z^\bk_{\underline{g},\underline{h}} \tilde U^\bk_{\underline{gh}}, \\
    &z_{\underline{g},\underline{h}}^\bk= z_{g,h} e^{-i \bk \cdot (\bt_{\underline{gh}}-p_h\bt_{\underline{h}}-\bt_{\underline{g}})}, 
\end{align}
for $\underline{g},\underline{h}\in G_\bk$.
This means that 
$\tilde U^\bk_{\underline{g}}$s form a projective representation of the group $G_\bk/\T$ with the factor system $z^\bk_{\underline{g},\underline{h}}$. 
We can always find the basis such that a Hamiltonian and symmetry representations are simultaneously block-diagonalized, where each block-diagonal element in symmetry representations is an irreducible representation (irrep).
Let $\alpha$ be an irrep of $G_\bk/\T$ with the factor system $z^\bk_{\underline{g},\underline{h}}$.
Let $\chi^\alpha_{\underline{g}}$ be the  character of irrep $\alpha$. 
We denote $\G\alpha$ the mapped irrep of $\alpha$ by $\G$, whose character is given by 
\begin{align}
    \chi^{\G\alpha}_{\underline{g}}
    =\frac{z^\bk_{\underline{g},\G}}{z^\bk_{\G,\G^{-1}\underline{g}\G}} \chi^\alpha_{\G^{-1}\underline{g}\G}
    =c_g \chi^\alpha_{\underline{g}}, \quad \underline{g} \in G_\bk/\T.
\end{align}
We can check whether the chiral operator $\tilde \G$ preserves the irrep $\alpha$ or not by the orthogonality test 
\begin{align}
    O_\alpha 
    &= \frac{1}{|G_\bk/\T|} \sum_{\underline{g} \in G_\bk/\T} [\chi^\alpha_{\underline{g}}]^*
    \chi^{\G\alpha}_{\underline{g}}\\ \nonumber 
    &= \frac{1}{|G_\bk/\T|} \sum_{\underline{g} \in G_\bk/\T} c_g |\chi^\alpha_{\underline{g}}|^2 \in \{0,1\}.
\end{align}
When $O_\alpha=0$, the mapped irrep $\G \alpha$ is not unitary equivalent to $\alpha$, and the chiral symmetry is absent within the irrep $\alpha$. 
Then, we introduce the following $\Z$ invariant 
\begin{align}
N_\bk^\alpha[\tilde H^0_\bk,\tilde H^1_\bk]:=
    N_\bk^\alpha[\tilde H^0_\bk] - N_\bk^\alpha[\tilde H^1_\bk] \in \Z, 
\label{eq:ZinvHH}
\end{align}
for a pair of Hamiltonians $\tilde H^0_\bk$ and $\tilde H^1_\bk$ defined on a common one-particle Hilbert space. Here, $N_\bk^\alpha[\tilde H_\bk]$ is the number of negative energy eigenstates of irrep $\alpha$, and $N_\bk^\alpha[\tilde H^0_\bk,\tilde H^1_\bk]$ measures how two Hamiltonians $\tilde H^0_\bk$ and $\tilde H^1_\bk$ are different. 
If $N_\bk^\alpha[\tilde H^0_\bk,\tilde H^1_\bk] \neq 0$, there is no continuous path connecting the two Hamiltonians $\tilde H^0_\bk$ and $\tilde H^1_\bk$, keeping the energy gap and symmetry.  
Notice that if $O_\alpha=1$, namely, if the chiral symmetry is preserved in irrep $\alpha$, the number $N_\alpha[\tilde H_\bk]$ of negative energy eigenstates is a constant determined solely from the one-particle Hilbert space, i.e., $N_\bk^\alpha[\tilde H^0_\bk,\tilde H^1_\bk]$ vanishes identically. 
The invariant Eq.~\eqref{eq:ZinvHH} will be used to define the symmetry indicator for chiral symmetric systems.

\subsection{Comments on topological invariant and $K$-group}
Let $(\tilde H^0_\bk,\tilde H^1_\bk)$ be a pair of gapped Hamiltonians defined on common atomic orbitals with symmetries representations  Eqs.~\eqref{eq:H_sym} and \eqref{eq:CS}. 
We ask if $\tilde H^0_\bk$ is continuously deformed into $\tilde H^1$ while keeping symmetries. 
This question is difficult to answer, and there is no general prescription.
It is also not necessarily an important question, depending on the physics in which we are interested. 

This question can be answered if all the definable topological invariants are known and can be computed: 
If there exists at least one different value of a topological invariant for two Hamiltonians, we know that continuous deformation between two Hamiltonians is forbidden. 
However, it is not easy to list all possible topological invariants, and it depends in general on the rank of the Hamiltonian and the concrete form of the symmetry representations. 
Under a weakening of the equivalence relation called ``stable equivalence~\cite{Kitaev:Periodic}," pairs of Hamiltonians can be classified more conveniently by $K$-theory~\cite{Karoubi:k-theory}. However, even in this case, to derive concrete expressions of topological invariants is a case-by-case problem and is not practical.

Also, nontrivial topological invariants in momentum space do not necessarily imply topological boundary states in real space.
It is necessary to define a suitable set of topological invariants in momentum space according to the physical properties of interest.
As we will see in the next section, the symmetry indicator can efficiently detect only nontrivial insulators with topological boundary states.

\subsection{Symmetry indicator}
\label{sec:SI_H}
The construction of the symmetry indicator in chiral symmetric systems is almost parallel to that in Ref.~\onlinecite{Ono_Po_Watababe:Refined_symmetry_indicators_SC}. 
Particle-hole symmetry in Ref.~\onlinecite{Ono_Po_Watababe:Refined_symmetry_indicators_SC} plays the same role as chiral symmetry in chiral symmetric systems. 
In this subsection, we briefly summarize how the symmetry indicator is formulated. 

The theory of symmetry indicators is composed of two free abelian groups in terms of topological invariants in momentum space, which are called ``lattice" in mathematics. One is the lattice of the band structures that satisfy all compatibility relations, and the other is a lattice of atomic insulators~\cite{Po_Vishwanath_Watanabe:Symmetry-based_indicators, Po:Symmetry_indicator_review}.

While it is not a pragmatic approach to list all possible topological invariants in momentum space, a subset of topological invariants can be used to show the absence of continuous deformations between two Hamiltonians. 
The $\Z$ invariant in Eq.~\eqref{eq:ZinvHH} defined at high symmetry points, which is a stable invariant in the sense of the $K$-theory, is an apparently topological invariant in momentum space and easy to compute. 
In chiral symmetric systems, we first define a free abelian group $E_1^{0,0}$ by the lattice formed by $\Z$ topological invariants in Eq.~\eqref{eq:ZinvHH} at high symmetry points called 0-cells in a symmetric cell decomposition. 
(We here use the notation of Atiyah-Hirzebruch spectral sequence~\cite{Shiozaki_Sato_Gomi:AHSS_band_topology}). 
Any pair of Hamiltonians $(\tilde H^0_\bk, \tilde H^1_\bk)$ that have finite energy gaps on 0-cells corresponds to a point in the lattice $E_1^{0,0}$ through the invariants in Eq.~\eqref{eq:ZinvHH}. 

The lattice $E_1^{0,0}$ is not unique as it depends on a choice of 0-cells. 
In addition, a pair of Hamiltonians $(\tilde H^0_\bk, \tilde H^1_\bk)$ representing a point of $E_1^{0,0}$ sometimes have an obvious obstacle in deforming each other on some lines connecting 0-cells while keeping an energy gap and symmetry due to a violation of the compatibility relation.
The compatibility relation is viewed as the first differential $d_1^{0,0}: E_1^{0,0} \to E_1^{1,0}$ of the Atiyah-Hirzebruch spectral sequence~\cite{Shiozaki_Sato_Gomi:AHSS_band_topology}, where $E_1^{1,0}$ is the lattice for $\Z$ invariants in Eq.~\eqref{eq:ZinvHH} on line segments connecting 0-cells called 1-cells, and $d_1^{0,0}$ is computed by the irreducible character~\cite{Kruthoff_deBoer_vanWezel_Kane_Slager:Compatibility_relation,Po_Vishwanath_Watanabe:Symmetry-based_indicators,Bradlyn_etal:Topological_quantum_chemistry,Shiozaki_Sato_Gomi:AHSS_band_topology}.
We define $E_2^{0,0} = \ker d_1^{0,0}$, which is a sublattice of $E_1^{0,0}$, as the lattice of band structures that satisfy all compatibility relations~\cite{Po_Vishwanath_Watanabe:Symmetry-based_indicators}\footnote{Indeed, $E_2^{0,0}$ is nothing but $\{\mathrm{BS}\}$ in Ref.~\onlinecite{Po_Vishwanath_Watanabe:Symmetry-based_indicators}.}.
It turns out that $E_2^{0,0}$ does not depend on a choice of 0-cells. 
For a pair of Hamiltonians $(H^0_\bk,H^1_\bk)$ corresponding to a point in $E_2^{0,0}$, $H^0_\bk$ can be deformed into $H^1_\bk$ on 1-cells with an energy gap and symmetry remained (in the sense of the stable equivalence). 

Next, we introduce the lattice of atomic insulators in chiral symmetric systems. 
In the absence of chiral symmetry, an atomic insulator is obtained by the following procedures: (i) choosing a Wyckoff position ${\sf x}$, (ii) choosing one irrep of the little group $G_{\sf x}:= \{g \in G| p_g{\sf x}+\bt_g = {\sf x}\}$ with the factor system $z_{g,h}$, and (iii) constructing the induced representation for the whole space group $G$. 
As a result, we have symmetry operators $\tilde U^{{\rm occ},\bk}_g$ of a Wyckoff position ${\sf x}$ and an irrep with the trivial fully occupied Hamiltonian $\tilde H^{\rm occ}_\bk\equiv -{\bf 1}$ (See Appendix C in Ref. \onlinecite{Po_Vishwanath_Watanabe:Symmetry-based_indicators} for the detail). 
Next, we enforce chiral symmetry on the Hamiltonian $\tilde H^{\rm occ}_\bk$ by doubling the degrees of freedom, which results in a chiral symmetric atomic insulator Hamiltonian $\tilde H^{\rm ai}_\bk = {\rm diag}(-{\bf 1},{\bf 1})$ with symmetry operators $\tilde U^\bk_g = {\rm diag}(U^{{\rm occ},\bk}_g,c_g U^{{\rm occ},\bk}_g)$ equipped with the chiral operator $\tilde \G = \s_x \otimes {\bf 1}$.
Are there any atomic insulator Hamiltonians defined on the same localized degrees of freedom? 
The only possibility is the negative one $-H^{\rm ai}_\bk = {\rm diag}(-{\bf 1},{\bf 1})$ with the same symmetry operators $\tilde U^\bk_g$. 
Conversely, given symmetry operators $\tilde U^\bk_g$ and the chiral operator $\tilde \G = \s_x \otimes {\bf 1}$ of localized orbitals, a constant Hamiltonian should be either of $\tilde H^{\rm ai}_\bk$ or $-\tilde H^{\rm ai}_\bk$. Therefore, we conclude that for given atomic degrees of freedom equipped with chiral symmetry, the only pair of Hamiltonian is $(\tilde H^{\rm ai}_\bk,-\tilde H^{\rm ai}_\bk)$. 
An atomic insulator $\tilde H^{\rm ai}_\bk$ corresponds to a point in $E_1^{0,0}$ through  $\Z$ invariants $N_\bk^\alpha[\tilde H^{\rm ai}_\bk,-\tilde H^{\rm ai}_\bk] = N_\bk^\alpha[\tilde H^{\rm ai}_\bk]-N_\bk^{\G \alpha}[\tilde H^{\rm ai}_\bk]$.
Here we have used the relation  $N_\bk^\alpha[-\tilde H_\bk] = N_\bk^{\G \alpha}[\tilde H_\bk]$.
Since the atomic insulator $\tilde H^{\rm ai}_\bk$ satisfies all compatibility conditions, the pair $(\tilde H^{\rm ai}_\bk,-\tilde H^{\rm ai}_\bk)$ is also a point in $E_2^{0,0}$. 
Collecting all possible atomic insulator Hamiltonians, we have a sublattice $K_{\rm ai} \subset E_2^{0,0}$, which we call the lattice of atomic insulators. 
We note that all Wyckoff positions for 230 space groups are available in Ref.~\onlinecite{Stokes_Hatch_Campbell}. 
Finally, the symmetry indicator group is defined by the quotient group $Q := E_2^{0,0}/K_{\rm ai}$. 

Our definition of atomic insulators is simply rephrased as follows. An atomic insulator in Hermitian chiral symmetric systems is a fully gapped Hamiltonian $\tilde H_{\bm{k}}$ that is deformable with keeping the band gap to a $k$-independent constant Hamiltonian $\tilde H_{\bm{k}} = {\rm diag}(-{\bf 1},{\bf 1})$ on the basis that the chiral symmetry operator is written as $\tilde{\Gamma} = \begin{pmatrix}{\bf 0}&{\bf 1}\\{\bf 1}&{\bf 0}\\ \end{pmatrix}$. We note that the Su-Schrieffer-Heeger (SSH) Hamiltonian, which is a prototypical topologically nontrivial model in chiral symmetric systems, is not an atomic insulator since the finite winning number prohibits the SSH Hamiltonian from being a $\bm{k}$-independent Hamiltonian.

Given a Hamiltonian $\tilde H_\bk$ defined on a localized atomic degrees of freedom, we have a trivial reference Hamiltonian $\pm \tilde H^{\rm ai}_\bk$. Thus the pair $(\tilde H^{\rm ai}_\bk,\pm \tilde H^{\rm ai}_\bk)$ is an element of the symmetry indicator group $Q$. 
Since the pair $(\tilde H^{\rm ai}_\bk, - \tilde H^{\rm ai}_\bk)$ is already subtracted in the quotient group $Q$, a choice of trivial reference Hamiltonian is arbitrary. 
Thus, a single Hamiltonian $\tilde H_\bk$ specifies an element of the symmetry indicator group $Q$. 
Suppose a Hamiltonian $\tilde H_\bk$ is a nontrivial entry of $Q$; In that case, it implies that (i) $\tilde H_\bk$ has a gapless point with codimension two or more or (ii) $\tilde H_\bk$ has a topological boundary state.

\section{Symmetry indicator in non-Hermitian systems}
\label{sec:NH}
In this section, we first formulate the definition of symmetry in non-Hermitian system by rewriting space group symmetry in Eq.~\eqref{eq:H_sym} for the off-diagonal part of the chiral Hamiltonian. 
There are some different caveats from the usual Hermitian systems. 

\subsection{Symmetry in non-Hermitian systems}
We move on to non-Hermitian systems. 
We can always choose the basis such that the chiral symmetry $\tilde \G$ takes the following form
\begin{align}
\tilde \G = \begin{pmatrix}
1&\\
&-1\\
\end{pmatrix}.
\end{align}
In the following, we fix $\tilde \G$ to this form.
In such a basis, a Hermitian Hamiltonian $\tilde H_\bk$ becomes 
\begin{align}
\tilde H_\bk
=
\begin{pmatrix}
&H_\bk\\
H_\bk^\dag\\
\end{pmatrix}. 
\label{eq:doubledHH}
\end{align}
Here, $H_\bk$ does not have to be Hermitian. 
Since the set of singular values of $H_\bk$ coincides with the set of positive eigenvalues of $\tilde H_\bk$, the existence of a finite energy gap of $\tilde H_\bk$ corresponds to the point gap condition on $H_\bk$. 
Depending on the sign of $c_g$, the unitary matrix $\tilde U^\bk_g$ is also written as 
\begin{align}
	\tilde U^\bk_g
	=\left\{\begin{array}{ll}
		\begin{pmatrix}
			A^\bk_g\\
			&B^\bk_g\\
		\end{pmatrix} & (c_g=1) \\
		\begin{pmatrix}
			&B^\bk_g\\
			A^\bk_g\\
		\end{pmatrix} & (c_g=-1)  \\
	\end{array}\right.,
	\label{eq:doubledAB}
\end{align}
with the factor system 
\begin{align}
	&z_{g,h} A^\bk_{gh} = 
	\left\{\begin{array}{ll}
		A^{p_h\bk}_gA^\bk_h&(c_h=1)\\
		B^{p_h\bk}_gA^\bk_h&(c_h=-1)\\
	\end{array}\right., \label{eq:NH_AB_1}\\
	&z_{g,h} B^\bk_{gh} = 
	\left\{\begin{array}{ll}
		B^{p_h\bk}_gB^\bk_h&(c_h=1)\\
		A^{p_h\bk}_gB^\bk_h&(c_h=-1)\\
	\end{array}\right. , \label{eq:NH_AB_2}
\end{align}
for $g,h \in G$. 
The unitarity of $\tilde U^\bk_g$ implies that $A^\bk_g$ and $B^\bk_g$ are also unitary. 
Substituting Eqs.~\eqref{eq:doubledHH} and \eqref{eq:doubledAB} into Eq.~\eqref{eq:H_sym}, we find that 
the symmetry relation for a non-Hermitian Hamiltonian $H_\bk$ is
\begin{align}
A^\bk_g H_\bk = 
\left\{\begin{array}{ll}
H_{p_g \bk} B^\bk_g &(c_g=1),\\ 
H_{p_g \bk}^\dag B^\bk_g &(c_g=-1). \\ 
\end{array}\right.
\label{eq:NH_sym}
\end{align}
Unlike Hermitian cases, $A^\bk_g = B^\bk_g$ does not hold but is subject to the relations in Eqs.~\eqref{eq:NH_AB_1} and \eqref{eq:NH_AB_2}. 
In particular, for order-two symmetries (i.e., $g^2=e$) such as $\Z_2$ onsite, reflection, two-fold rotation, and inversion symmetries, we have $A_g^{p_g\bk} B^\bk_g=z_{g,g}$ for $c_g=-1$. 

Unlike Hermitian systems, a basis change of the one-particle Hilbert space is a simultaneous transformation
\begin{align}
    &H_\bk \mapsto u H_\bk v^{-1}, \label{eq:NH_UT}\\
    &A_\bk \mapsto u A^\bk_g u^{-1},\quad B_g^\bk \mapsto v B_g^\bk v^{-1},\quad (c_g=1), \label{eq:NH_UT_c1}\\
    &A_\bk \mapsto v A^\bk_g u^{-1},\quad B_g^\bk \mapsto u B_g^\bk v^{-1},\quad (c_g=-1), \label{eq:NH_UT_c-1}
\end{align}
with $u$ and $v$ {\it independent} constant unitary matrices. 
It should be compared with Hermitian systems where $v = u$ always holds. 
If $v^{-1} u$ is not proportional to the identity matrix, the above transformation \eqref{eq:NH_UT} on the Hamiltonian generally changes the energy spectrum. 
Even when $H_\bk$ has a real- or imaginary-line gap, the transformed Hamiltonian $u H_\bk v^{-1}$ may not have a line gap, which is a unique feature of non-Hermitian systems. 
More importantly, the transformation \eqref{eq:NH_UT} does not change the point-gap topological invariants, but may change a line-gap topological invariant, which implies that the transformation \eqref{eq:NH_UT} may changes the topological boundary state. 
We will see in Sec.~\ref{sec:p4} an example where this is the case. 

Although the unitary matrices $A^\bk_g$ and $B^\bk_g$ with constraints in Eqs.~\eqref{eq:NH_AB_1} and \eqref{eq:NH_AB_2} satisfy the desired algebra \eqref{eq:cg_deg} in the doubled Hermitian systems, they may be meaningless symmetry. 
To see this, let us focus on a high-symmetry momentum $p_g\bk =\bk + {}^\exists \bG$ and elements with $c_g=1$, where the symmetry constraint in Eq.~\eqref{eq:NH_sym} is $A_g^\bk H_\bk = H_\bk B^\bk_g$. 
It means that if $H_\bk$ is invertible, then $A^\bk_g$ and $B^\bk_g$ belong to a common representation of the little group $G^0_\bk = \{g \in G | p_g \bk \equiv \bk + {}^\exists \bG, c_g=1\}$.
In other words, if $A^\bk_g$ and $B^\bk_g$ belong to different representations of $G^0_\bk$, the Hamiltonian $H_\bk$ should not be invertible. 
Namely, the point gap closes at $\bk$. 
Since we are interested in fully point-gapped Hamiltonians, at least in high-symmetry momenta, we further assume that there exists a {\it constant} unitary matrix $u_H$ such that $u_H$ satisfies the symmetry relation 
\begin{align}
    A_g^\bk u_H = \left\{
    \begin{array}{ll}
        u_H B_g^\bk & (c_g=1), \\
        u_H^\dag B^\bk_g & (c_g=-1),
    \end{array}\right.
\end{align}
i.e., the matrices $B^\bk_g$ are related with $A^\bk_g$ by 
\begin{align}
    B^\bk_g = \left\{
    \begin{array}{ll}
    u_H^\dag A^\bk_g u_H & (c_g=1),  \\
    u_H A^\bk_g u_H & (c_g=-1).
    \end{array} \right.
\end{align}
The constant unitary matrix $u_H$ plays the role of the atomic insulator in non-Hermitian systems. 
The trivial atomic insulator Hamiltonian in the doubled Hermitianized system is given by 
\begin{align}
    \tilde H^{\rm ai}_\bk 
    = \pm \begin{pmatrix}
        &u_H\\
        u_H^\dag
    \end{pmatrix}.
\end{align}
We refer to $u_H$ as the reference unitary matrix.

\subsection{On Hermitianization}
\label{sec:Hermitianization}
When $A_g^\bk = B_g^\bk$ holds for all $g \in G$, we find that if $H_\bk$ has a real- (imaginary-) line gap, $H_\bk$ can be deformed into an Hermitian Hamiltonian while keeping the line gap and symmetry \eqref{eq:NH_sym}, which can be shown by the same way as in Appendix D in Ref.~\onlinecite{Ashida_Gong_Ueda:NH_review}.
However, for generic non-Hermitian symmetry as $A^\bk_g \neq B^\bk_g$, the symmetry constraint in Eq.~\eqref{eq:NH_sym} may prevent the Hamiltonian $H_\bk$ from being Hermite with the line gap kept~\cite{Kawabata_Sato_Shiozaki:HigherOrder_nonHermitian_skin_effect}.

\subsection{Examples of non-Hermitian symmetry}
We provide a few examples of symmetries in non-Hermitian systems. 
We see that a $\Z$ invariant is indeed defined from the structure of symmetries, which is instructive even when the generic formula in Eq.~\eqref{eq:ZinvHH} is available. 
We note that the following relation holds:
\begin{align}
A^{p_g^2\bk}_g B_g^{p_g\bk} A_g^\bk H_\bk=A_g^{p_g^2\bk} H_{p_g^2\bk}A_g^{p_g\bk} B_g^\bk
\label{eq:ABAH}
\end{align}
for $c_g=-1$. 
At high-symmetry points satisfying $p_g\bk = \bk + {}^\exists \bG$, Eq.~\eqref{eq:ABAH} reads 
\begin{align}
[A^{\bk}_g B_g^{\bk}, A_g^\bk H_\bk]=0. 
\label{eq:ABAH_2}
\end{align}
This implies that the Hamiltonian $H_\bk$ is block-diagonalized by the unitary matrix $A_g^\bk B_g^\bk$.

\subsubsection{$\Z_2$ symmetry}
As a simple example of nontrivial symmetries in non-Hermitian systems, let us consider the following $\Z_2$ onsite symmetry 
\begin{align}
&AH=H^\dag B, \\
&AB=1, 
\end{align}
with $A,B$ unitary matrices. 
This is equivalent to $(AH)^\dag = AH$. 
If there is a point gap $\det H\neq 0$, the Hermitian matrix $AH$ has a finite energy gap. Then, we can define a $\Z$ invariant $N[AH]$ as the number of negative eigenvalues of $AH$ that is stable unless the point gap of $H$ closes. 
The corresponding symmetry of the doubled Hamiltonian \eqref{eq:doubledHH} is $\tilde U \tilde H \tilde U^{-1} = \tilde H$ with 
\begin{align}
    \tilde U = \begin{pmatrix}
        &B\\
        A\\
    \end{pmatrix}.
    \label{eq:tU_AB}
\end{align}
The equality $N^1[\tilde H] = N[AH]$ holds, where $N^1[\tilde H]$ is the number of negative eigenstates of $\tilde H$ with $\tilde U=1$. 

\subsubsection{$\Z_4$ symmetry}
Let us discuss the following $\Z_4$ symmetry 
\begin{align}
    AH=H^\dag B,\quad 
(AB)^2=-1. 
\end{align}
From Eq.~\eqref{eq:ABAH_2}, the matrix $AH$ splits into sectors of $AB=\pm i$. 
For $AB=\pm i$, $(e^{\mp \frac{\pi}{4}i} AH)^\dag=e^{\mp \frac{\pi}{4}i}AH$ and $N[e^{\mp \frac{\pi}{4}i}AH|_{AB=\pm i}]$, the numbers of negative eigenvalues of $e^{\mp \frac{\pi}{4}i}AH$, are $\Z$ invariants. 

One can directly connect eigenstates of $AH$ with ones of the doubled Hermitianized Hamiltonian \eqref{eq:doubledHH} with $\tilde U$ defined by \eqref{eq:tU_AB}. 
Suppose that $\ket{\psi}$ belongs to the sector $AB=i$ and is an eigenstate $e^{-i\frac{\pi}{4}} AH\ket{\psi} =\la \ket{\psi}$ with $\la \in \R$. 
Then, we find that $\ket{\tilde \psi} := (e^{-\frac{\pi}{4}i}B \ket{\psi},\ket{\psi})^T$ satisfies $\tilde H \ket{\tilde \psi} = \la \ket{\tilde \psi}$ and $\tilde U \ket{\tilde \psi} = e^{\frac{\pi}{4}i} \ket{\tilde \psi}$. 
Similarly, if $\ket{\psi}$ belongs to the sector $AB=-i$ and is an eigenstate $e^{\frac{\pi}{4}i}AH= \la \ket{\psi}$ with $\la \in \R$, then  $\ket{\tilde \psi} := ( e^{\frac{\pi}{4}i}B \ket{\psi},\ket{\psi})^T$ satisfies $\tilde H \ket{\tilde \psi} = \la \ket{\tilde \psi}$ and $\tilde U \ket{\tilde \psi} = e^{-\frac{\pi}{4}i} \ket{\tilde \psi}$. 
Therefore, 
\begin{align}
    &N^{\tilde U=e^{\frac{\pi}{4}i}}[\tilde H] = N[e^{-\frac{\pi}{4}i}AH|_{AB=i}], \label{eq:AB1}\\
    &N^{\tilde U=e^{-\frac{\pi}{4}i}}[\tilde H] = N[e^{\frac{\pi}{4}i}AH|_{AB=-i}].\label{eq:AB2}
\end{align}

\subsection{Symmetry indicator in non-Hermitian systems}
Given a non-Hermitian Hamiltonian $H_\bk$ with symmetries \eqref{eq:NH_sym}, one can define the doubled Hermitianized Hamiltonian \eqref{eq:doubledHH} with symmetries \eqref{eq:doubledAB}. 
One can straightforwardly apply the symmetry indicator in Hermitian chiral symmetric systems introduced in Sec.~\ref{sec:SI_H} to non-Hermitian systems to get the symmetry indicator in non-Hermitian systems.

If a non-Hermitian Hamiltonian $H_\bk$ has a nontrivial value of symmetry indices, then either one of the followings is true: 
(i) $H_\bk$ has a {\it point-gap gapless point} with codimension two or more where the point gap closes. 
(ii) $H_\bk$ has a finite point gap in the entire BZ, and it can not be deformed into a constant Hamiltonian while keeping symmetry \eqref{eq:NH_sym}. 
Unlike Hermitian cases, the physical implication of nontrivial symmetry indices for the case (ii), namely symmetry indices for fully point-gapped Hamiltonians, is more subtle. 
A nontrivial symmetry index implies a topological boundary gapless state as Hermitian systems, or a non-Hermitian skin effect if the Hamiltonian $H_\bk$ is put on the open boundary condition. 
These two physically different phenomena are interchanged by closing the line gap with keeping the point gap. 
Therefore, the symmetry indicator is sometimes insufficient to fully characterize the topological boundary states, which is a common feature of the symmetry indicator in Hermitian systems.

\section{Case studies}
\label{sec:ex}
This section discusses two examples to illustrate the use of the symmetry indicator in non-Hermitian systems.

\subsection{$p4$ with $B$ representation}
\label{sec:p4}
Let us consider the following four-fold rotation-type symmetry in two-dimensional systems~\cite{Kawabata_Sato_Shiozaki:HigherOrder_nonHermitian_skin_effect}
\begin{align}
    &A^\bk H_\bk = H_{c_4\bk}^\dag B^\bk,
    \label{eq:NH_C4} \\
    &A^{-c_4\bk}B^{-\bk}A^{c_4\bk}B^\bk = -1.
\end{align}
Here, $\bk = (k_x,k_y)$ and $c_4\bk = (-k_y,k_x)$. 
The corresponding Hermitianized Hamiltonian \eqref{eq:doubledHH} and symmetry representations \eqref{eq:doubledAB} satisfy
\begin{align}
    &\tilde U^\bk \tilde H_\bk [\tilde U^\bk]^{-1} = H_{c_4\bk}, \\
    &\tilde U^{-c_4\bk} \tilde U^{-\bk} \tilde U^{c_4\bk} \tilde U^{\bk}=-1, \\
    &\tilde \G \tilde U^\bk = - \tilde U^\bk \tilde \G. 
\end{align}
At high-symmetry points $\G = (0,0)$ and $M = (\pi,\pi)$, there are four irreps labeled by $\alpha \in \{1,3,5,7\}$ with the eigenvalue $\tilde U^\bk = e^{i\frac{\alpha}{4}\pi}$ for $\bk \in \{\G,M\}$, and the chiral operator $\tilde \G$ changes the sign of $\tilde U^\bk$ and the energy eigenvalue. 
As a result, the effective Altland-Zirnbauer (EAZ) symmetry classes at $\G$ and $M$ are class A for each pair $(\alpha=1,\alpha=5), (\alpha=7,\alpha=3)$ of irreps, implying that we have $\Z^{\times 4}$ invariant in total. 
For a given pair $(\tilde H^0_\bk,\tilde H^1_\bk)$ of Hamiltonians, the $\Z$ invariants are defined by 
\begin{align}
    &N_\bk^\alpha[\tilde H^0_\bk,\tilde H^1_\bk]= N_\bk^\alpha[\tilde H^0_\bk]-N_\bk^\alpha[\tilde H^1_\bk],\\
    &\bk \in \{\G,M\},\quad \alpha \in \{1,7\}.
\end{align}
Here, $N_\bk^\alpha[\tilde H_\bk]$ is the number of negative energy eigenstates of the irrep $U^\bk= e^{i\frac{\alpha}{4}\pi}$. 
Because the chiral symmetry is defined on each irrep at other points, there are no other topological invariants.
The $\Z$ invariants $N_\bk^\alpha[\tilde H^0_\bk, \tilde H^1_\bk]$ with $\bk \in \{\G,M\}$ and $\alpha\in\{1,7\}$ span the lattice $E_1^{0,0} = \Z^{\times 4}$. 
Since there is no compatibility relation in this symmetry setting, the lattice of band structures that satisfy all compatibility relations, denoted by $E_2^{0,0}$, is the completely same as $E_1^{0,0}=\Z^{\times 4}$. 

Next, we compute the lattice of atomic insulators denoted by $K_{\rm ai}$ in Sec.~\ref{sec:SI_H}. 
The set of atomic insulators is generated by localized orbitals at Wyckoff positions ${\sf a}=(0,0)$ and ${\sf b} = (\frac{1}{2},\frac{1}{2})$ in unit cell. 
We have four independent atomic insulators labeled by $({\sf a}, \alpha=1)$, $({\sf a}, \alpha=7)$, $({\sf b}, \alpha=1)$, and $({\sf b}, \alpha=7)$, respectively, with a localized orbital at the Wyckoff position ${\sf a}$ or ${\sf b}$ with an irrep $\alpha=1$ or $\alpha=7$. 
According to the construction of the $k$-space Hamiltonian illustrated in Sec.~\ref{sec:SI_H}, corresponding Hamiltonians are written as  
\begin{align}
    \tilde H^{\rm ai}_\bk = \begin{pmatrix}
        -1\\
        &1\\
    \end{pmatrix}
\end{align}
with the symmetry representations
\begin{align}
    &U_{({\sf a},\alpha=1)}^{\bk} = \begin{pmatrix}
        e^{\frac{\pi}{4}i}\\
        &-e^{\frac{\pi}{4}i}\\
    \end{pmatrix}, \\
    &U_{({\sf a},\alpha=7)}^{\bk} = \begin{pmatrix}
        e^{\frac{7\pi}{4}i}\\
        &-e^{\frac{7\pi}{4}i}\\
    \end{pmatrix}, \\
    &U_{({\sf b},\alpha=1)}^{\bk} = \begin{pmatrix}
        e^{\frac{\pi}{4}i}e^{-ik_y}\\
        &-e^{\frac{\pi}{4}i}e^{-ik_y}\\
    \end{pmatrix}, \\
    &U_{({\sf b},\alpha=7)}^{\bk} = \begin{pmatrix}
        e^{\frac{7\pi}{4}i}e^{-ik_y}\\
        &-e^{\frac{7\pi}{4}i}e^{-ik_y}\\
    \end{pmatrix},
\end{align}
in the basis so that 
\begin{align}
    \tilde \G = \begin{pmatrix}
        &1\\
        1\\
    \end{pmatrix}.
\end{align}
They have the $\Z^{\times 4}$ invariants $N_\bk^\alpha[H^{\rm ai}_\bk,-H^{\rm ai}_\bk]$ as follows. 
\begin{align}
    \begin{array}{l|cccc}
         & N_\G^{\alpha=1}&N_\G^{\alpha=7}&N_M^{\alpha=1}&N_M^{\alpha=7} \\
         \hline
        ({\sf a},\alpha=1) & 1&0&1&0\\
        ({\sf a},\alpha=7) & 0&1&0&1\\
        ({\sf b},\alpha=1) & 1&0&-1&0\\
        ({\sf b},\alpha=7) & 0&1&0&-1\\
    \end{array}. 
\end{align}
Therefore, $K_{\rm ai}$ is spanned by vectors $\{ (1,0,1,0), (0,1,0,1), (1,0,-1,0), (0,1,0,-1)\}$ in $E_2^{0,0}$. 
We have the symmetry indicator group $Q = E_2^{0,0}/K_{\rm ai} = \Z_2^{\times 2}$ (Layer group number 49 with $B$ representation. See Appendix~\ref{tab:Layer_Spinful}). 
The two $\Z_2$ indices are given as
\begin{align}
    &\mu_{C_4} = 
    N_\G^{\alpha=1}[\tilde H_\G,\tilde H^{\rm ai}_\G]
    +N_\G^{\alpha=7}[\tilde H_\G,\tilde H^{\rm ai}_\G] \nonumber \\
    &\quad
    +N_M^{\alpha=1}[\tilde H_M,\tilde H^{\rm ai}_M]
    +N_M^{\alpha=7}[\tilde H_M,\tilde H^{\rm ai}_M] \quad {\rm mod}\ 2, \\
    &z_{C_4} = N_\G^{\alpha=1}[\tilde H_\G,\tilde H^{\rm ai}_\G]+N_M^{\alpha=1}[\tilde H_M,\tilde H^{\rm ai}_M]\quad {\rm mod}\ 2.
\end{align}
(Recall that $\tilde H^{\rm ai}_\bk$ is the atomic insulator Hamiltonian on which the Hamiltonian $\tilde H_\bk$ defined.)
The index $\mu_{C_4}$ detects a gapless point since near a high-symmetry point $\G$ or $M$, the following Hamiltonian creates a gapless point in generic momenta
\begin{align}
    &\tilde H_\bk = (k_x^2+k_y^2-\mu)\s_z + k_xk_y \s_y,\\
    &\tilde \G=\s_x,\quad
    \tilde U^\bk = e^{\frac{\pi}{4}i}\s_z {\rm\ or\ } e^{\frac{7\pi}{4}i} \s_z, 
\end{align}
with changing $\mu_{C_4}$ by 1 when the chemical potential $\mu$ passes zero.
Under the condition that $\mu_{C_4} \equiv 0$, the second index $z_{C_4}$ detects the second-order topological insulator for Hermitianized chiral systems. 

It turns out that the index $z_{C_4}$ is the same as the $\Z_2$ quantized Wess-Zumino (WZ) term of non-Hermitian Hamiltonian $H_\bk$~\cite{Kawabata_Sato_Shiozaki:HigherOrder_nonHermitian_skin_effect},
\begin{align}
    {\rm WZ}[H_\bk]
    :=\frac{1}{24\pi^2} \int_{T^2 \times [0,1]} \tr[H_{\bk,t}^{-1} d H_{\bk,t}]^3 \in \R/\Z. 
\end{align}
Here, $H_{\bk,t}$ for $t \in [0,1]$ is an extension of $H_\bk$ to a constant Hamiltonian ({\it without} keeping the symmetry \eqref{eq:NH_C4}). 
In the presence of the four-fold rotation-like symmetry in Eq.~\eqref{eq:NH_C4}, ${\rm WZ}[H_\bk]$ is quantized to $\{0,\frac{1}{2}\}$. 
Comparing explicit generator models, we find that 
\begin{align}
2 {\rm WZ}[H_\bk] \equiv z_{C_4} \quad{\rm mod}\  2.     
\end{align}
As discussed in Ref.~\onlinecite{Kawabata_Sato_Shiozaki:HigherOrder_nonHermitian_skin_effect}, ${\rm WZ}[H_\bk]=\frac{1}{2}$ yields a 2nd-order skin effect in the absence of a gapless edge state. 
In the rest of this section, we present examples of indicators for gapless and gapped phases, respectively. 

\subsubsection{$\mu_{C_4}$}
Let us consider the four-fold rotation-like symmetry 
\begin{align}
    H_\bk = i H_{c_4\bk}^\dag.
\end{align}
This is equivalent to the case with $A^\bk=1, B^\bk=i$. 
We find a reference unitary matrix $u_H=e^{\frac{\pi}{4}i}$.
At $\bk=\G$ and $M$, $e^{-\frac{\pi}{4}i} H$ is a Hermitian matrix so that the numbers $N[e^{-\frac{\pi}{4}i}H_{\bk \in \{\G,M\}}]$ of negative energy eigenstates of $e^{-\frac{\pi}{4}i} H$ are $\Z$ invariants unless the point gap closes. 
By using relations Eqs.~\eqref{eq:AB1} and \eqref{eq:AB2}, the index $\mu_{C_4}$ is recast as 
\begin{align}
    \mu_{C_4}= N[e^{-\frac{\pi}{4}i}H_\G] + N[e^{-\frac{\pi}{4}i}H_M] \quad {\rm mod\ }2.
\end{align}
(The $\bk$-independence of both $A^\bk$ and $B^\bk$ implies that $N_\G^\alpha[\tilde H^{\rm ai}_\G] = N_M^\alpha[\tilde H^{\rm ai}_M]$.)
Indeed, $\mu_{C_4}$ indicates the existence of a gapless point (in the sense of the point gap) with a winding number in generic momenta.

For example, consider the following $1 \times 1$ Hamiltonian 
\begin{align}
    H_\bk = e^{\frac{\pi}{4}i}(t \cos k_x + t \cos k_y - \mu + i \psi_\bk). 
\end{align}
Here, $\psi_\bk$ is a function satisfying $\psi_{c_4 \bk} = -\psi_{\bk}$. 
If $2t-\mu$ and $-2t-\mu$ have opposite signs, $\mu_{C_4}=1$ and there should be points $\bk_*$ with a winding number $w_1 = \frac{1}{2\pi i} \oint_{|\delta \bk|=\epsilon} d \log \det H_{\bk_*+\delta \bk} = \pm 1$ around a small circle enclosing the point $\bk_*$ in a four-fold symmetric manner.

\subsubsection{$z_{C_4}$}
\label{sec:zc4}
The next example is the four-fold rotation-like symmetry discussed in Ref.~\onlinecite{Kawabata_Sato_Shiozaki:HigherOrder_nonHermitian_skin_effect}, 
\begin{align}
    &A^\bk H_\bk=H_{c_4\bk}^\dag B^\bk, \\
    &A^\bk =1, \quad B^\bk=-i\s_y.
    \label{eq:BBH_sym}
\end{align}
We find a reference unitary matrix $u_H=e^{-\frac{\pi}{4}i\s_y}$.
It was shown that under this symmetry, the Wess-Zumino term ${\rm WZ}[H_\bk]$ is $\Z_2$-quantized and the nontrivial WZ term relates the 2nd-order skin effect that exhibits the skin modes localized at corners for the full open boundary condition. 
By using relations Eqs.~\eqref{eq:AB1} and  \eqref{eq:AB2}, the indices  are recast into
\begin{align}
    &\mu_{C_4} = N[e^{-\frac{\pi}{4}i}H_\G|_{\s_y=-1}]+N[e^{\frac{\pi}{4}i}H_\G|_{\s_y=1}]\nonumber \\
    &\quad +N[e^{-\frac{\pi}{4}i} H_M|_{\s_y=-1}]+N[e^{\frac{\pi}{4}i} H_M|_{\s_y=1}]\quad {\rm mod}\ 2, \\
    &z_{C_4} = N[e^{-\frac{\pi}{4}i}H_\G|_{\s_y=-1}]+N[e^{-\frac{\pi}{4}i} H_M|_{\s_y=-1}]\quad {\rm mod}\ 2.
\end{align}

Let us discuss a model with nearest-neighbor hopping given by
\begin{align}
    H_\bk
    &=\la_0 \sin k_x \s_z + \la_0 \sin k_y \s_x \nonumber\\
    &-\frac{i}{\sqrt{2}} (2\g+\la_1 \cos k_x+\la_1\cos k_y ) e^{\frac{\pi}{4}i \s_y} \nonumber \\
    &-\frac{i\la_2}{\sqrt{2}} (\cos k_x-\cos k_y) e^{-\frac{\pi}{4}i\s_y}, 
    \label{eq:NH_BBH}
\end{align}
where $\g,\la_0,\la_1$, and $\la_2$ are real parameters, and $\s_\mu (\mu=x,y,z)$ are Pauli matrices. 
For $\la_0=\la_1=\la_2$, the model \eqref{eq:NH_BBH} is the the non-Hermitian counterpart of the Benalcazar-Bernevig-Hughes (BBH)
model~\cite{Benalcazar_Bernevig_Hughes:Quantized_electric_multipole_insulators}.
When $\la_0 \neq 0$, the point gap closes at $|\g|=|\la_1|$. 
The first index $\mu_{C_4}$ vanishes identically. 
One can see that the second index is  $z_{C_4}=1$ if $|\g|<|\la_1|$, and for the full open boundary condition, the model \eqref{eq:NH_BBH} exhibits the 2nd-order skin effect for $\la_0=\la_1=\la_2$, where the skin modes appear at the corners~\cite{Kawabata_Sato_Shiozaki:HigherOrder_nonHermitian_skin_effect}.

As noticed in Ref.~\onlinecite{Kawabata_Sato_Shiozaki:HigherOrder_nonHermitian_skin_effect}, the nontrivial WZ term, or the index $z_{C_4}$, does not always imply the 2nd-order skin effect. 
This can be well illustrated by introducing the Chern number ${\rm Ch}$ associated with the real-line gap. 
If the real-line gap is open, one can define the Chern number by 
\begin{align}
    {\rm Ch}
    =\frac{1}{24\pi^2} \int_{[0,2\pi]^2\times [-\infty,\infty]} 
    \tr [GdG^{-1}]^3, 
\end{align}
with $G(\omega,\bk)^{-1} = i\omega-H_\bk$ the inverse of Green's function. 
When the real-line gap closes, the Chern number ${\rm Ch}$ can change with the point gap opened. 

\begin{figure}[t]
\centering
\includegraphics[width=\linewidth, trim=0cm 1cm 6cm 1cm]{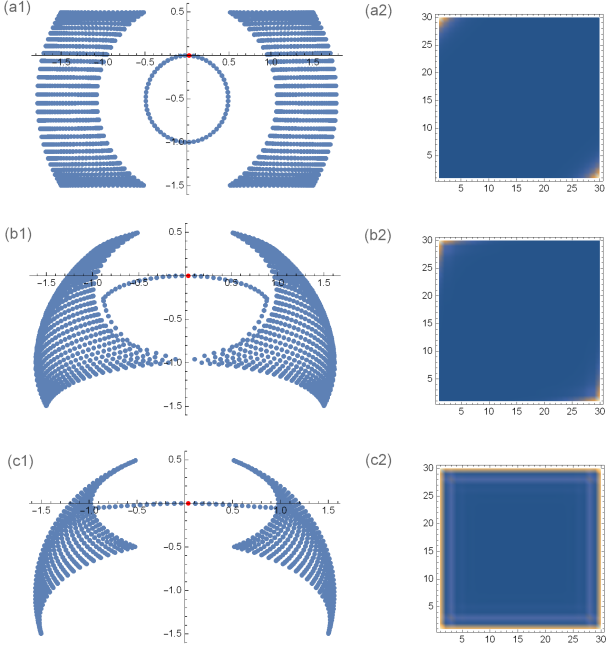}
\caption{
A phase transition between the 2nd-order non-Hermitian skin effect and a Chern insulator. 
The complex spectra of the model \eqref{eq:NH_BBH} are shown for 30 $\times$ 30 sites in the full open boundary condition. 
The parameters are given as $\la_0=\la_1=1.0, \g=0.5$, as well as (a1, a2) $\la_2=1.0$, (b1, b2) $\la_2=0.5$, or (c1, c2) $\la_2=0$. 
Each (a2), (b2), and (c2) show the densities of right eigenvectors of the eigenvalues specified by the red dots in (a1), (b1), and (c1), respectively.  
}
\label{fig1}
\end{figure}

Figure~\ref{fig1} shows the energy spectra with $\la_0=\la_1=1.0$ and $\g=0.5$ for different $\la_2$s in the open boundary conditions for $x$ and $y$ directions. 
At $\la_2=1.0$ (Fig.~\ref{fig1} (a1) and (a2)), ${\rm Ch}=0$ and the corner skin modes appear and they form a closed loop detached from the bulk spectrum. 
The line gap closes near $\la_1=0.5$ (Fig.~\ref{fig1} (b1)), and at the same time a part of the skin modes are absorbed by the bulk. 
At $\la_2=0$ (Fig.~\ref{fig1} (c1) and (c2)), ${\rm Ch}=1$ and an edge mode appears and they form an interval and bridge the bulk spectrum. 
We emphasize that while $\la_2$ is varying, the index $z_{C_4}$ unchanged. 

\begin{figure}[t]
\centering
\includegraphics[width=\linewidth, trim=0cm 3cm 3cm 1cm]{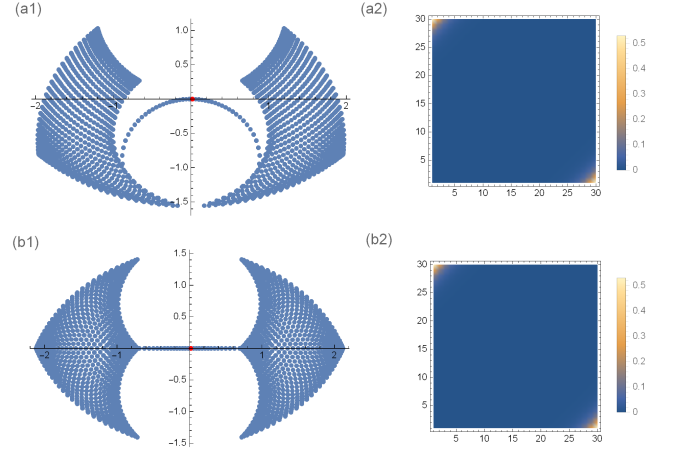}
\caption{
The complex spectra of the model in Eq.~\eqref{eq:BBH_tr1} are shown for 30 $\times$ 30 sites in the full open boundary condition.
The parameters are given as $\la_0=\la_1=\la_2=1.0, \g=0.5$, as well as (a1, a2) $\theta=\frac{\pi}{8}$, or (b1, b2) $\theta=\frac{\pi}{4}$. 
(a2) and (b2) show the densities of right eigenvectors of the eigenvalues specified by the red dots in (a1) and (b1), respectively.  
}
\label{fig2}
\end{figure}

It is instructive to consider how the spectrum changes against a continuous deformation of basis defended by Eqs.~\eqref{eq:NH_UT} and \eqref{eq:NH_UT_c-1}. 
Let us consider the following simultaneous transformation parametrized by $\theta$, 
\begin{align}
    H_{\bk,\theta} = H_\bk e^{i\s_y \theta}, 
    \label{eq:BBH_tr1}
\end{align}
\begin{align}
    A^{\bk,\theta} = e^{-i\s_y\theta},\quad
    B^{\bk,\theta}=e^{i\s_y(\theta-\frac{\pi}{2})}.
    \label{eq:BBH_tr2}
\end{align}
This does not affect the index $z_{C_4}$, since the simultaneous transformation in Eqs.~\eqref{eq:BBH_tr1} and \eqref{eq:BBH_tr2} is a unitary transformation for the doubled Hermitianized system. 
However, since \eqref{eq:BBH_tr1} is not a similarity transformation, the spectrum of the Hamiltonian $H_{\bk,\theta}$ changes, and it may exhibit different topological boundary modes as $\theta$ varies. 
Figure~\ref{fig2} shows the energy spectra for $\theta=\frac{\pi}{8}$ (Fig.~\ref{fig2} (a1, a2)) and $\theta=\frac{\pi}{4}$ (Fig.~\ref{fig2} (b1, b2)) in the full open boundary condition. 
The other parameters are fixed as $\la_0=\la_1=\la_2=1.0$ and $\g=0.5$.
We find that ${\rm Ch}=0$ at $\theta=0$ (Fig.~\ref{fig1} (a1, a2)), and ${\rm Ch}=1$ at $\theta=\frac{\pi}{4}$.
Near $\theta=\frac{\pi}{8}$ (Fig.~\ref{fig2} (a1, a2)) the line gap closes, and the Chern number is changed by one. 
We emphasize again that while the index $z_{C_4}$ is unchanged, the Chern number is changed. 
Let us focus on $\theta=\frac{\pi}{4}$, where ${\rm Ch}=1$. 
Remarkably, even when the system is in the real-line gap topological phase of the Chern insulator with ${\rm Ch}=1$, the mid-gap state is localized at the corners as shown in Fig.~\ref{fig2} (b2), due to a non-Hermitian term, like the 2nd-order skin effect. 

The parameter $\theta = \frac{\pi}{4}$ is a particular point, where $A^{\bk,\theta=\frac{\pi}{4}}= B^{\bk,\theta=\frac{\pi}{4}}$ holds, and in such a case, we have the equality ${\rm Ch} \equiv 2 {\rm WZ}$.~\cite{Kawabata_Sato_Shiozaki:HigherOrder_nonHermitian_skin_effect} 
The Hamiltonian reads 
\begin{align}
    H_{\bk,\theta=\frac{\pi}{4}}
    &=
    \la \sin k_x \frac{\s_x+\s_z}{\sqrt{2}}
    +\la \sin k_y \frac{\s_x-\s_z}{\sqrt{2}}\nonumber\\
    &+ 
    \frac{1}{\sqrt{2}}(2\g +\la_1 \cos k_x+\la_1\cos k_y)\s_y\nonumber\\
    &
    -\frac{i\la_2}{\sqrt{2}} (\cos k_x-\cos k_y).
\end{align}
This can be viewed as a model of a  Chern insulator with non-Hermite perturbation. 
The imaginary term $\la_2$ can be turned off without closing the point gap, resulting in a Hermitian Chern insulator.

\subsection{$P\bar 1$, $A_u$ representation}
\label{sec:P-1}
Inversion symmetry $I$ in non-Hermitian Hamiltonian with $c_I = -1$ (the $A_u$ representation) is written as
\begin{align}
    A^\bk H_\bk = H_{-\bk}^\dag B^\bk,
    \quad
    A^{-\bk}B^\bk=1.
    \label{eq:inv_sym}
\end{align}
At high-symmetry momenta $\bk$ satisfying $-\bk\equiv \bk+{}^\exists \bG$, $A^\bk H_\bk$ is a Hermitian matrix, and thus the numbers of negative eigenvalues of $A^\bk H_\bk$ are denoted by $N[A^\bk H_\bk]$. 
For three space dimensions with primitive lattice, the symmetry indicator group is $Q = \Z_8 \times \Z_4^{\times 3} \times \Z_2^{\times 3}$. 
(Space group number 2 with $A_u$ representation. See Appendix~\ref{tab:Space}). 
In Appendix~\ref{app:p1bar}, we present the index formulas for the doubled Hermitianized Hamiltonian, which are written in the non-Hermitian Hamiltonian as 
\begin{align}
    &z_{2,I,\mu \in \{x,y,z\}} \nonumber\\
    &= \sum_{k_{\nu \neq \mu}=\pi,k_\mu\in\{0,\pi\}} (N[A^\bk H_\bk]-N[A^\bk u_H]) \quad {\rm mod\ }2, \\
    &z_{4,I,\mu \in \{x,y,z\}} \nonumber\\
    &= \sum_{k_\mu=\pi,k_{\nu \neq \mu}\in\{0,\pi\}} (N[A^\bk H_\bk]-N[A^\bk u_H]) \quad {\rm mod\ }4, \\
    &z_{8,I} = \sum_{k_x,k_y,k_z \in \{0,\pi\}} (N[A^\bk H_\bk]-N[A^\bk u_H]) \quad {\rm mod\ }8.
\end{align}
Here, $u_H$ is a reference unitary matrix. 
A part of roles of indices was discussed in the literature~\cite{Okugawa_Takahashi_Yokomizo:second_order_nonHermitian_skin_effects,Okugawa_Takahashi_Yokomizo:generalized_inversion_symmetry,Vecsei_Denner_Neupert_Schindler:Symmetry_indicator_nonHermitian}.
The index $z_{2,I,\mu} \equiv 1$ implies the 1st-order skin effect in one-dimension.
$z_{4,I,\mu}\equiv 1$ mod 2 corresponds to the existence of a pair of inversion symmetric exceptional lines.
In the same way as Sec.~\ref{sec:p4}, one can show that $z_{4,\mu}  \equiv 2$ indicates either a 2nd-order skin effect or a Chern insulator in two dimensions. 
For three dimensions, $z_{8,I} \equiv 1$ mod 2 indicates the existence of a single exceptional ring surrounding an inversion symmetric point. 
For a fully point-gapped non-Hermitian Hamiltonian, the relation $2 z_{8,I} \equiv W_3$ mod 2 holds, where $W_3 = \frac{1}{24\pi^2} \int \tr[H_\bk^{-1}dH_\bk]^3$ is the three-dimensional winding number.
Therefore, $z_{8,I}\equiv 2$ mod 4 means a 3D intrinsic non-Hermitian topological phase~\cite{Okuma_Kawabata_Shiozaki_Sato:NonHermitian_skin_effect} that was discussed to exhibit the chiral magnetic effect~\cite{Bessho_Sato:topological_duality, Kawabata_Shiozaki_Ryu:TFT_nonHermitian_systems} and host the boundary mode with a single exceptional point~\cite{Denner_etal:exceptional_topological_insulators}.

\begin{figure}[t]
\centering
\includegraphics[width=\linewidth, trim=1cm 0cm 1cm 0cm]{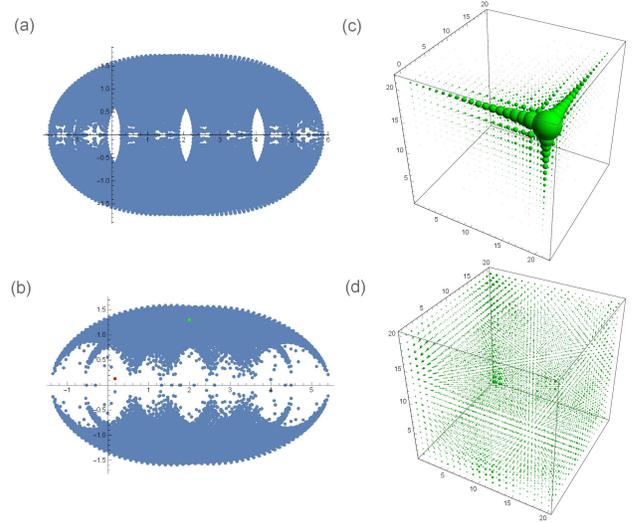}
\caption{
The complex spectra and density plots of right eigenstates of Eq.~\eqref{eq:NH_3rd_TI}. 
The parameters are given as $\la_{x4}=0.32, \la_{y4}=0.03, \la_{z4}=-0.26, \la_{x5}=0.04, \la_{y5}=0.3, \la_{z5}=0.25, \eta_x=0.05, \eta_y=-0.04, \eta_z=0.02$~\cite{Dai_Wang_Yang_Zhao:Takagi_Topological_Insulator}.
(a) The complex spectrum for $50 \times 50 \times 50$ sites with the periodic boundary conditions for $x,y,z$ directions. 
(b) The complex spectrum for $20 \times 20 \times 20$ sites with full open boundary condition. 
(c) and (d) show the densities of right eigenstates of the eigenvalues specified by the red and green dots in (b), respectively.
}
\label{fig3}
\end{figure}

Let us see the physical implication of $z_{8,I} \equiv 4$ in more detail, which has not been discussed in the literature. 
For the doubled Hermitianized system, the phase with $z_{8,I} \equiv 4$ turns out to be the 3rd-order topological phases with chiral zero modes localized at the corners~\cite{Khalaf:Higher-order, Trifunovic_Brouwer:Higher-Order}.
We can expect the corresponding non-Hermitian model exhibits the 3rd-order skin effect. 
Rewriting the cubic lattice model discussed in Ref.~\onlinecite{Dai_Wang_Yang_Zhao:Takagi_Topological_Insulator} in terms of the non-Hermitian Hamiltonian, we have the following $4 \times 4$ model 
\begin{align}
    &H_\bk=H_\bk^0+V_\bk,
    \label{eq:NH_3rd_TI}\\
    &H_\bk^0=
    -it\sum_{i=x,y,z} \sin k_i \tau_x \s_i
    +m-\la \sum_{i=x,y,z} \cos k_i, \\
    &V_\bk= 
    \sum_{i=x,y,z}\{ \la_{i 4} (-\tau_z\s_i)
    +\la_{i5} (\tau_y\s_i)
    +\eta_i (-\tau_x \s_i)\}.
\end{align}
Here, $t,m,\la$, $\la_{i4}$s, $\la_{i5}$s, and $\eta_i$s are real parameters, and $\tau_i,\s_i (i=x,y,z)$ are Pauli matrices. 
The model in Eq.~\eqref{eq:NH_3rd_TI} has the inversion symmetry with $A^\bk=B^\bk=1$ and a reference unitary matrix is given by $u_H=1$. 
In the following, we set $t=\la=1$. 
The parameter $m$ plays the role of the reference energy. 
For the unperturbed Hamiltonian $H_\bk^0$, we find that the index $z_{8,I}$ takes values as $z_{8,I} \equiv 4$ for $1<|m|<3$, and $z_{8,I} \equiv 0$ for $|m|<1$. 
In order to have corner-localized skin modes, we need appropriate perturbation.
The perturbation $V_\bk$ is designed to induce the mass in the boundary mode for the corresponding doubled Hermitianized system~\cite{Dai_Wang_Yang_Zhao:Takagi_Topological_Insulator}.
We employ the same parameters as in Ref.~\onlinecite{Dai_Wang_Yang_Zhao:Takagi_Topological_Insulator} (See the caption of Fig.~\ref{fig3} for the detail). 
Figure~\ref{fig3} (a) shows the complex spectrum with the full periodic boundary condition. In the spectrum, a finite point gap is open around the reference energies $E=0, 2, 4$. 
Figure~\ref{fig3} (b) shows the complex spectrum with the full open boundary condition for $20 \times 20 \times 20$ sites. 
One can see a finite set of eigenvalues that seems to be outside the bulk spectrum around $E=0,4$. 
The distribution of a right eigenstate picked from this set is shown in Fig.~\ref{fig3} (c), where one can find that the eigenstate is localized at a corner. 
On the one hand, Fig.~\ref{fig3} (d) shows the distribution of a right eigenstate of the bulk spectrum. As shown in the figure, the states are delocalized. 
These features are consistent with the 3rd-order skin effect. 

Similar to the $z_{C_4}$ index in Sec.~\ref{sec:zc4}, if the real-line gap open, the $z_8\equiv 4$ phase may be understood as a Hermitian 2nd-order topological phase. 
To see this, let us consider the following simultaneous transformation 
\begin{align}
    &H'^{0}_\bk = H^0_\bk \tau_y, \nonumber \\
    &\quad
    =\sum_{\mu=x,y,z} \sin k_\mu \s_\mu \tau_z+(m-\sum_{\mu = x,y,z} \cos k_\mu)\tau_y, \\
    &A'^\bk = \tau_y A^\bk = \tau_y, \quad
    B'^\bk = B^\bk \tau_y^\dag = \tau_y.
\end{align}
This transformation does not affect the indices. 
Remarkably, $A'_\bk=B'_\bk$ holds. 
In such a case, as noticed in Sec.\ref{sec:Hermitianization}, if a real-line gap open, a non-Hermitian Hamiltonian can be deformed into a Hermitian one while keeping symmetry and real-line gap. 
In fact, the model $H'^0_\bk$ is Hermitian and has an energy gap unless $m=-3,-1,1,3$. 
The model $H'^0_\bk$ is known to be a 2nd-order topological insulator with inversion symmetry, which shows a chiral hinge mode with suitable symmetry-preserving perturbation.

\section{Conclusion and outlook}
\label{sec:con}

In this paper, we developed a theory of the symmetry indicator for non-Hermitian systems, especially for detecting topological phases under the point-gap condition, and computed the symmetry indicator groups. 
Our strategy is based on the fact that the point-gap condition in non-Hermitian systems corresponds to the usual gap condition in the corresponding doubled Hermitian Hamiltonian (Eq.\eqref{eq:doubledHH}) with chiral symmetry. 
The commutative or anticommutative relation between the space group symmetry operator and the chiral operator is specified by a sign representation of the point group. 
The corresponding symmetry in a non-Hermitian system is a kind of space group symmetry involving the Hermitian conjugate for a nontrivial sign representation. 
Under such a generalized space group symmetry, the symmetry indicators are defined. 
In Appendix~\ref{app1}, the results of symmetry indicator groups for spinless and spinful factor systems are shown. 
The explicit formulas of topological indices for spinful factor systems are also discussed in Appendix~\ref{KeySGs} in the notation of doubled Hermitianized systems. 
In Sec~\ref{sec:ex}, we discuss the layer group $p4$ and the space group $P\bar 1$ as examples and show how the symmetry indices are defined and the corresponding physical phenomena. 

We close this paper with some perspectives.
In this paper, we did not take internal symmetry into account.
It is known that there are 38 internal symmetry classes in non-Hermitian systems~\cite{Kawabata_Shiozaki_Ueda_Sato:NonHermite}.
For example, the symmetry indicator of non-Hermitian systems with reciprocity, which is a kind of time-reversal symmetry, is interesting. In corresponding doubled Hermitian systems, time-reversal symmetry is added to chiral symmetry, and the symmetry indicators for superconductors~\cite{Ono_Po_Shiozaki:Z2_symmetry_indicator_SC} are applicable.
In this paper, we have focused on the spectral properties and the boundary modes of the system described by the non-Hermitian Hamiltonian, but it can also be applied to the detection of nontrivial unitary operators that are not continuously deformable into trivial unitary operators. 
The application to the Floquet operators is also interesting.

\begin{acknowledgements}
K.~S. thanks Wataru Ishizaki, Kohei Kawabata, Daichi Nakamura, Masatoshi Sato, and Mayuko Yamashita for helpful discussions. S.O. thanks Hoi Chun Po and Haruki Watanabe for earlier collaborations on related topics.
The work of K.~S. was supported by PRESTO, JST (Grant No. JPMJPR18L4) and CREST, JST (Grant No. JPMJCR19T2). The work of S.O. is supported by The ANRI Fellowship and  KAKENHI Grant No. JP20J21692 from the JSPS. 
\end{acknowledgements}

\bibliography{Band_Topology, NonHermite}
\clearpage
\appendix
\onecolumngrid
\section{Symmetry indicator groups in layer and space groups}
\label{app1}
Here, we list symmetry indicator groups for all layer and space groups. For simplicity, every symmetry indicator group ${X_{\text{BS}}} = \mZ_p \times \mZ_q \times\mZ_r\times\ldots$ is denoted by $\{p, q,r,\ldots\}$ in the following tables. 
\clearpage
\subsection{Space group}
\label{tab:Space}
\subsubsection{spinful systems}
\small \selectfont
\input{SG_Spinful}
\normalsize\selectfont
\clearpage
\subsubsection{spinless systems}
\small \selectfont
\input{SG_Spinless}
\normalsize\selectfont
\clearpage
\subsection{Layer group}
\subsubsection{spinful systems}
\label{tab:Layer_Spinful}
\small \selectfont
\input{LG_Spinful}
\normalsize\selectfont
\clearpage
\subsubsection{spinless systems}
\small \selectfont
\input{LG_Spinless}
\normalsize\selectfont
\clearpage

\section{Fundamental space groups for spinful systems}
\label{KeySGs}
\twocolumngrid
In this section, we introduce topological indices that characterize symmetry indicator groups for several fundamental symmetry settings.
Importantly, they cover various space groups, i.e., $\XBS$ in most symmetry settings can be computed from these topological indices. For convenience, the band label $N_\alpha[\tilde{H}^{0}_{\bk},\tilde{H}^{1}_{\bk}]:=
N_\alpha[\tilde{H}^{0}_{\bk}] - N_\alpha[\tilde{H}^{1}_{\bk}] \in \Z$ is denoted by $N_{\bk}^{\alpha}$ in this section. 

There are two remarks on indices that will be defined in the following. The first one is about expressions of the topological indices. In general, an irreducible representation is related to another irreducible representation by the chiral symmetry. Correspondingly, the formulas can be written using  `independent' ones. However, to make indices valid for various symmetry settings, we do not simplify their expressions and formulate them using characters of irreducible representations. 

The second remark is about the topology indicated by topological indices. While connections between several indices and their topologies are revealed, we do not derive such relations in all symmetry settings.  

\begin{table}[h]
	\begin{center}
		\caption{\label{tab:elem} $\XBS$ for spinful systems in fundamental symmetry settings.}
		\scalebox{0.9}[0.9]{
		\begin{tabular}{c|c|c|c}
			\hline\hline
			Key SG  (rep. of $\{c_g\}$) & 1D & 2D & 3D \\
			\hline
			$P\bar{1}$  ($A_u$)  &$\mathbb{Z}_2$ &$(\mathbb{Z}_2)^2 \times \mZ_4$ &$(\mathbb{Z}_2)^3 \times (\mathbb{Z}_4)^3 \times \mZ_8$ \\
			\hline
			$P2$  ($B$)  &$-$ &$(\mathbb{Z}_2)^2 \times \mZ_4$ &$(\mathbb{Z}_2)^2 \times \mZ_4$ \\
			\hline
			$Pm$  ($A''$)  &$\mZ_2$ &$\mZ_1$ &$\mZ_2$ \\
			\hline
			$P2/m$ ($B_u$) &$-$ &$(\mathbb{Z}_2)^4 \times (\mathbb{Z}_{4})^2$&$(\mathbb{Z}_2)^6 \times  (\mathbb{Z}_{4})^3$ \\
			$P2/m$ ($A_u$) &$(\mathbb{Z}_2)^2$ &$\mathbb{Z}_{2} $&$(\mathbb{Z}_2)^5 \times \mathbb{Z}_{4}$ \\
			\hline
			$P4 (B)$ & $-$ & $(\mathbb{Z}_2)^2$ & $(\mathbb{Z}_2)^2$\\
			\hline
			$P\bar{4} (B)$ & $(\mathbb{Z}_2)^2$ & $(\mathbb{Z}_2)^2$ & $(\mathbb{Z}_2)^4 \times (\mathbb{Z}_{4})^2$\\
			\hline
			$I\bar{4} (B)$ & $-$ & $-$ & $(\mathbb{Z}_2)^2 \times (\mathbb{Z}_{4})^3$\\
			\hline
			$P4/m$ ($B_u$) &$-$ &$\mathbb{Z}_{4}$&$  (\mathbb{Z}_{2})^2\times \mathbb{Z}_4$ \\
			$P4/m$ ($B_g$) &$-$ &$(\mathbb{Z}_2)^4 $&$ (\mathbb{Z}_{2})^6$ \\
			$P4/m$ ($A_u$) &$(\mathbb{Z}_2)^4$ &$\mathbb{Z}_{4}$&$(\mathbb{Z}_2)^7 \times \mathbb{Z}_{4} \times \mZ_8$ \\
			\hline
			$P4/mmm$ ($B_{1u},B_{2u}$) &$-$ &$\mathbb{Z}_{2}\times\mZ_8$&$\mathbb{Z}_2 \times  (\mathbb{Z}_{4})^2\times \mathbb{Z}_8$ \\
			$P4/mmm$ ($A_{1u},A_{2u}$) &$(\mathbb{Z}_2)^2$ &$\mathbb{Z}_{2}\times\mZ_8$&$(\mathbb{Z}_2)^4 \times \mathbb{Z}_{4} \times \mZ_8 \times \mZ_{16}$ \\
			\hline
			$P\bar{3} (A_u)$& $(\mZ_2)^3$ & $\mZ_4$ & $(\mZ_2)^3 \times \mZ_4 \times \mZ_8$\\
			\hline
			$P\bar{6}(B)$& $(\mZ_2)^3$ & $\mZ_3$ & $ (\mZ_2)^5\times (\mZ_6)^2$\\
			\hline
			$P6/m (B_g)$& $-$ & $\mZ_6$ & $\mZ_2\times(\mZ_6)^2$\\
			$P6/m (A_u)$& $(\mZ_2)^6$ & $\mZ_6$ & $ (\mZ_2)^8\times\mZ_6 \times \mZ_{12}$\\
			$P6_3/m (A_u)$& $(\mZ_2)^3$ & $-$ & $ (\mZ_2)^4\times \mZ_{12}$\\
			$P6/mmm$ ($A_{1u},A_{2u}$) &$(\mathbb{Z}_2)^3$ &$\mZ_{12}$&$(\mathbb{Z}_2)^4 \times \mZ_{12} \times \mZ_{24}$ \\
			\hline
			$Pm\bar{3}$ &$-$ &$-$&$  (\mathbb{Z}_{2})^3\times \mathbb{Z}_4 \times \mZ_8$ \\
			\hline\hline
		\end{tabular}
	}
	\end{center}
\end{table}

\subsection{$P\bar{1}$ with $A_u$ representation}
\label{app:p1bar}
In this subsection, we discuss inversion symmetric systems in one-, two-, and three-dimensions, where the inversion symmetry anticommutes with the chiral symmetry. 
\subsubsection{1D}
For 1D systems, the symmetry indicator group $\XBS$ is $\mZ_2$. We define the $\mZ_2$ index by
\begin{align}
	\label{eq:z2_inv}
	z_{2,I} \equiv \frac{1}{2}\sum_{k=0,\pi}\left(N_{k}^{+} - N_{k}^{-}\right)\mod 2,
\end{align}
where $N_{k}^{+ (-)}$ represents the $\mZ$-band label with the positive (negative) parity at $k=0$ and $\pi$. This index is related to the 1D winding number. 

\subsubsection{2D}
We then proceed to 2D systems. The symmetry indicator group is given by  $\XBS=(\mZ_2)^2 \times \mZ_4$. The  two $\mZ_2$ factors originate from 1D systems. The $\mZ_4$ factor is detected by 
\begin{align}
	\label{eq:z4_inv}
	z_{4,I}\equiv \frac{1}{2}\sum_{k \in \text{2D TRIMs}}\left(N_{k}^{+} - N_{k}^{-}\right) \mod 4,
\end{align}
where $N_{k}^{+ (-)}$ represents the $\mZ$-band label with the positive (negative) parity at two-dimensional time-reversal invariant momenta (2D TRIMs). When $z_{4,I} = 1,3$, the system should have gapless points. This can be understood as follows~\cite{Ono_Watanabe:Unified_understanding}. For Hermitian systems, the product of inversion parities indicates the Chern number modulo two if the system is fully gapped. However, the Chern number must be zero in the presence of the chiral symmetry (class AIII). To these two facts be consistent with each other, the system must possess gapless points. 

\subsubsection{3D}
Finally, we consider 3D systems, and the symmetry indicator group is $\XBS=(\mZ_2)^3 \times (\mZ_4)^3 \times \mZ_8$. The three $\mZ_2$ and $\mZ_4$ factors originate from 1D and 2D indices. The remaining $\mZ_8$ factor is characterized by
\begin{align}
	\label{eq:z8_inv}
	z_{8,I}\equiv \frac{1}{2}\sum_{k \in \text{3D TRIMs}}\left(N_{k}^{+} - N_{k}^{-}\right) \mod 8,
\end{align}
For systems with odd $z_{8,I}$, a gapless line encircled by a TRIM should appear. 
For $z_{8,I} = 2, 6$ without any weak index, the system should have $w_{3d} \mod 2 =1$, where $w_{3d}$ denotes the three-dimensional winding number. This can be understood by the following two facts: (i) $w_{3d}\mod2$ equals the magnetoelectric polarization; (ii) the magnetoelectric polarization can be computed from the product of inversion parities at TRIMs. As mentioned in the main text, we can construct a system with $z_{8,I} =4$ that exhibits the non-Hermitian third-order skin effect. 

Note that irreducible representations at TRIMs are often degenerate in the presence of additional symmetries. For example, there are two-dimensional irreducible representations at all TRIMs in $Pmmm$. Then, we substitute the pre-factor $\tfrac{1}{4}$ for $\tfrac{1}{2}$ in Eqs.~\eqref{eq:z2_inv}-\eqref{eq:z8_inv}. In the following, when we need to divide the indices by two, $\tilde{z}$ indicates divided ones.

\subsection{$P2$ with $B$ representation}
The space group $P2$ is generated by the two-fold rotation symmetry $C_2$. For $B$ representation, the chiral symmetry anticommutes with the chiral symmetry.  
In this symmetry setting, the symmetry indicator group in 3D is the same as that in 2D due to the existence of compatibility relations at the rotation axis. Therefore, we discuss only the 2D case here. We define two $\mZ_2$ and a $\mZ_4$ indices by 
\begin{align}
	z_{2, C_2, i=x\text{ or }y} \equiv \frac{1}{2}\sum_{k \in \text{TRIMs at $k_i = 0$}}\left(N_{k}^{+} - N_{k}^{-}\right) \mod 2, \\
	z_{4, C_2} \equiv \frac{1}{2}\sum_{k \in \text{2D TRIMs}}\left(N_{k}^{+} - N_{k}^{-}\right)  \mod 4,
\end{align}
where $N_{k}^{+ (-)}$ denote the $\mZ$-valued band label with the rotation eigenvalue $+i\ (-i)$. 

\subsection{$Pm$ with $A''$ representation}
This space group has the mirror symmetry, which anticommutes with the chiral symmetry. 
As shown in Table~\ref{tab:elem}, for 1D systems perpendicular to mirror plane, the symmetry indicator group is $\XBS=\mZ_2$. This is because the mirror symmetry is almost the same as the inversion. Then, we define the index like Eq.~\eqref{eq:z2_inv} by
\begin{align}
	z_{2,M_y} \equiv \frac{1}{2}\sum_{k=0,\pi}\left(N_{k}^{+} - N_{k}^{-}\right)\mod 2,
\end{align}
where $N_{k}^{+ (-)}$ represents the $\mZ$-valued band label with the mirror eigenvalue $+i\ (-i)$ at $k=0$ and $\pi$. For 2D systems in the mirror plane, $\XBS$ is trivial due to the compatibility relations, which results in the fact that $\XBS$ in 3D is the same as that in 1D. 

\subsection{$P2/m$ with $B_u$ representation}
This space group has the inversion and the two-fold rotation symmetries as generators. 
For this symmetry setting, the chiral symmetry commutes with the mirror symmetry $M_y$. In other words, the chiral symmetry does not change the mirror sector, and therefore the one-dimensional winding number can be defined for each mirror sector. Correspondingly, we can define topological indices in Eqs.~\eqref{eq:z2_inv} and \eqref{eq:z4_inv} for each mirror sector. Then, four $\mZ_2$ and three $\mZ_4$ factors in $\XBS$ are characterized by
\begin{align}
	z_{2,I, M_y}^{\pm} &\equiv \frac{1}{2}\sum_{\alpha= \pm 1}\sum_{k=0,\pi}\alpha N_{k}^{\alpha, \pm} \mod 2,\\
	z_{4,I, M_y}^{\pm} &\equiv \frac{1}{2}\sum_{\alpha= \pm 1}\sum_{k \in \text{2D TRIMs}}\alpha N_{k}^{\alpha, \pm}  \mod 4,
\end{align}
where $\alpha$ represents the parity and $N_{k}^{\alpha, \pm}$ signify $\mZ$-band label with the parity $\alpha$ and the mirror eigenvalue $\pm i$. The two remaining $\mZ_2$ parts are detected by $z_{4, I, k_x = \pi}$ and $z_{4, I, k_z = \pi}$. Note that, since $z_{4, I, k_x = \pi}$ and $z_{4, I, k_z = \pi}$ are always even because of the compatibility relations, they become $\mZ_2$ indices. 

\subsection{$P2/m$ with $A_u$ representation}
\subsubsection{1D}
As is the case for $B_u$ representation, the chiral symmetry commute with the rotation symmetry $C_{2}^{y}$. Again, the one-dimensional winding number and topological indices can be defined for each rotation sector. Then, we introduce two $\mZ_2$ indices by 
\begin{align}
	z_{2,I, C_2}^{\pm} &\equiv \frac{1}{2}\sum_{\alpha= \pm 1}\sum_{k=0,\pi}\alpha N_{k}^{\alpha, \pm} \mod 2,
\end{align}
where $N_{k}^{\alpha, \pm}$ means $\mZ$-band label with the parity $\alpha$ and the rotation eigenvalue $\pm i$. 
\subsubsection{2D}
Unlike the case for $B_u$, the chiral symmetry anticommute with the mirror symmetry $M_y$. Due to the anticommutation relation, there exist compatibility relations in the mirror plane. As a result, the $\mZ_2$ index in Eq.~\ref{eq:z2_inv} always vanishes, and also $\mZ_4$ index in Eq.~\ref{eq:z4_inv} should be even. Consequently, the symmetry indicator group is reduced to $\XBS=\mZ_2$. 

\subsubsection{3D}
Since the same logic holds for any mirror plane, the $\mZ_8$ index is also even, which indicates $\mZ_8$ is demoted to $\mZ_4$. Together with the above consideration, $z_{2,I}^{\pm}$ at the rotation axes, $z_{4, I}$ in the mirror plane, and $z_{8,I}$ fully characterize the symmetry indicator group $\XBS = (\mZ_2)^5 \times \mZ_4$. 

\subsection{$P4$ with $B$ representation}
This space group is generated by the four-fold rotation symmetry, which anticommutes with the chiral symmetry. 
As is the case of $P2$, the symmetry indicator group in 3D is the same as that in 2D. Again, we discuss only the 2D case here.
\begin{align}
	\label{eq:2Dmu4}
	\mu_{C_4} &\equiv \frac{1}{\sqrt{2}}\sum_{k = (0,0), (\pi, \pi)} \sum_{\alpha=1,3,5,7}e^{i \frac{\alpha}{4}\pi}N_{k}^{\alpha}, \\
	\label{eq:z2-C4-1}
	z_{2,C_4} &\equiv \frac{1}{2}\left(\Re[\mu_{C_4}] + \Im[	\mu_{C_4}]\right)\mod 2,\\
	\label{eq:z2-C4-2}
	z'_{2,C_4} &\equiv \frac{1}{2}\sum_{k = (0,0), (\pi, \pi)} \left(N_{k}^{1} - N_{k}^{7} + 3N_{k}^{3} - 3N_{k}^{5}  \right) \mod 2,
\end{align}
where $N_{k}^{\alpha}$ represents the $\mZ$-valued label with the four-fold rotation eigenvalue $e^{i \frac{\alpha}{4}\pi}$. Note that $z'_{2,C_4}$ appears only in $P4$. In other words, $z_{2,C_4}$ can characterize all $\mZ$ factors in space groups associated with $P4$. 

\subsection{$P\bar{4}$ with $B$ representation}
This space group has the four-fold rotoinversion symmetry $S_4$ as a generator. For $B$ representation, the chiral symmetry anticommutes with $S_4$. 
\subsubsection{1D}
Let us begin with 1D $S_4$-invariant systems along the $k_z$-direction. Since $(S_{4})^2 = C_2$ commutes with the chiral symmetry, the one-dimensional winding number can be defined in each rotation sector. Then, we introduce two $\mZ_2$ indices 
\begin{align}
	z_{2,S_4}^{+} &\equiv  \frac{1}{2}\sum_{k_z =0,\pi}\left(N_{k_z}^{1}-N_{k_z}^{5}\right)\mod 2,\\
	z_{2,S_4}^{-} &\equiv  \frac{1}{2}\sum_{k_z =0,\pi}\left(N_{k_z}^{3}-N_{k_z}^{7}\right)\mod 2.
\end{align}
where $N_{k}^{\alpha}$ represents the $\mZ$-valued label with the four-fold rotoinversion eigenvalue $e^{i \frac{\alpha}{4}\pi}$.

\subsubsection{2D}
Next, we consider 2D cases. Essentially, for 2D systems, $S_4$ is the same as $C_4$. Then, we define similar indices by
\begin{align}
	\label{eq:muS4}
	\mu_{S_4}^{\text{2D}} &\equiv \frac{1}{\sqrt{2}}\sum_{k = (0,0), (\pi, \pi)} \sum_{\alpha=1,3,5,7}e^{i \frac{\alpha}{4}\pi}N_{k}^{\alpha}, \nonumber \\
	z_{2,S_4} &\equiv \frac{1}{2}\left(\Re[\mu_{S_4}^{\text{2D}}] + \Im[	\mu_{S_4}^{\text{2D}}]\right)\mod 2,\\
	z'_{2,S_4} &\equiv  \frac{1}{2}\left(\Re[\mu_{S_4}^{\text{2D}}] - \Im[	\mu_{S_4}^{\text{2D}}]\right)\mod 2.
\end{align}
Note that, when irreducible representations are degenerate, we use the prefactor $\frac{1}{2\sqrt{2}}$ in Eq.~\eqref{eq:2Dmu4} instead of $\frac{1}{\sqrt{2}}$

\subsubsection{3D}
Finally, we discuss 3D systems. The symmetry indicator group is $\XBS= (\mathbb{Z}_2)^4 \times (\mathbb{Z}_{4})^2$. Indeed, all $\mZ_2$ factors originate from lower dimensions. The remaining $\mZ_4$ parts is explained by 
\begin{align}
	\mu_{S_4} &\equiv \frac{1}{\sqrt{2}}\sum_{k \in K_4}\sum_{\alpha=1,3,5,7}e^{i \frac{\alpha}{4}\pi}N_{k}^{\alpha}, \nonumber \\
	\label{eq:3D-S4-1}
	z_{4,S_4} &\equiv \frac{1}{2}\left(\Re[\mu_{S_4}] + \Im[	\mu_{S_4}]\right)\mod 4,\\
	\label{eq:3D-S4-2}
	z'_{4,S_4} &\equiv  \frac{1}{2}\left(\Re[\mu_{S_4}] - \Im[	\mu_{S_4}]\right)\mod 4,
\end{align}
where $K_4$ denotes the set of $S_4$-invariant momenta. 

\subsection{$P4/m$ with $B_g$ representation}
This space group is generated by the inversion $I$ and the four-fold rotation symmetries $C_{4}$. For this symmetry setting, the chiral symmetry commutes with $I$ but anticommutes with $C_4$. In the mirror plane, we define the topological indices in Eqs.~\eqref{eq:z2-C4-1} and \eqref{eq:z2-C4-2} for each parity
\begin{align}
	\mu_{C_4}^{\pm} &\equiv \frac{1}{\sqrt{2}}\sum_{k = (0,0), (\pi, \pi)} \sum_{\alpha=1,3,5,7}e^{i \frac{\alpha}{4}\pi}N_{k}^{\alpha, \pm}, \\
	z_{2,C_4}^{\pm} &\equiv \frac{1}{2}\left(\Re[\mu^{\pm}_{C_4}] + \Im[\mu^{\pm}_{C_4}]\right)\mod 2,\\
		{z'_{2}}_{,C_4}^{\pm} &\equiv \frac{1}{2}\left(\Re[\mu^{\pm}_{C_4}] - \Im[\mu^{\pm}_{C_4}]\right)\mod 2,
\end{align}
where $N_{k}^{\alpha, \beta}$ denotes the $\mZ$-valued label with the four-fold rotoinversion eigenvalue $e^{i \frac{\alpha}{4}\pi}$ and the inversion parity $\beta$. In fact, these characterize four $\mZ_2$ factors in 2D. Furthermore, two additional $\mZ_2$ factors exist in 3D systems, which can be understood by  Eqs.~\eqref{eq:3D-S4-1} and \eqref{eq:3D-S4-2}.

\subsection{$P4/m$ with $B_u$ representation}
Here, we discuss another symmetry class in $P4/m$ where the chiral symmetry anticommutes with both $I$ and $C_4$. Since $\XBS$ in 1D is trivial because of compatibility relations, let us begin with 2D cases. 

\subsubsection{2D}
In the mirror plane, we define the following index
\begin{align}
	\label{eq:mirror-C4}
	z_{4}^{m} &\equiv -\frac{3}{2} N^{5, +}_{\Gamma} +\frac{3}{2} N^{3, -}_{\Gamma} + \frac{1}{2}N^{1, +}_{\Gamma} - \frac{1}{2}N^{7, -}_{\Gamma}\notag\\
	&\quad\quad\quad-\frac{3}{2} N^{5, +}_{M} +\frac{3}{2} N^{3, -}_{M} + \frac{1}{2}N^{1, +}_{M}\notag\\
	& \quad\quad\quad- \frac{1}{2}N^{7, -}_{M}  - N^{2, +}_{X} + N^{-2, -}_{X},
\end{align}
where $N_{\bm{k}}^{\alpha,\beta}$ represents the $\mZ$-valued band labels with the rotation eigenvalue $e^{i \frac{\pi\beta}{n}}$ ($n$ depends on high-symmetry points) and the inversion parity $\beta$ at $\bk$. In addition, the labels of high-symmetry points follow Ref.~\onlinecite{Bilbao}. 
\subsection{$P4/mmm$ with $B_{1u, 2u}$ representation}
In addition to symmetries in $P4/m$, this space group has additional mirror and rotation symmetries. 
For $B_{1u, 2u}$ representations, the chiral symmetry anticommutes with both $I$ and $C_4$. As shown in Table~\ref{tab:elem}, the symmetry indicator group in 1D is trivial, and therefore let us start with 2D cases.
\subsubsection{2D}
While $\XBS=\mZ_4$ in $P4/m$, the symmetry indicator group is promoted to $\mZ_2\times \mZ_8$. 
Unlike the case for $P4/m$, irreducible representations at high-symmetry points are two-fold degenerate due to additional symmetries. Correspondingly, compatibility relations in the mirror plane become trivial. Then, we define the $\mZ_2$ index by Eq.~\eqref{eq:z2_inv} along the $k_x$-direction. The remaining $\mZ_8$ factor is characterized by
\begin{align}
	z_{8} \equiv z_{4}^{m} - 2\tilde{z}_{2,C_4}. 
\end{align}
This is based on the observation that $z_{4}^{m}=2\tilde{z}_{2,C_4} \mod 8$ for any atomic insulator. 

\subsubsection{3D}
For 3D systems, the symmetry indicator group is  $(\mZ_2)^2 \times \mZ_4 \times \mZ_8$. Indeed, the $\mZ_2\times\mZ_8$ part is explained by the above indices. The remaining $(\mZ_4)^2$ factors are characterized by $\tilde{z}_{4,I, k_x=0}$ and $\tilde{z}_{8,I}$. 

\subsection{$P4/m$ with $A_u$ representation}
In this subsection, we consider a different symmetry class in $P4/m$, where the chiral symmetry anticommutes with $I$ but commutes with $C_4$. As is the case for $P2/m$ with $A_u$ representation, the winding numbers at rotation symmetric lines are characterized by rotation eigenvalues. Here we define the following $\mZ_2$ indices
\begin{align}
		z_{2,I}^{\beta} &\equiv \frac{1}{2}\sum_{\alpha= \pm 1}\sum_{k=0,\pi}\alpha N_{k}^{\alpha, \beta} \mod 2,
\end{align}
where $N_{k}^{\alpha, \beta}$ represents the $\mZ$-valued label with the four-fold rotation eigenvalue $e^{i \frac{\beta}{4}\pi}$ and the parity $\alpha$. 
For the mirror plane, the formula in Eq.~\eqref{eq:mirror-C4} is a $\mZ_4$ index. Since the rotoinversion $S_4$ is the product of $I$ and $C_4$, $S_4$ also anticommutes with the chiral symmetry. In addition to $z_{8,I}$ in Eq.~\eqref{eq:z8_inv}, we can the define the index in Eq.~\eqref{eq:muS4}. Interestingly, we find that $z_{8,I}$ is always even and $z_{8,I}/2 = \Re[\mu_{4,S_4}] \mod 8$, and then we can define $\mZ_8$ index by
\begin{align}
	z'_{8} \equiv \tilde{z}_{8,I} - \Re[\mu_{4,S_4}] \mod 8,
	\intertext{where}
	\tilde{z}_{8,I}= \frac{1}{4}\sum_{k \in \text{3D TRIMs}}\left(N_{k}^{+} - N_{k}^{-}\right).
\end{align}

Together with $z_{4,I}/2$ for the $yz$-plane ($z_{4,I}$ is also even), the above indices explain the symmetry indicator group $\XBS = (\mathbb{Z}_2)^7 \times \mathbb{Z}_{4} \times \mZ_8$.

\subsection{$P4/mmm$ with $A_{1u, 2u}$ representation}
When the chiral symmetry anticommutes with $I$ but commutes with $C_4$, most of indices are in common with $P4/m$ with $A_u$ representation. However, there exist the following three differences: (i)
Since there are no compatibility relation at $k_x$- or $k_y$-axes due to the degeneracy originating from additional symmetries , we can define $z_{2,I}$ and $z_{4, I}$ for the $k_y$-axis and $yz$-plane; (ii) $z_{2,I}^{\beta}$ for $\beta=1,7$ ($\beta=3,5$) are the same because of the degeneracy originating from the additional symmetries; (iii) $z_{4}^{m}$ and $z'_{8}$ are promoted to $\mZ_8$ and $\mZ_{16}$ indices.

\subsection{$I\bar{4}$ with $B$ representation}
Symmetry operations in this space group are the same as $P\bar{4}$, but primitive lattice vectors in $I\bar{4}$ are different from those in $P\bar{4}$. While the symmetry indicator group for $P\bar{4}$ is $(\mathbb{Z}_2)^4 \times (\mathbb{Z}_{4})^2$, that for $I\bar{4}$ is $(\mathbb{Z}_2)^2 \times (\mathbb{Z}_{4})^3$. It is obvious that an additional $\mZ_4$ index needs for this space group. We define the $\mZ_4$ index by
\begin{align}
	\label{eq:S4-z4}
	\eta_{k} &\equiv \frac{1}{2\sqrt{2}}\left(\Re\left[\sum_{\alpha=1,3,5,7}e^{i \frac{\alpha}{4}\pi}N_{k}^{\alpha}\right] + \Im\left[\sum_{\alpha=1,3,5,7}e^{i \frac{\alpha}{4}\pi}N_{k}^{\alpha}\right]\right),\nonumber\\
	\xi_{k} &\equiv \frac{1}{2\sqrt{2}}\left(\Re\left[\sum_{\alpha=1,3,5,7}e^{i \frac{\alpha}{4}\pi}N_{k}^{\alpha}\right] - \Im\left[\sum_{\alpha=1,3,5,7}e^{i \frac{\alpha}{4}\pi}N_{k}^{\alpha}\right]\right),\nonumber\\
	z_{4, S_4, B} &\equiv \eta_{\Gamma} - \eta_{\text{M}} -\xi_{\text{P}} + \xi_{\text{PA}},
\end{align}
where the labels of high-symmetry points follow Ref.~\onlinecite{Bilbao}. 
Although $z_{2,S_4}^{\pm}$ are defined at $\Gamma-\text{M}$ and $\text{P}-\text{PA}$ lines, they span the only $\mZ_{2}$ part. Then, we introduce a new $\mZ_2$ index 
\begin{align}
	\label{eq:S4-z2_body}
	z_{2, S_4, B} \equiv \sqrt{2}\Re\left[\sum_{k = \text{M}, \text{P}} \sum_{\alpha=1,3,5,7}e^{i \frac{\alpha}{4}\pi}N_{k}^{\alpha}\right].
\end{align}
\subsection{$P\bar{3}$ with $A_u$ representation}
This space group is generated by the three-fold rotation $C_3$ and the inversion symmetries. For $A_u$ representation, the chiral symmetry anticommutes with $I$ but commutes with $C_3$. Again, the winding number at the rotation symmetric lines are well-defined for each rotation sector. Then, we can define the following two indices
\begin{align}
	\label{eq:z2-C3}
	z_{2,I, C_3}^{\beta} &\equiv \frac{1}{2}\sum_{\alpha= \pm 1}\sum_{k=0,\pi}\alpha N_{k}^{\alpha, \beta} \mod 2,
\end{align}
where $N_{k}^{\alpha, \beta}$ represents $\mZ$-band label with the parity $\alpha$ and the rotation eigenvalue $e^{i \frac{\beta}{3}\pi}$. Other factors in $\XBS$ can be understood by $z_{4,I}$ and $z_{8, I}$. 

\subsection{$P\bar{6}$ with $A''$ representation}
This space group is generated by the six-fold rotoinversion symmetry $S_6$ that anticommutes with the chiral symmetry. Since $C_3 = (S_6)^2$ commutes with the chiral symmetry, we define three $\mZ_2$ indices in the same way as the case for $P\bar{4}$
\begin{align}
	\label{eq:S6-1D-1}
	z_{2,S_6}^{1} &\equiv  \frac{1}{2}\sum_{k_z =0,\pi}\left(N_{k_z}^{1}-N_{k_z}^{5}\right)\mod 2,\\
	z_{2,S_6}^{2} &\equiv  \frac{1}{2}\sum_{k_z =0,\pi}\left(N_{k_z}^{3}-N_{k_z}^{7}\right)\mod 2,\\
	\label{eq:S6-1D-3}
	z_{2,S_6}^{3} &\equiv  \frac{1}{2}\sum_{k_z =0,\pi}\left(N_{k_z}^{3}-N_{k_z}^{7}\right)\mod 2,
\end{align}
where $N_{k}^{\alpha}$ represents the $\mZ$-valued label with the six-fold rotoinversion eigenvalue $e^{i \frac{\alpha}{6}\pi}$. 
To characterize $\XBS$, we use the fact that $\XBS =(\mZ_2)^5\times (\mZ_6)^2$ is isomorphic to $(\mZ_2)^7 \times (\mZ_3)^2$. 
For 3D systems, we have three $C_3$ symmetric lines, and then there are nine indices. In fact, seven of them can explain all $\mZ_2$ factors in $(\mZ_2)^7 \times (\mZ_3)^2$. The remaining $\mZ_3$ parts can be understood by
\begin{align}
	z_{3, k_z = 0}^{m} &\equiv -\frac{1}{2}N_{K}^{5} +\frac{1}{2}N_{K}^{1} + \frac{3}{2}N_{K}^{9} -\frac{1}{2}N_{K'}^{5}  \nonumber\\
	&\quad\ \ +\frac{1}{2}N_{K'}^{1}+ \frac{3}{2}N_{K'}^{9} + N_{\Gamma}^{5}- N_{\Gamma}^{1} \quad\mod 3, \\
	z_{3, k_z = \pi}^{m} &\equiv -\frac{1}{2}N_{H}^{5} +\frac{1}{2}N_{H}^{1} + \frac{3}{2}N_{H}^{9} -\frac{1}{2}N_{H'}^{5}  \nonumber\\
	&\quad\ \ +\frac{1}{2}N_{H'}^{1}+ \frac{3}{2}N_{H'}^{9} + N_{A}^{5}- N_{A}^{1} \quad\mod 3,
\end{align}
where the labels of high-symmetry points follow Ref.~\onlinecite{Bilbao}. Indeed, these formulas are obtained by rewriting the formulas for mirror Chern numbers~\cite{Fang_Gilbert_Bernevig:topological_invariant_point_group}. 

\subsection{$P6/m$ with $B_g$ representation}
This space group is generated by the inversion and the six-fold rotation $C_6$ symmetries. For $B_g$ representation, the chiral symmetry commutes with the inversion but anticommutes with $C_6$. In the mirror plane, we can define topological indices by
\begin{align}
	z_{6, k_z = 0}^{m} &\equiv \frac{3}{2} (-N^{-3, +}_{\Gamma} +N^{ 3, -}_{\Gamma} )+ \frac{5}{2} (N^{ 5, +}_{\Gamma} - N^{ -5, -}_{\Gamma}) \nonumber \\
	&\quad+\frac{1}{2}(N^{ 1, +}_{\Gamma} - N^{ -1, -}_{\Gamma})+N_{K}^{1} - N_{K}^{-1}+3N_{K}^{-3} \nonumber\\
	&\quad\quad\quad+\frac{3}{2}(-N_{M}^{2,+}+N_{M}^{-2,-})\quad \mod 6, \\
	z_{6, k_z = \pi}^{m} &\equiv \frac{3}{2} (-N^{-3, +}_{\Gamma} +N^{ 3, -}_{\Gamma} )+ \frac{5}{2} (N^{ 5, +}_{\Gamma} - N^{ -5, -}_{\Gamma}) \nonumber \\
	&\quad+\frac{1}{2}(N^{ 1, +}_{A} - N^{ -1, -}_{A})+N_{H}^{1} - N_{H}^{-1}+3N_{H}^{-3} \nonumber\\
	&\quad\quad\quad+\frac{3}{2}(-N_{L}^{2,+}+N_{L}^{-2,-})\quad \mod 6, 
\end{align}
where the labels of high-symmetry points follow Ref.~\onlinecite{Bilbao}. Here, $N_{\bm{k}}^{\alpha,\beta}$ represents the $\mZ$-valued band labels with the rotation eigenvalue $e^{i \frac{\pi\beta}{n}}$ ($n$ depends on high-symmetry points) and the inversion eigenvalue $\beta$ at $\bk$. 
One might think that these can explain two $\mZ_6$ factors in $\XBS$, but this is untrue. In fact, they span $\mZ_3 \times \mZ_6$. 
Since $\mZ_6 \simeq \mZ_2 \times \mZ_3$, the remaining part is $(\mZ_2)^2$, which can be understood by $z_{2,S_6}^{\beta}$ in Eqs.~\eqref{eq:S6-1D-1}-\eqref{eq:S6-1D-3} due to the presence of $S_6 = IC_6$.

\subsection{$P6/m$ with $A_u$ representation}
Unlike the case for $B_g$ representation, the chiral symmetry anticommutes with the inversion but commutes with $C_6$ for $A_u$ representation. Then, as is the case for $P2/m$ and $P4/m$, the winding numbers are defined for each rotation sector. Accordingly, $z_{2,I}$ in Eq.~\eqref{eq:z2_inv} can be defined for each sector 
\begin{align}
	z_{2,I, C_6}^{\beta} &\equiv \frac{1}{2}\sum_{\alpha= \pm 1}\sum_{k=0,\pi}\alpha N_{k}^{\alpha, \beta} \mod 2,
\end{align}
where $N_{k}^{\alpha, \beta}$ represents the $\mZ$-valued label with the six-fold rotation eigenvalue $e^{i \frac{\beta}{6}\pi}$ and the parity $\alpha$. 
Since there also exist $C_2$- and $C_3$- symmetric lines, $z_{2,S_6}^{\beta}$ and $z_{2,I,C_2}^{\pm}$ are well-defined. As is the case for $B_g$, $(z_{6, k_z = 0}^{m} ,z_{6, k_z = \pi}^{m})$ again spans $\mZ_3 \times \mZ_6$. Since $\mZ_{12}\simeq \mZ_{3}\times\mZ_{4}$, we infer that a $\mZ_{12}$ index can be constructed by the $\mZ_3$ and the $\mZ_4$ indices. In fact, we can define the $\mZ_{12}$ index by
\begin{align}
	z_{12} \equiv 3\tilde{z}_{8,I} - 4z_{6,k_z=0}^{m} \mod 12. 
\end{align}
These indices explain all factors in $\XBS$. 

\subsection{$P6/mmm$ with $A_{1u}$ and $A_{2u}$ representation}
In addition to symmetries in $P6/m$, this space group has additional mirror and rotation symmetries. For $A_{1u}$ and $A_{2u}$ representations, the chiral symmetry anticommutes with $I$ but commutes with $C_6$. Similar to the relation between $P4/m$ and $P4/mmm$, there exist the following three differences: (i) Since there are no compatibility relation at $k_x$- or $k_y$-axes due to the degeneracy originating from additional symmetries, we can define $z_{2,I}$ and $z_{4, I}$; (ii) $z_{2,I, C_6}^{\beta}$ for $\beta=1,11$, $\beta=3,9$, and $\beta=5,7$ are the same because of the additional symmetries; (iii) $\XBS$ in 2D is $\mZ_{12}$, not $\mZ_6$; and (iv) $z_{12}$ is promoted to the $\mZ_{24}$ index. In particular, as for 2D systems, $z_{6}^{m}$ and $\tilde{z}_{4,I}$ are well-defined. However, they hold the relation $2z_{6}^{m}=3\tilde{z}_{4,I} \mod 12$ for any atomic insulator. As a result, we can construct the $\mZ_{12}$ index from $z_{6}^{m}$ and $\tilde{z}_{4,I}$ such that
\begin{align}
	z'_{12}\equiv 2z_{6}^{m}-3\tilde{z}_{4,I}\quad \mod 12.
\end{align}


\subsection{$P6_3/m$ with $A_u$ representation}
In this space group, we consider the six-fold screw symmetry $S = T_{z/2}C_6$, where $T_{z/2}: (x,y,z) \rightarrow (x,y,z+1/2)$. Although the $\mZ_{12}$ factor in $\XBS$ is in common with $P6/m$, other parts are affected by the screw symmetry. In the following, we explain the changes. 
Since $S^6 = T_{3z}$ and the representation satisfies $[U_{\bk}(C_{6})]^{6} = -1$, irreducible representations at the six-fold screw line can be expressed by $U^{\alpha}_{\bk}(S) = e^{i\frac{\alpha}{6}\pi}e^{-i \frac{k_z}{2}}$. When we change $k_z$ to $k_z + 2\pi$, $U^{\alpha}_{\bk+2\pi \bm{e}_z}(S) = U^{\alpha+6}_{\bk}(S)$, which implies that irreducible representations $U^{\alpha}_{\bk}(S)$ and $U^{\alpha+6}_{\bk}(S)$ should appear at the same time. Indeed, their three-fold rotation eigenvalues are in common. Then, the $\mZ_2$ indices in Eq.~\eqref{eq:z2-C3} are well-defined. 

While $z_{2, S_6}^{\beta}$ does not change, the same thing happens in $z_{2,I,C_2}^{\pm}$, and therefore $z_{2,I}$ in Eq.~\eqref{eq:z2_inv} at $C_2$-symmetric line is the $\mZ_2$ index. Note that $z_{6,k_z=\pi}^{m}$ is not topological indices due to the screw symmetry. As a result, the $\mZ_6$ factor vanishes in this symmetry setting. 

\subsection{Point group $T_h$ and $O_h$}
For space groups which have $T_h$ and $O_h$ as their point groups, various indices defined in preceding sections account for most factors in $\XBS$. Here, we introduce some indices to detect a few remaining ones. 
\subsubsection{three-fold rotation along (111) direction}
These space groups have the three-fold rotation along (111)-direction. As a result, $\mZ_2$-valued indices defined in Eq.~\eqref{eq:z2-C3} exist for this line, which are denoted by $z_{2,I, C_3}^{(111), \beta}$. Note that $\beta=1,2$ are the same and $\beta=3$ is always even because of additional symmetries. 

\subsubsection{Face-centered lattice}
For face-centered lattices, indices introduced in preceding sections cannot characterize a $\mZ_2$ factor. Then, we define
\begin{align}
	z_{2,F} \equiv \frac{1}{2\sqrt{2}}\sum_{\alpha}\chi_{X}^{\alpha}(S_{4}^{x})N_{X}^{\alpha} + \frac{1}{2\sqrt{2}}\sum_{\beta=1,3,5,7}\chi_{W}^{\beta}(S_{4}^{z})N_{W}^{\beta},
\end{align}
where $X=(1,0,0)$ and $W=(1,0,\frac{1}{2})$. Note that the rotoinversion symmetries at $X$ and $W$ are not the same. More concretely, $X$ and $W$ are invariant under $S_{4}^{x}$ and $S_{4}^{z}$, respectively.

\subsubsection{Body-centered lattice}
As is the case for face-centered lattices, some factors in $\XBS$ for SG $Im\bar{3}m$ are characterized by the following new indices
\begin{align}
	z_{2, B} &\equiv N_{H}^{6} + N_{H}^{9} + N_{P}^{6}\quad\mod 2,\\
	z'_{2, B} &\equiv N_{N}^{5} + N_{P}^{6}\quad\mod 2,
\end{align}
where $H=(1,1,1), N=(\tfrac{1}{2}, \tfrac{1}{2}, 0)$, and $P=(\tfrac{1}{2}, \tfrac{1}{2}, \tfrac{1}{2})$. The labels of irreducible representations follow Ref.~\onlinecite{Bilbao}. 
\subsection{Other space groups}
While most symmetry settings can be understood by indices introduced in the preceding sections, they cannot explain a few space groups. Here, we introduce a technique to systematically construct new indices and derive formulas for these symmetry settings.
 
 Suppose that we have a complete set of $\{\mathrm{AI}\} = \mathrm{span}_{\mZ}\left\{\ba_i\right\}_{i=1}^{d_{\mathrm{AI}}}$, which forms a $(n \times d_{\mathrm{AI}})$-dimensional matrix 
\begin{align}
	A = \left(\ba_1, \ba_2, \cdots, \ba_{d_{\mathrm{AI}}}\right).
\end{align}
We can always find unimodular matrices $U$ and $V$ such that
\begin{align}
	\label{eq:Smith}
	A &= U \begin{pmatrix}
		\Sigma & 0 \\
		0 & 0
	\end{pmatrix} V^{-1},\\
	\Sigma &= \text{diag}(s_1, s_2, \cdots s_{d_{\mathrm{AI}}}),
\end{align}
where $s_i$ can be positive integers~\cite{Po:Symmetry_indicator_review}. Indeed, there exist the following three facts: (i) $A'=AV = \left(\ba'_1, \ba'_2, \cdots, \ba'_{d_{\mathrm{AI}}}\right)$ is a new set of the basis vectors of $\{\mathrm{AI}\}$; (ii) Each column vector in $U$ is the basis of $\{\mathrm{BS}\}$ (denoted by $\bb'_i$); (iii) the set of $s_i$ corresponds to the symmetry indicator group. Furthermore, from Eq.~\eqref{eq:Smith}, we have the following relation
\begin{align}
	\begin{pmatrix}
		\Sigma & 0 \\
		0 & 0
	\end{pmatrix} = U^{-1}A'.
\end{align}
When each row vector of $U^{-1}$ is denoted by $(\bm{u}_j)^T$, we find $(\bm{u}_j)^T\bm{a}'_i = 0 \mod s_i$ and $(\bm{u}_j)^T\bm{b}'_i = \delta_{ji}$.
This implies that each row vector of $U^{-1}$ results in a topological index. Using this fact, we list topological indices for some complicated symmetry settings. In the following, the labels of high-symmetry points and irreducible representations follow Ref.~\onlinecite{Bilbao}.

We obtain formulas by applying this strategy to several symmetry settings: $127(B_{1u}, B_{2u})$, $128(B_{1u}, B_{2u})$, $140(B_{1u}, B_{2u})$,  $210(A_2)$, and $227(A_{2u})$. 
\begin{itemize}
	\item SG 127 ($B_{1u}, B_{2u}$)
	\begin{align}
		z_{4, \text{SG 127}} &\equiv -N_{\Gamma}^{7} - N_{Z}^{8} - N_{M}^{8} + N_{A}^{8} \quad\mod 4,  \\
		z'_{4, \text{SG 127}} &\equiv -N_{\Gamma}^{7} - N_{Z}^{6} + N_{M}^{6} - N_{A}^{8} \quad\mod 4.
	\end{align}
	
	\item SG 128 ($B_{1u}, B_{2u}$)
	\begin{align}
		z_{4, \text{SG 128}} &\equiv N_{\Gamma}^{7} +3N_{Z}^{5}-4N_{Z}^{7} \nonumber\\
		&\quad \quad+2N_{R}^{6}- N_{M}^{6} + 3N_{A}^{8} \quad\mod 4.
	\end{align}

	\item SG 140 ($B_{1u}, B_{2u}$)
	\begin{align}
		z_{4, \text{SG 140}} &\equiv N_{\Gamma}^{6}-2N_{\Gamma}^{7} +N_{X}^{6}-3N_{M}^{8}\ \ \mod 4,\\
		z_{2, \text{SG 140}} &\equiv -N_{\Gamma}^{7} - N_{M}^{8}\quad\mod 2.
	\end{align}

	\item SG 210 ($A_2$)
	\begin{align}
		z_{2, \text{SG 210}} &\equiv N_{L}^{5} \quad\mod 2.
	\end{align}
	\item SG 227 ($A_{2u}$)
	\begin{align}
		z_{2, \text{SG 227}} &\equiv -2N_{\Gamma}^{7}-N_{L}^{4}-N_{L}^{9} \quad\mod 2,\\
		z'_{2, \text{SG 227}} &\equiv N_{\Gamma}^{11}+N_{\Gamma}^{7}+N_{W}^{4} \quad\mod 2,\\
		z''_{2, \text{SG 227}} &\equiv -N_{\Gamma}^{7}-N_{L}^{4}+N_{L}^{6}-N_{L}^{9} +N_{W}^{4} \ \ \mod 2.
	\end{align}
\end{itemize}
\onecolumngrid
\begin{table}[H]
	\begin{center}
		\caption{\label{tab:XBS} The list of key space groups and their super groups. The key space groups are underlined.}
		\scalebox{0.8}[0.8]{
		\begin{tabular}{cc|cc}
			\hline \hline
			$G_0$&Key Indices&$\XBS$&Space Groups\\
			\hline
			\multirow{25}{*}{$P\bar{1}$ ($A_u$)} & \multirow{25}{*}{$z_{2,I, i}, z_{4,I, i}, z_{8,I}$} & $(\mathbb{Z}_2)^3\times(\mathbb{Z}_4)^3\times\mathbb{Z}_8$ &  \underline{$2(A_u)$}, $47(A_{u}, B_{1u}, B_{2u}, B_{3u})$, \\&\\
			&&\multirow{1}{*}{$(\mathbb{Z}_2)^2\times\mathbb{Z}_4\times\mathbb{Z}_8$} & $65 (A_{u}, B_{1u}),$ \\
			&&\multirow{1}{*}{$\mathbb{Z}_2\times\mathbb{Z}_4\times\mathbb{Z}_8$} & $ 191(B_{1u}, B_{2u}).$ \\
			&&$(\mathbb{Z}_2)^3\times\mathbb{Z}_8$ &  $69 (B_{1u}, B_{2u},B_{3u}, A_{u}), 71 (B_{1u}, B_{2u},B_{3u}, A_{u})$.\\
			&&$(\mathbb{Z}_2)^2\times(\mathbb{Z}_4)^2$ & $12(A_u), 15(B_u)$ \\
			&&$\mathbb{Z}_2\times(\mathbb{Z}_4)^2$ &  $15(A_u), 131(B_{2u},B_{1u})$\\
			&&$(\mathbb{Z}_2)^3\times\mathbb{Z}_4$ & $11(A_u)$ \\
			&&\multirow{2}{*}{$(\mathbb{Z}_2)^2\times\mathbb{Z}_4$} &  $11(B_u), 13(A_u), 14(B_u, A_u), 49(B_{2u},B_{3u}), 51(B_{2u}, A_u), 55(B_{1u},A_u),$\\ 
			&&& $67 (B_{1u}, B_{2u},B_{3u}, A_{u}),139 (B_{1u}, B_{2u})$\\
			&&&\\
			&&\multirow{5}{*}{$\mathbb{Z}_2\times\mathbb{Z}_4$} &  $48 (B_{1u}, B_{2u},B_{3u}, A_{u}), 50 (B_{1u}, B_{2u},B_{3u}, A_{u}), 52 (B_{1u}, B_{3u}, A_{u}), $\\
			&&&$53(A_u), 54 (B_{2u},A_{u}), 56 (B_{1u}, B_{2u},B_{3u}, A_{u}),57 (B_{3u}, A_{u}),  $\\
			&&&$58 (B_{1u}, A_{u}), 59(B_{1u}, B_{2u},B_{3u}, A_{u}), 60(B_{1u},B_{3u}), 62(B_{1u}, B_{2u},B_{3u}, A_{u}),$ \\
			&&&$63(B_{3u}, A_{u}), 64(B_{3u}, A_{u}), 68(B_{1u}, B_{2u},B_{3u}, A_{u}), 70(B_{1u}, B_{2u},B_{3u}, A_{u}), $\\ 
			&&&$72(B_{1u}, B_{2u},B_{3u}, A_{u}),73(B_{1u}, B_{2u},B_{3u}, A_{u}), 74(A_u), 132(B_{1u}, B_{2u})$\\
			&&&\\
			&&$(\mathbb{Z}_2)^3$ & $87(B_u)$ \\
			&&\multirow{2}{*}{$\mathbb{Z}_4$} &  $52(A_u), 60(B_{2u}, A_u), 61(B_{1u}, B_{2u},B_{3u}, A_{u}), 88(B_u), 136(B_{2u}, B_{1u}), $\\
			&&&$176(B_u), 192(B_{2u}, B_{1u}), 193(B_{1u}), 194(B_{2u})$\\
			&&$(\mathbb{Z}_2)^2$ &  $84(B_u), 85(B_u), 125(B_{2u}, B_{1u}), 129(B_{2u}, B_{1u})$\\
			&&\multirow{2}{*}{$\mathbb{Z}_2$} &  $86(B_u), 126(B_{2u}, B_{1u}), 130(B_{2u}, B_{1u}), 133(B_{2u}, B_{1u}), 134(B_{2u},B_{1u}),$\\
			&&&$135(B_{2u}, B_{1u}), 137(B_{2u}, B_{1u}), 138(B_{2u}, B_{1u}), 141(B_{1u}), 142(B_{2u}, B_{1u})$\\
			\hline
			\multirow{6}{*}{$P2 (B)$} & \multirow{6}{*}{$z_{2, C_2, i}, z_{4, C_2}$} & $(\mathbb{Z}_2)^2 \times \mathbb{Z}_{4}$ & \underline{$3 (B)$}\\
			&& $(\mZ_2)^2$ & $28 (B_1), 91(A_2), 95(A_2)$ \\ 
			&& \multirow{2}{*}{$\mathbb{Z}_2$} & $10(B_g), 13(B_g), 17(B_2, B_3), 18(B_2,B_3), 21(B_2,B_3), 28(B_2), 30(B_1,B_2),32(B_1,B_2), 34(B_1,B_2), 35(B_1,B_2), $ \\
			&&&$37(B_1,B_2), 51(B_{1g},B_{3g}), 52(B_{2g}), 53(B_{1g}), 54(B_{1g},B_{3g}), 55(B_{2g},B_{3g}), 57(B_{1g}, B_{2g}), 91(B_2), 95(B_2)$\\
			&&$\mathbb{Z}_{4}$ & $168 (B), 171 (B), 172(B)$\\
			\hline
			$C2(B)$& $z_{2,C_2}$  & $\mZ_2$ & $5(B), 20(B_1), 24(B_1,B_2, B_3), 40(B_1), 46(B_1,B_2),98(B_2,A_2)$ \\
			\hline
			$Pm(A'')$& $z_{2,M_y}$  & $\mZ_2$ & $6(A'')$ \\
			\hline
			\multirow{7}{*}{$P2/m (B_u)$} &\multirow{7}{*}{$z_{2,I}^{\pm}, z_{4,I}^{\pm}, z_{4,I, i}$} & $(\mathbb{Z}_2)^6 \times (\mathbb{Z}_{4})^{3}$ & \underline{$10 (B_u)$}\\
			&& $ (\mathbb{Z}_{2})^{3} \times (\mathbb{Z}_{4})^{2}$ & $51(B_{1u}, B_{3u}), 55(B_{2u}, B_{3u})$\\
			&& $ (\mathbb{Z}_{2})^{4} \times \mathbb{Z}_{4}$ & $53(B_{1u})$\\
			&& $ (\mathbb{Z}_{2})^{3} \times \mathbb{Z}_{4}$ & $53(B_{2u}), 58(B_{2u}, B_{3u}), 66(B_{2u}, B_{3u})$\\
			&&$ (\mathbb{Z}_{4})^{3}$ &$175 (B_u)$\\
			\multirow{3}{*}{$P2/m (A_u)$}& \multirow{3}{*}{$z_{2,I}^{\pm}, z_{4, I}, z_{8, I}$} & $(\mathbb{Z}_2)^5 \times \mathbb{Z}_{4}$ & \underline{$10 (A_u)$}\\
			&& $(\mathbb{Z}_2)^3 \times \mathbb{Z}_{4}$ & $49(A_u, B_{1u})$\\
			&& $(\mathbb{Z}_2)^2 \times \mathbb{Z}_{4}$ & $53(B_{3u}), 66(A_u, B_{1u})$\\
			\multirow{2}{*}{$P2/c (B_u)$}& \multirow{2}{*}{$z_{2,I}^{\pm},z_{2,I}, z_{4,I}, \tilde{z}_{8,I}$}  & $(\mathbb{Z}_2)^3 \times (\mathbb{Z}_4)^2 $ & \underline{$13 (B_u)$}\\
			&& $(\mathbb{Z}_2)^2 \times \mathbb{Z}_{4}$ & $52(B_{2u}), 54(B_{1u},B_{3u}), 57(B_{1u},B_{2u})$\\
			\hline
			\multirow{6}{*}{$C2/m(B_u)$} & & $(\mathbb{Z}_2)^4 \times (\mathbb{Z}_{4})^2$ & $\underline{12 (B_u)}$\\
			&\multirow{1}{*}{$z_{2,I, M}^{\pm}, \tilde{z}_{2,I},$}&$(\mathbb{Z}_2)^2 \times \mathbb{Z}_{4}$ & $63 (B_{1u}, B_{2u}), 64 (B_{1u}, B_{2u}), 74(B_{2u}, B_{3u})$\\
			&\multirow{1}{*}{$\tilde{z}_{4,I},\tilde{z}_{8,I}$}& $(\mathbb{Z}_2)^4 \times \mathbb{Z}_{4} \times \mZ_8$ & $65 (B_{2u}, B_{3u})$\\
			&&$(\mathbb{Z}_2)^2 $ &$141(B_{2u})$\\
			\hline
			$P222_1 (B_1)$& $z_{2, C_{2}^{x}}, z_{4, C_{2}^{x}},z_{4, C_{2}^{y}}$ &$(\mZ_2)^3$  & $17(B_1)$\\
			\hline
			\multirow{2}{*}{$Imma(B_{1u})$} &\multirow{1}{*}{$z_{2,I, M_x}^{\pm}, z_{2,I, M_y}^{\pm}$} & \multirow{2}{*}{$(\mathbb{Z}_2)^4 \times \mathbb{Z}_{4}$} & \multirow{2}{*}{$\underline{74(B_{1u})}$}\\
			& $\tilde{z}_{4,I,k_3=0},\tilde{z}_{8,I}$ && \\
			\hline\hline
		\end{tabular}
	}
	\end{center}
\end{table}
\clearpage
\begin{table*}
	\begin{center}
		\label{tab:XBS2}(Continued from the previous page)
		\scalebox{0.8}[0.8]{
			\begin{tabular}{cc|cc}
				\hline\hline
				\multirow{2}{*}{$P4 (B)$} &\multirow{2}{*}{$z_{2, C_4}, z'_{2,C_4}$}&$(\mZ_2)^2$&\underline{$75(B)$}\\
				&&$\mZ_2$& $89 (B_1,B_2), 90(B_1,B_2), 99(B_1,B_2), 100(B_1,B_2), 103(B_1,B_2), 104(B_1,B_2), 207(A_2)$\\
				\hline
				\multirow{11}{*}{$P\bar{4} (B)$} && $(\mathbb{Z}_2)^4 \times (\mathbb{Z}_{4})^2$&\underline{$81(B)$}\\
				&&$(\mathbb{Z}_2)^4$ & $85(B_g)$ \\
				&& \multirow{2}{*}{$(\mathbb{Z}_2)^2 \times \mathbb{Z}_{4}$} & $111(B_1,B_2),112(B_1,B_2),113(B_1,B_2),114(B_1,B_2),115(B_1,B_2),$\\
				&$z_{4,S_4}, z'_{4,S_4}$&&$116(B_1,B_2),117(B_1,B_2),118(B_1,B_2),215(A_2),218(A_2)$\\
				&&&\\
				&\multirow{1}{*}{$z_{2, S_4}^{\pm}, z_{2,S_4}, z'_{2,S_4},$}&$(\mZ_2)^2$ & $84(B_g), 86(B_g), 125(B_{1g}, B_{2g}), 126(B_{1g}, B_{2g}), 129(B_{1g}, B_{2g}), 130(B_{1g}, B_{2g}), 222(A_{2g})$\\
				&&&\\
				&&\multirow{2}{*}{$\mZ_2$} & $131(B_{1g}, B_{2g}), 132(B_{1g}, B_{2g}), 133(B_{1g}, B_{2g}),134(B_{1g}, B_{2g}),135(B_{1g}, B_{2g}),$\\
				&&& $136(B_{1g}, B_{2g}), 137(B_{1g}, B_{2g}), 138(B_{1g}, B_{2g}), 223(A_{2g}), 224(A_{2g})$ \\
				\hline
				\multirow{7}{*}{$I\bar{4} (B)$} && $(\mathbb{Z}_2)^2 \times (\mathbb{Z}_{4})^3$&\underline{$82(B)$}\\
				&&$(\mathbb{Z}_2)^4$ & $87(B_g)$ \\
				&\multirow{1}{*}{$z_{2, S_4}^{\pm}, z_{2,S_4, B},z_{4,S_4}, $}&$(\mathbb{Z}_2)^2\times\mZ_4$ & $119(B_1,B_2), 120(B_1,B_2), 121(B_2), 122(B_2), 216(A_2), 217(A_2), 219(A_2)$ \\
				&$\Re[\mu_{4,S_4}],z_{4, S_4, B}$& $\mZ_2 \times \mZ_4$ & $88(B_g), 122 (B_1), 220(A_2)$\\
				&& \multirow{1}{*}{$(\mathbb{Z}_2)^2$} & $139(B_{2g},B_{1g}), 140(B_{2g}, B_{1g}), 225(A_{2g}), 226(A_{2g}), 229(A_{2g})$\\
				&&\multirow{1}{*}{$\mZ_2$} & $141(B_{1g}, B_{2g}), 142(B_{1g}, B_{2g}), 227(A_{2g}), 228(A_{2g}), 230(A_{2g}),$\\
				\hline
				$P4/m (B_u)$ & $z_{4}^{m}, z_{4,I}, \tilde{z}_{8,I}$  & $  (\mathbb{Z}_{2})^2\times \mathbb{Z}_4$& \underline{$83 (B_u)$}\\&\\
				\multirow{2}{*}{$P4/m (B_g)$} & $z_{2,C_4}^{\pm}, {z'_{2}}_{,C_4}^{\pm},$ &$ (\mathbb{Z}_{2})^6$ & \underline{$83 (B_g)$}\\
				&$z_{4,S_4}, z'_{4,S_4}$& $(\mathbb{Z}_{2})^3$ & $123(B_{1g}, B_{2g}), 124(B_{1g}, B_{2g}), 127(B_{1g}, B_{2g}), 128(B_{1g}, B_{2g}), 221 (A_{2g})$ \\&\\
				\multirow{4}{*}{$P4/m (A_u)$} & & $(\mathbb{Z}_2)^7 \times \mathbb{Z}_{4} \times \mZ_8$& \underline{$83 (A_u)$}\\
				& $z_{2,I}^{\beta}, z_{4, I},$ & $(\mathbb{Z}_2)^3 \times \mathbb{Z}_{4} \times \mZ_8$& $127 (A_{1u},A_{2u})$\\
				& $ z_{4}^{m}, z_{8, P4/m}$& $(\mathbb{Z}_2)^4 \times \mZ_8$& $124 (A_{1u},A_{2u})$\\
				& & $(\mathbb{Z}_2)^3 \times  \mZ_8$& $128 (A_{1u},A_{2u})$\\
				\hline
				\multirow{6}{*}{$P4_2/m (A_u)$} & & $(\mathbb{Z}_2)^5 \times \mathbb{Z}_{4}$& \underline{$84 (A_u)$}\\
				& $z_{2,I, C_2}^{\pm}, z_{2, I},$ & $(\mathbb{Z}_2)^3 \times \mathbb{Z}_{4} \times \mZ_8$& $131 (A_{1u},A_{2u})$\\
				& $ z_{4, I}, z_{2,S_4}, \tilde{z}_{8}$& $(\mathbb{Z}_2)^3 \times \mZ_8$& $132 (A_{1u},A_{2u})$\\
				& & $(\mathbb{Z}_2)^2 \times  \mZ_4$& $135 (A_{1u},A_{2u})$\\
				& & $(\mathbb{Z}_2)^2 \times  \mZ_8$& $136 (A_{1u},A_{2u})$\\
				\multirow{3}{*}{$P4/n (A_u)$} & & $(\mathbb{Z}_2)^3 \times \mathbb{Z}_{4}$& \underline{$85 (A_u)$}\\
				& $z_{2,I}^{\beta},\tilde{z}_{4, I}, \tilde{z}_{8}$ & $(\mathbb{Z}_2)^2 \times \mathbb{Z}_{4}$& $125 (A_{1u},A_{2u}),129 (A_{1u},A_{2u})$\\
				&& $\mathbb{Z}_2 \times \mathbb{Z}_{4}$& $126 (A_{1u},A_{2u}),130 (A_{1u},A_{2u})$\\
				\multirow{3}{*}{$P4_2/n (A_u)$} & $z_{2,I, C_2}^{\pm}, z_{4, I},$& $(\mathbb{Z}_2)^3 \times \mathbb{Z}_{4}$& \underline{$86 (A_u)$}\\
				&  $ \tilde{z}_{4,S_4},\tilde{z}'_{4,S_4}, \tilde{z}_{8}$& $(\mathbb{Z}_2)^2 \times \mathbb{Z}_{4}$& $134 (A_{1u},A_{2u}),138 (A_{1u},A_{2u})$\\
				& & $\mathbb{Z}_2 \times \mathbb{Z}_{4}$& $133 (A_{1u},A_{2u}),137 (A_{1u},A_{2u})$\\
				\hline
				\multirow{4}{*}{$I4/m (A_u)$} & & $(\mathbb{Z}_2)^6 \times\times \mZ_8$& \underline{$87 (A_u)$}\\
				& $z_{2,I}^{\beta}, z_{4, I},$ & $(\mathbb{Z}_2)^4 \times \mZ_{16}$& $139 (A_{1u},A_{2u})$\\
				& $z_{8, P4/m}$& $(\mathbb{Z}_2)^3 \times \mZ_8$& $140 (A_{1u},A_{2u})$\\
				\multirow{2}{*}{$I4_1/a (A_u)$} & $z_{2,I, C_2}^{\pm}, $ & $(\mZ_2)^2\times(\mZ_4)^2$ & \underline{$88 (A_u)$}\\
				&&$(\mZ_2)^2\times \mZ_4$ & \underline{$141 (A_{1u}), 142(A_{1u}, A_{2u})$}\\
				\hline
				$P4/mmm (B_{1u})$ & $\tilde{z}_{2,I}, z_{8}, \tilde{z}_{4,I}, \tilde{z}_{8,I}$ & $  (\mathbb{Z}_{2})^2\times \mathbb{Z}_4\times \mZ_8$& \underline{$123 (B_{1u}, B_{2u})$}\\
				$P4/mmm (A_{1u})$ & $z_{2,I}^{\beta}, z_{4, I},z_{4}^{m}, z_{16}$ & $(\mathbb{Z}_2)^4 \times \mathbb{Z}_{4} \times \mZ_8\times \mZ_{16}$& \underline{$123 (A_{1u}, A_{2u})$}\\
				$P4/mcc (B_{1u})$ & $\tilde{z}_{2,I}, z_{8}$ & $  \mathbb{Z}_{2}\times \mathbb{Z}_4\times \mZ_8$& $124 (B_{1u}, B_{2u})$\\
				$I4_1/amd  (A_{2u})$ & $z_{2, I, C_2}^{\pm}, z_{2, I, M_x}^{\pm}, z_{4, S_4, B}, \tilde{z}_{8,I}$ &$(\mathbb{Z}_2)^3 \times \mathbb{Z}_{4}$&  $141(A_{2u})$\\
				\hline\hline
			\end{tabular}
		}
	\end{center}
\end{table*}

\begin{table*}
	\begin{center}
		\label{tab:XBS3}(Continued from the previous page)
		\scalebox{0.8}[0.8]{
		\begin{tabular}{cc|cc}
			\hline\hline
			\multirow{3}{*}{$P\bar{3}$ ($A_{u}$)} &\multirow{2}{*}{$z_{2,I, C_3}^{\beta}, z_{4,I}, z_{8,I}$}& $(\mathbb{Z}_2)^3\times\mathbb{Z}_4\times\mathbb{Z}_ 8$ & \underline{$147(A_u)$}\\
			&& $(\mathbb{Z}_ 2)^2 \times \mathbb{Z}_4$ &$162(A_{1u}), 163(A_{2u}), 164(A_{1u}), 165(A_{2u}).$\\
			&&$\mathbb{Z}_ 2 \times \mathbb{Z}_4$ &$163(A_{1u}), 165(A_{1u}).$\\
			\multirow{3}{*}{$R\bar{3}$ ($A_{u}$)} &\multirow{2}{*}{$z_{2,I, C_3}^{\beta}, z_{4,I}, z_{8,I}$}& $(\mathbb{Z}_2)^3\times\mathbb{Z}_4\times\mathbb{Z}_ 8$ & $148(A_u)$\\
			&& $(\mathbb{Z}_2)^2\times\mathbb{Z}_ 4$ & $166(A_{1u}), 167(A_{2u}).$\\
			&& $\mathbb{Z}_2\times\mathbb{Z}_ 4$ & $167(A_{1u})$\\
			\hline
			\multirow{2}{*}{$P312$, $P321$ ($A_{2}$)} & \multirow{2}{*}{$z_{2,C_2}$} & \multirow{2}{*}{$\mZ_2$} & $149(A_2), 150(A_2), 151(A_2), 152(A_2), 153(A_2),$\\ 
			&&& $154(A_2), 178(A_2), 179(A_2), 182(A_2).$ \\
			$R321$ ($A_{2}$) & $z_{2,C_2}$ & $\mZ_2$& $155(A_2)$ \\
			$P\bar{3}1m$, $P\bar{3}m1$ ($A_{1u}$) & $z_{2, I}^{\pm},z_{2,I, C_3}^{\beta}, \tilde{z}_{8,I}$ &$(\mathbb{Z}_2)^4\times\mathbb{Z}_4$& $162 (A_{2u}), 164 (A_{2u}).$\\
			$R\bar{3}m1$ ($A_{1u}$) & $z_{2, I}^{\pm},z_{2,I, C_3}^{\beta}, \tilde{z}_{8,I}$ &$(\mathbb{Z}_2)^4\times\mathbb{Z}_4$& $166 (A_{2u})$\\
			\hline
			\multirow{8}{*}{$P\bar{6}$} &\multirow{8}{*}{$z_{2,S_6}^{\beta}, z_{3, k_z = 0, \pi}^{m}$}& $(\mathbb{Z}_2)^5\times(\mathbb{Z}_6)^2$ & \underline{$174(A'')$}\\
			&& $\mZ_2 \times(\mathbb{Z}_6)^2$ & $187(A''_1, A''_2),189(A''_1, A''_2)$\\
			&& $(\mathbb{Z}_2)^3\times\mathbb{Z}_6$ & $188(A''_2), 190(A''_2)$\\
			&& $(\mathbb{Z}_2)^2\times\mathbb{Z}_6$ & $188(A''_1), 190(A''_1)$\\
	        && $\mZ_3 \times \mZ_6$ & $191 (B_{1g}, B_{2g})$\\
	        && $\mZ_2 \times \mZ_6$ & $176 (B_{g})$\\
			&& $\mathbb{Z}_ 6$ & $192 (B_{1g}, B_{2g}), 193 (B_{1g}, B_{2g}),194 (B_{1g}, B_{2g})$\\
			\hline
			\multirow{2}{*}{$P6/m$ ($A_{u}$)}& $z_{2,I,C_6}^{\beta} z_{2,S_6}^{\beta} z_{2,I,C_2}^{\beta}, z_{6,k_z=\pi}^{m}, z_{12} $ & $(\mZ_2)^8\times\mZ_6 \times \mZ_{12}$& \underline{$175(A_u)$}\\
			&& $(\mathbb{Z}_2)^{4}\times\mathbb{Z}_{12}$ & $192(A_{1u}, A_{2u}).$ \\
			\multirow{1}{*}{$P6/m$ ($B_{g}$)}& $z_{2,S_6}^{\beta}, z_{6, k_z = 0, \pi}^{m}$ & $ \mathbb{Z}_2\times(\mathbb{Z}_{6})^2$ &  \underline{$175(B_g)$}\\
			\multirow{2}{*}{$P6_3/m$ ($A_{u}$)}&\multirow{2}{*}{$z_{2,I, C_3}^{\beta}, z_{2, S_6}^{\beta}, z_{2,I}, z_{12}$}& $(\mathbb{Z}_2)^{4}\times\mathbb{Z}_{12}$ & \underline{$176(A_u)$} \\
			&& $(\mathbb{Z}_2)^{2}\times\mathbb{Z}_{12}$ & $193(A_{1u}),  194(A_{1u})$\\
		$P6/mmm$ ($A_{1u}, A_{2u}$)	&$z_{2,I,C_6}^{\beta} z_{2,S_6}^{\beta} z_{2,I,C_2}^{\beta}, z'_{12}, z_{12} $& $(\mathbb{Z}_2)^{4}\times\mathbb{Z}_{12}\times\mathbb{Z}_{24}$ & \underline{$191(A_{1u}, A_{2u})$} \\
		$P6_3/mcm$, $P6_3/mmc$ ($A_{2u}$)	&$z_{2,I, M}^{\pm}, z_{2,I,C_3}^{\beta} z_{2,S_6}^{\beta}, z_{12} $& $(\mathbb{Z}_2)^{3}\times\mathbb{Z}_{12}$ & $193(A_{2u}), 194(A_{2u})$ \\
		$P6_3/mcm (B_{2u})$, $P6_3/mmc(B_{1u})$	& $z_{2,I, M}^{\pm}, \tilde{z}_{8,I}$& $\mathbb{Z}_2\times\mathbb{Z}_{4}$ & $193(B_{2u}), 194(B_{1u})$ \\
			\hline\hline
		\end{tabular}
	}
	\end{center}
\end{table*}

\begin{table*}
	\begin{center}
		\caption{\label{tab:XBS4}The list of topological indices for point group $T_h$ and $O_h$.}
		\scalebox{0.9}[0.9]{
			\begin{tabular}{cc|cc}
				\hline \hline
				$G_0$&Indices&$\XBS$&Space Groups\\
				\hline
				\multirow{6}{*}{$Pm\bar{3}$} & \multirow{6}{*}{$z^{(1,1,1), \beta}_{2, I, C_3}, \tilde{z}_{2, I}, \tilde{z}_{4, I}, \tilde{z}_{8,I}$} & $(\mathbb{Z}_{2})^3\times \mathbb{Z}_4 \times \mZ_8$ & $200(A_u)$\\
				&& $(\mZ_2)^3\times \mZ_4$ & $201 (A_u)$\\ 
				&& $(\mZ_2)^2\times \mZ_4$ & $205 (A_u)$\\ 
				&& $(\mZ_2)^2$ & $222 (A_{2u}), 224(A_{2u})$\\ 
				&& $\mZ_2\times \mZ_4$ & $223 (A_{2u})$\\ 
				\multirow{6}{*}{$Fm\bar{3}$} & \multirow{6}{*}{$z_{2,I, C_3}^{(111), \beta}, \tilde{z}_{2, I}, \tilde{z}_{8,I}$} & $(\mathbb{Z}_{2})^3\times \mZ_8$ & $202(A_u)$\\
				&& $(\mZ_2)^3\times \mZ_4$ & $203 (A_u)$\\ 
				&& $(\mZ_2)^2\times \mZ_4$ & $225 (A_{2u})$\\ 
				&& $\mZ_2\times \mZ_4$ & $226 (A_{2u})$\\ 
				&& $(\mZ_2)^2$ & $228 (A_{2u})$\\ 
				\multirow{5}{*}{$Im\bar{3}$} & \multirow{5}{*}{$z_{2,I, C_3}^{(111), \beta}, \tilde{z}_{2, I}, \tilde{z}_{8,I}$} & $(\mathbb{Z}_{2})^3\times \mZ_8$ & $204(A_u)$\\
				&& $(\mZ_2)^3\times \mZ_4$ & $206 (A_u)$\\ 
				&& $(\mZ_2)^2\times \mZ_4$ & $229 (A_{2u})$\\ 
				&& $(\mZ_2)^2$ & $230 (A_{2u})$\\ 
				\hline
				\multirow{2}{*}{$Pm\bar{3}m (A_{1u})$} & \multirow{1}{*}{$z_{2,C_4}^{\beta},z_{2,I, C_3}^{(111), \beta}, z_{8}^{m}, z_{16}$} & \multirow{1}{*}{$(\mathbb{Z}_{2})^4\times \mathbb{Z}_8 \times \mZ_{16}$} & \multirow{1}{*}{$221(A_{1u})$}\\
				&&&\\
				\multirow{2}{*}{$Pm\bar{3}m (A_{2u})$} & \multirow{1}{*}{$z_{2,I, C_3}^{(111), \beta}, z_{8}, \tilde{z}_{8,I}$} & \multirow{1}{*}{$\mathbb{Z}_{2}\times \mathbb{Z}_4 \times \mZ_{8}$} & \multirow{1}{*}{$221(A_{2u})$}\\
				&&&\\
				\multirow{2}{*}{$Pn\bar{3}n$} & \multirow{1}{*}{$z_{2,C_4}^{\beta},z_{2,I, C_3}^{(111), \beta}, \tilde{z}_{8,I}$} & \multirow{1}{*}{$(\mathbb{Z}_{2})^2\times \mathbb{Z}_4$} & \multirow{1}{*}{$222(A_{1u})$}\\
				&&&\\
				\multirow{3}{*}{$Pn\bar{3}m, Pm\bar{3}n$} & \multirow{2}{*}{$z_{2,C_2}^{\beta},z_{2,I, C_3}^{(111), \beta}, \tilde{z}_{8,I}$} & \multirow{1}{*}{$(\mathbb{Z}_{2})^3\times \mathbb{Z}_8$} & \multirow{1}{*}{$223(A_{1u})$}\\
				& & \multirow{1}{*}{$(\mathbb{Z}_{2})^3\times \mathbb{Z}_4$} & \multirow{1}{*}{$224(A_{1u})$}\\
				&&&\\
				\hline
					\multirow{2}{*}{$Fm\bar{3}m$} & \multirow{1}{*}{$z_{2,C_4}^{\beta},z_{2,I, C_3}^{(111), \beta}, z_{2,F}, z_{16}$} & \multirow{1}{*}{$(\mathbb{Z}_{2})^4 \times \mZ_{16}$} & \multirow{1}{*}{$225(A_{1u})$}\\
				&&&\\
				\multirow{2}{*}{$Fm\bar{3}c$} & \multirow{1}{*}{$z_{2,C_4}^{\beta},z_{2,I, C_3}^{(111), \beta}, z_{2,F}, z_{4,S_4}, \tilde{z}_{8,I}$} & \multirow{1}{*}{$(\mathbb{Z}_{2})^3\times \mathbb{Z}_8$} & \multirow{1}{*}{$226(A_{1u})$}\\
				&&&\\
				\multirow{2}{*}{$Fd\bar{3}m$} & \multirow{2}{*}{$z_{2,C_2}^{\beta},z_{2,I, C_3}^{(111), \beta},z_{4,S_4},\tilde{z}_{8,I}$} & \multirow{1}{*}{$(\mathbb{Z}_{2})^2\times \mathbb{Z}_4$} & \multirow{1}{*}{$227(A_{1u})$}\\
				&&&\\
				\multirow{2}{*}{$Fd\bar{3}c$} & \multirow{2}{*}{$z_{2,C_2}^{\beta},z_{2,I, C_3}^{(111), \beta},  z_{4,S_4},\tilde{z}_{8,I}$} & \multirow{2}{*}{$(\mathbb{Z}_{2})^2\times \mathbb{Z}_4$} & \multirow{2}{*}{$228(A_{1u})$}\\
				&&&\\
				\hline
				\multirow{2}{*}{$Im\bar{3}m$} & \multirow{1}{*}{$z_{2,C_4}^{\beta},z_{2,I, C_3}^{(111), \beta}, z_{2,B}, z'_{2,B} z_{16}$} & \multirow{1}{*}{$(\mathbb{Z}_{2})^4 \times \mZ_{16}$} & \multirow{1}{*}{$229(A_{1u})$}\\
				&&&\\
				\multirow{2}{*}{$Ia\bar{3}d$} & \multirow{2}{*}{$z_{2,I, C_3}^{(111), \beta},  \tilde{z}_{4,S_4},\tilde{z}_{8,I}$} & \multirow{2}{*}{$(\mathbb{Z}_{2})^2\times \mathbb{Z}_4$} & \multirow{2}{*}{$230(A_{1u})$}\\
				&&&\\
				\hline\hline
			\end{tabular}
		}
	\end{center}
\end{table*}

\end{document}

%% file: SG_Spinful.tex
\begin{align*}
\begin{array}{lll}
		\text{SG} & \text{Irrep} & X_{\text{BS}} \\ \hline
		1 & \text{A} & \{\} \\
		2 & A_u & \{2,2,2,4,4,4,8\} \\
		2 & A_g & \{\} \\
		3 & \text{B} & \{2,2,4\} \\
		3 & \text{A} & \{\} \\
		4 & \text{B} & \{\} \\
		4 & \text{A} & \{\} \\
		5 & \text{B} & \{2\} \\
		5 & \text{A} & \{\} \\
		6 & \text{A''} & \{2\} \\
		6 & \text{A'} & \{\} \\
		7 & \text{A''} & \{\} \\
		7 & \text{A'} & \{\} \\
		8 & \text{A''} & \{\} \\
		8 & \text{A'} & \{\} \\
		9 & \text{A''} & \{\} \\
		9 & \text{A'} & \{\} \\
		10 & B_u & \{2,2,2,2,2,2,4,4,4\} \\
		10 & B_g & \{2\} \\
		10 & A_u & \{2,2,2,2,2,4\} \\
		10 & A_g & \{\} \\
		11 & B_u & \{2,2,4\} \\
		11 & B_g & \{\} \\
		11 & A_u & \{2,2,2,4\} \\
		11 & A_g & \{\} \\
		12 & B_u & \{2,2,2,2,4,4\} \\
		12 & B_g & \{\} \\
		12 & A_u & \{2,2,4,4\} \\
		12 & A_g & \{\} \\
		13 & B_u & \{2,2,2,4,4\} \\
		13 & B_g & \{2\} \\
		13 & A_u & \{2,2,4\} \\
		13 & A_g & \{\} \\
		14 & B_u & \{2,2,4\} \\
		14 & B_g & \{\} \\
		14 & A_u & \{2,2,4\} \\
		14 & A_g & \{\} \\
		15 & B_u & \{2,2,4,4\} \\
		15 & B_g & \{\} \\
		15 & A_u & \{2,4,4\} \\
		15 & A_g & \{\} \\
		16 & B_1 & \{\} \\
		16 & B_2 & \{\} \\
		16 & B_3 & \{\} \\
		16 & A_1 & \{\} \\
		17 & B_1 & \{2,2,2\} \\
		17 & B_2 & \{2\} \\
		17 & B_3 & \{2\} \\
		17 & A_1 & \{\} \\
	\end{array}&&
	\begin{array}{lll}
		\text{SG} & \text{Irrep} & X_{\text{BS}} \\ \hline
		18 & B_1 & \{\} \\
		18 & B_2 & \{2\} \\
		18 & B_3 & \{2\} \\
		18 & A_1 & \{\} \\
		19 & B_1 & \{\} \\
		19 & B_2 & \{\} \\
		19 & B_3 & \{\} \\
		19 & A_1 & \{\} \\
		20 & B_1 & \{2\} \\
		20 & B_2 & \{\} \\
		20 & B_3 & \{\} \\
		20 & A_1 & \{\} \\
		21 & B_1 & \{\} \\
		21 & B_2 & \{2\} \\
		21 & B_3 & \{2\} \\
		21 & A_1 & \{\} \\
		22 & B_1 & \{\} \\
		22 & B_2 & \{\} \\
		22 & B_3 & \{\} \\
		22 & A_1 & \{\} \\
		23 & B_1 & \{\} \\
		23 & B_2 & \{\} \\
		23 & B_3 & \{\} \\
		23 & A_1 & \{\} \\
		24 & B_1 & \{2\} \\
		24 & B_2 & \{2\} \\
		24 & B_3 & \{2\} \\
		24 & A_1 & \{\} \\
		25 & B_1 & \{\} \\
		25 & B_2 & \{\} \\
		25 & A_2 & \{\} \\
		25 & A_1 & \{\} \\
		26 & B_1 & \{\} \\
		26 & B_2 & \{\} \\
		26 & A_2 & \{\} \\
		26 & A_1 & \{\} \\
		27 & B_1 & \{\} \\
		27 & B_2 & \{\} \\
		27 & A_2 & \{\} \\
		27 & A_1 & \{\} \\
		28 & B_1 & \{2,2\} \\
		28 & B_2 & \{2\} \\
		28 & A_2 & \{\} \\
		28 & A_1 & \{\} \\
		29 & B_1 & \{\} \\
		29 & B_2 & \{\} \\
		29 & A_2 & \{\} \\
		29 & A_1 & \{\} \\
		30 & B_1 & \{2\} \\
	\end{array}&&
	\begin{array}{lll}
		\text{SG} & \text{Irrep} & X_{\text{BS}} \\ \hline
		30 & B_2 & \{2\} \\
		30 & A_2 & \{\} \\
		30 & A_1 & \{\} \\
		31 & B_1 & \{\} \\
		31 & B_2 & \{\} \\
		31 & A_2 & \{\} \\
		31 & A_1 & \{\} \\
		32 & B_1 & \{2\} \\
		32 & B_2 & \{2\} \\
		32 & A_2 & \{\} \\
		32 & A_1 & \{\} \\
		33 & B_1 & \{\} \\
		33 & B_2 & \{\} \\
		33 & A_2 & \{\} \\
		33 & A_1 & \{\} \\
		34 & B_1 & \{2\} \\
		34 & B_2 & \{2\} \\
		34 & A_2 & \{\} \\
		34 & A_1 & \{\} \\
		35 & B_1 & \{2\} \\
		35 & B_2 & \{2\} \\
		35 & A_2 & \{\} \\
		35 & A_1 & \{\} \\
		36 & B_1 & \{\} \\
		36 & B_2 & \{\} \\
		36 & A_2 & \{\} \\
		36 & A_1 & \{\} \\
		37 & B_1 & \{2\} \\
		37 & B_2 & \{2\} \\
		37 & A_2 & \{\} \\
		37 & A_1 & \{\} \\
		38 & B_1 & \{\} \\
		38 & B_2 & \{\} \\
		38 & A_2 & \{\} \\
		38 & A_1 & \{\} \\
		39 & B_1 & \{\} \\
		39 & B_2 & \{\} \\
		39 & A_2 & \{\} \\
		39 & A_1 & \{\} \\
		40 & B_1 & \{2\} \\
		40 & B_2 & \{\} \\
		40 & A_2 & \{\} \\
		40 & A_1 & \{\} \\
		41 & B_1 & \{\} \\
		41 & B_2 & \{\} \\
		41 & A_2 & \{\} \\
		41 & A_1 & \{\} \\
		42 & B_1 & \{\} \\
		42 & B_2 & \{\} \\
	\end{array}&&
	\begin{array}{ccc}
		\text{SG} & \text{Irrep} & X_{\text{BS}} \\ \hline
		42 & A_2 & \{\} \\
		42 & A_1 & \{\} \\
		43 & B_1 & \{\} \\
		43 & B_2 & \{\} \\
		43 & A_2 & \{\} \\
		43 & A_1 & \{\} \\
		44 & B_1 & \{\} \\
		44 & B_2 & \{\} \\
		44 & A_2 & \{\} \\
		44 & A_1 & \{\} \\
		45 & B_1 & \{\} \\
		45 & B_2 & \{\} \\
		45 & A_2 & \{\} \\
		45 & A_1 & \{\} \\
		46 & B_1 & \{2\} \\
		46 & B_2 & \{2\} \\
		46 & A_2 & \{\} \\
		46 & A_1 & \{\} \\
		47 & B_{1u} & \{2,2,2,4,4,4,8\} \\
		47 & B_{1g} & \{\} \\
		47 & B_{2u} & \{2,2,2,4,4,4,8\} \\
		47 & B_{2g} & \{\} \\
		47 & B_{3u} & \{2,2,2,4,4,4,8\} \\
		47 & B_{3g} & \{\} \\
		47 & A_u & \{2,2,2,4,4,4,8\} \\
		47 & A_g & \{\} \\
		48 & B_{1u} & \{2,4\} \\
		48 & B_{1g} & \{\} \\
		48 & B_{2u} & \{2,4\} \\
		48 & B_{2g} & \{\} \\
		48 & B_{3u} & \{2,4\} \\
		48 & B_{3g} & \{\} \\
		48 & A_u & \{2,4\} \\
		48 & A_g & \{\} \\
		49 & B_{1u} & \{2,2,2,4\} \\
		49 & B_{1g} & \{\} \\
		49 & B_{2u} & \{2,2,4\} \\
		49 & B_{2g} & \{\} \\
		49 & B_{3u} & \{2,2,4\} \\
		49 & B_{3g} & \{\} \\
		49 & A_u & \{2,2,2,4\} \\
		49 & A_g & \{\} \\
		50 & B_{1u} & \{2,4\} \\
		50 & B_{1g} & \{\} \\
		50 & B_{2u} & \{2,4\} \\
		50 & B_{2g} & \{\} \\
		50 & B_{3u} & \{2,4\} \\
		50 & B_{3g} & \{\} \\
		50 & A_u & \{2,4\} \\
	\end{array}
\end{align*}
\begin{align*}
	\begin{array}{lll}
		\text{SG} & \text{Irrep} & X_{\text{BS}} \\ \hline
		50 & A_g & \{\} \\
		51 & B_{1u} & \{2,2,2,4,4\} \\
		51 & B_{1g} & \{2\} \\
		51 & B_{2u} & \{2,2,4\} \\
		51 & B_{2g} & \{\} \\
		51 & B_{3u} & \{2,2,2,4,4\} \\
		51 & B_{3g} & \{2\} \\
		51 & A_u & \{2,2,4\} \\
		51 & A_g & \{\} \\
		52 & B_{1u} & \{2,4\} \\
		52 & B_{1g} & \{\} \\
		52 & B_{2u} & \{2,2,4\} \\
		52 & B_{2g} & \{2\} \\
		52 & B_{3u} & \{2,4\} \\
		52 & B_{3g} & \{\} \\
		52 & A_u & \{4\} \\
		52 & A_g & \{\} \\
		53 & B_{1u} & \{2,2,2,2,4\} \\
		53 & B_{1g} & \{2\} \\
		53 & B_{2u} & \{2,2,2,4\} \\
		53 & B_{2g} & \{\} \\
		53 & B_{3u} & \{2,2,4\} \\
		53 & B_{3g} & \{\} \\
		53 & A_u & \{2,4\} \\
		53 & A_g & \{\} \\
		54 & B_{1u} & \{2,2,4\} \\
		54 & B_{1g} & \{2\} \\
		54 & B_{2u} & \{2,4\} \\
		54 & B_{2g} & \{\} \\
		54 & B_{3u} & \{2,2,4\} \\
		54 & B_{3g} & \{2\} \\
		54 & A_u & \{2,4\} \\
		54 & A_g & \{\} \\
		55 & B_{1u} & \{2,2,4\} \\
		55 & B_{1g} & \{\} \\
		55 & B_{2u} & \{2,2,2,4,4\} \\
		55 & B_{2g} & \{2\} \\
		55 & B_{3u} & \{2,2,2,4,4\} \\
		55 & B_{3g} & \{2\} \\
		55 & A_u & \{2,2,4\} \\
		55 & A_g & \{\} \\
		56 & B_{1u} & \{2,4\} \\
		56 & B_{1g} & \{\} \\
		56 & B_{2u} & \{2,4\} \\
		56 & B_{2g} & \{\} \\
		56 & B_{3u} & \{2,4\} \\
		56 & B_{3g} & \{\} \\
		56 & A_u & \{2,4\} \\
		56 & A_g & \{\} \\
	\end{array}&&
	\begin{array}{lll}
		\text{SG} & \text{Irrep} & X_{\text{BS}} \\ \hline
		57 & B_{1u} & \{2,2,4\} \\
		57 & B_{1g} & \{2\} \\
		57 & B_{2u} & \{2,2,4\} \\
		57 & B_{2g} & \{2\} \\
		57 & B_{3u} & \{2,4\} \\
		57 & B_{3g} & \{\} \\
		57 & A_u & \{2,4\} \\
		57 & A_g & \{\} \\
		58 & B_{1u} & \{2,4\} \\
		58 & B_{1g} & \{\} \\
		58 & B_{2u} & \{2,2,2,4\} \\
		58 & B_{2g} & \{\} \\
		58 & B_{3u} & \{2,2,2,4\} \\
		58 & B_{3g} & \{\} \\
		58 & A_u & \{2,4\} \\
		58 & A_g & \{\} \\
		59 & B_{1u} & \{2,4\} \\
		59 & B_{1g} & \{\} \\
		59 & B_{2u} & \{2,4\} \\
		59 & B_{2g} & \{\} \\
		59 & B_{3u} & \{2,4\} \\
		59 & B_{3g} & \{\} \\
		59 & A_u & \{2,4\} \\
		59 & A_g & \{\} \\
		60 & B_{1u} & \{2,4\} \\
		60 & B_{1g} & \{\} \\
		60 & B_{2u} & \{4\} \\
		60 & B_{2g} & \{\} \\
		60 & B_{3u} & \{2,4\} \\
		60 & B_{3g} & \{\} \\
		60 & A_u & \{4\} \\
		60 & A_g & \{\} \\
		61 & B_{1u} & \{4\} \\
		61 & B_{1g} & \{\} \\
		61 & B_{2u} & \{4\} \\
		61 & B_{2g} & \{\} \\
		61 & B_{3u} & \{4\} \\
		61 & B_{3g} & \{\} \\
		61 & A_u & \{4\} \\
		61 & A_g & \{\} \\
		62 & B_{1u} & \{2,4\} \\
		62 & B_{1g} & \{\} \\
		62 & B_{2u} & \{2,4\} \\
		62 & B_{2g} & \{\} \\
		62 & B_{3u} & \{2,4\} \\
		62 & B_{3g} & \{\} \\
		62 & A_u & \{2,4\} \\
		62 & A_g & \{\} \\
		63 & B_{1u} & \{2,2,4\} \\
	\end{array}&&
	\begin{array}{lll}
		\text{SG} & \text{Irrep} & X_{\text{BS}} \\ \hline
		63 & B_{1g} & \{\} \\
		63 & B_{2u} & \{2,2,4\} \\
		63 & B_{2g} & \{\} \\
		63 & B_{3u} & \{2,4\} \\
		63 & B_{3g} & \{\} \\
		63 & A_u & \{2,4\} \\
		63 & A_g & \{\} \\
		64 & B_{1u} & \{2,2,4\} \\
		64 & B_{1g} & \{\} \\
		64 & B_{2u} & \{2,2,4\} \\
		64 & B_{2g} & \{\} \\
		64 & B_{3u} & \{2,4\} \\
		64 & B_{3g} & \{\} \\
		64 & A_u & \{2,4\} \\
		64 & A_g & \{\} \\
		65 & B_{1u} & \{2,2,4,8\} \\
		65 & B_{1g} & \{\} \\
		65 & B_{2u} & \{2,2,2,2,4,8\} \\
		65 & B_{2g} & \{\} \\
		65 & B_{3u} & \{2,2,2,2,4,8\} \\
		65 & B_{3g} & \{\} \\
		65 & A_u & \{2,2,4,8\} \\
		65 & A_g & \{\} \\
		66 & B_{1u} & \{2,2,4\} \\
		66 & B_{1g} & \{\} \\
		66 & B_{2u} & \{2,2,2,4\} \\
		66 & B_{2g} & \{\} \\
		66 & B_{3u} & \{2,2,2,4\} \\
		66 & B_{3g} & \{\} \\
		66 & A_u & \{2,2,4\} \\
		66 & A_g & \{\} \\
		67 & B_{1u} & \{2,2,4\} \\
		67 & B_{1g} & \{\} \\
		67 & B_{2u} & \{2,2,4\} \\
		67 & B_{2g} & \{\} \\
		67 & B_{3u} & \{2,2,4\} \\
		67 & B_{3g} & \{\} \\
		67 & A_u & \{2,2,4\} \\
		67 & A_g & \{\} \\
		68 & B_{1u} & \{2,4\} \\
		68 & B_{1g} & \{\} \\
		68 & B_{2u} & \{2,4\} \\
		68 & B_{2g} & \{\} \\
		68 & B_{3u} & \{2,4\} \\
		68 & B_{3g} & \{\} \\
		68 & A_u & \{2,4\} \\
		68 & A_g & \{\} \\
		69 & B_{1u} & \{2,2,2,8\} \\
		69 & B_{1g} & \{\} \\
	\end{array}&&
	\begin{array}{ccc}
		\text{SG} & \text{Irrep} & X_{\text{BS}} \\ \hline
		69 & B_{2u} & \{2,2,2,8\} \\
		69 & B_{2g} & \{\} \\
		69 & B_{3u} & \{2,2,2,8\} \\
		69 & B_{3g} & \{\} \\
		69 & A_u & \{2,2,2,8\} \\
		69 & A_g & \{\} \\
		70 & B_{1u} & \{2,4\} \\
		70 & B_{1g} & \{\} \\
		70 & B_{2u} & \{2,4\} \\
		70 & B_{2g} & \{\} \\
		70 & B_{3u} & \{2,4\} \\
		70 & B_{3g} & \{\} \\
		70 & A_u & \{2,4\} \\
		70 & A_g & \{\} \\
		71 & B_{1u} & \{2,2,2,8\} \\
		71 & B_{1g} & \{\} \\
		71 & B_{2u} & \{2,2,2,8\} \\
		71 & B_{2g} & \{\} \\
		71 & B_{3u} & \{2,2,2,8\} \\
		71 & B_{3g} & \{\} \\
		71 & A_u & \{2,2,2,8\} \\
		71 & A_g & \{\} \\
		72 & B_{1u} & \{2,4\} \\
		72 & B_{1g} & \{\} \\
		72 & B_{2u} & \{2,4\} \\
		72 & B_{2g} & \{\} \\
		72 & B_{3u} & \{2,4\} \\
		72 & B_{3g} & \{\} \\
		72 & A_u & \{2,4\} \\
		72 & A_g & \{\} \\
		73 & B_{1u} & \{2,4\} \\
		73 & B_{1g} & \{\} \\
		73 & B_{2u} & \{2,4\} \\
		73 & B_{2g} & \{\} \\
		73 & B_{3u} & \{2,4\} \\
		73 & B_{3g} & \{\} \\
		73 & A_u & \{2,4\} \\
		73 & A_g & \{\} \\
		74 & B_{1u} & \{2,2,2,2,4\} \\
		74 & B_{1g} & \{\} \\
		74 & B_{2u} & \{2,2,4\} \\
		74 & B_{2g} & \{\} \\
		74 & B_{3u} & \{2,2,4\} \\
		74 & B_{3g} & \{\} \\
		74 & A_u & \{2,4\} \\
		74 & A_g & \{\} \\
		75 & \text{B} & \{2,2\} \\
		75 & \text{A} & \{\} \\
		76 & \text{B} & \{\} \\
	\end{array}
\end{align*}
\begin{align*}
	\begin{array}{lll}
		\text{SG} & \text{Irrep} & X_{\text{BS}} \\ \hline
		76 & \text{A} & \{\} \\
		77 & \text{B} & \{\} \\
		77 & \text{A} & \{\} \\
		78 & \text{B} & \{\} \\
		78 & \text{A} & \{\} \\
		79 & \text{B} & \{\} \\
		79 & \text{A} & \{\} \\
		80 & \text{B} & \{\} \\
		80 & \text{A} & \{\} \\
		81 & \text{B} & \{2,2,2,2,4,4\} \\
		81 & \text{A} & \{\} \\
		82 & \text{B} & \{2,2,4,4,4\} \\
		82 & \text{A} & \{\} \\
		83 & B_u & \{2,2,4\} \\
		83 & B_g & \{2,2,2,2,2,2\} \\
		83 & A_u & \{2,2,2,2,2,2,2,4,8\} \\
		83 & A_g & \{\} \\
		84 & B_u & \{2,2\} \\
		84 & B_g & \{2,2\} \\
		84 & A_u & \{2,2,2,2,2,4\} \\
		84 & A_g & \{\} \\
		85 & B_u & \{2,2\} \\
		85 & B_g & \{2,2,2,2\} \\
		85 & A_u & \{2,2,2,4\} \\
		85 & A_g & \{\} \\
		86 & B_u & \{2\} \\
		86 & B_g & \{2,2\} \\
		86 & A_u & \{2,2,2,4\} \\
		86 & A_g & \{\} \\
		87 & B_u & \{2,2,2\} \\
		87 & B_g & \{2,2,2,2\} \\
		87 & A_u & \{2,2,2,2,2,2,8\} \\
		87 & A_g & \{\} \\
		88 & B_u & \{4\} \\
		88 & B_g & \{2,4\} \\
		88 & A_u & \{2,2,4,4\} \\
		88 & A_g & \{\} \\
		89 & B_2 & \{2\} \\
		89 & B_1 & \{2\} \\
		89 & A_2 & \{\} \\
		89 & A_1 & \{\} \\
		90 & B_2 & \{2\} \\
		90 & B_1 & \{2\} \\
		90 & A_2 & \{\} \\
		90 & A_1 & \{\} \\
		91 & B_2 & \{2\} \\
		91 & B_1 & \{\} \\
		91 & A_2 & \{2,2\} \\
		91 & A_1 & \{\} \\
	\end{array}&&
	\begin{array}{lll}
		\text{SG} & \text{Irrep} & X_{\text{BS}} \\ \hline
		92 & B_2 & \{\} \\
		92 & B_1 & \{\} \\
		92 & A_2 & \{\} \\
		92 & A_1 & \{\} \\
		93 & B_2 & \{\} \\
		93 & B_1 & \{\} \\
		93 & A_2 & \{\} \\
		93 & A_1 & \{\} \\
		94 & B_2 & \{\} \\
		94 & B_1 & \{\} \\
		94 & A_2 & \{\} \\
		94 & A_1 & \{\} \\
		95 & B_2 & \{2\} \\
		95 & B_1 & \{\} \\
		95 & A_2 & \{2,2\} \\
		95 & A_1 & \{\} \\
		96 & B_2 & \{\} \\
		96 & B_1 & \{\} \\
		96 & A_2 & \{\} \\
		96 & A_1 & \{\} \\
		97 & B_2 & \{\} \\
		97 & B_1 & \{\} \\
		97 & A_2 & \{\} \\
		97 & A_1 & \{\} \\
		98 & B_2 & \{2\} \\
		98 & B_1 & \{\} \\
		98 & A_2 & \{2\} \\
		98 & A_1 & \{\} \\
		99 & B_2 & \{2\} \\
		99 & B_1 & \{2\} \\
		99 & A_2 & \{\} \\
		99 & A_1 & \{\} \\
		100 & B_2 & \{2\} \\
		100 & B_1 & \{2\} \\
		100 & A_2 & \{\} \\
		100 & A_1 & \{\} \\
		101 & B_2 & \{\} \\
		101 & B_1 & \{\} \\
		101 & A_2 & \{\} \\
		101 & A_1 & \{\} \\
		102 & B_2 & \{\} \\
		102 & B_1 & \{\} \\
		102 & A_2 & \{\} \\
		102 & A_1 & \{\} \\
		103 & B_2 & \{2\} \\
		103 & B_1 & \{2\} \\
		103 & A_2 & \{\} \\
		103 & A_1 & \{\} \\
		104 & B_2 & \{2\} \\
	\end{array}&&
	\begin{array}{lll}
		\text{SG} & \text{Irrep} & X_{\text{BS}} \\ \hline
		104 & B_1 & \{2\} \\
		104 & A_2 & \{\} \\
		104 & A_1 & \{\} \\
		105 & B_2 & \{\} \\
		105 & B_1 & \{\} \\
		105 & A_2 & \{\} \\
		105 & A_1 & \{\} \\
		106 & B_2 & \{\} \\
		106 & B_1 & \{\} \\
		106 & A_2 & \{\} \\
		106 & A_1 & \{\} \\
		107 & B_2 & \{\} \\
		107 & B_1 & \{\} \\
		107 & A_2 & \{\} \\
		107 & A_1 & \{\} \\
		108 & B_2 & \{\} \\
		108 & B_1 & \{\} \\
		108 & A_2 & \{\} \\
		108 & A_1 & \{\} \\
		109 & B_2 & \{\} \\
		109 & B_1 & \{\} \\
		109 & A_2 & \{\} \\
		109 & A_1 & \{\} \\
		110 & B_2 & \{\} \\
		110 & B_1 & \{\} \\
		110 & A_2 & \{\} \\
		110 & A_1 & \{\} \\
		111 & B_2 & \{2,2,4\} \\
		111 & A_2 & \{\} \\
		111 & B_1 & \{2,2,4\} \\
		111 & A_1 & \{\} \\
		112 & B_2 & \{2,2,4\} \\
		112 & A_2 & \{\} \\
		112 & B_1 & \{2,2,4\} \\
		112 & A_1 & \{\} \\
		113 & B_2 & \{2,2,4\} \\
		113 & A_2 & \{\} \\
		113 & B_1 & \{2,2,4\} \\
		113 & A_1 & \{\} \\
		114 & B_2 & \{2,2,4\} \\
		114 & A_2 & \{\} \\
		114 & B_1 & \{2,2,4\} \\
		114 & A_1 & \{\} \\
		115 & B_2 & \{2,2,4\} \\
		115 & A_2 & \{\} \\
		115 & B_1 & \{2,2,4\} \\
		115 & A_1 & \{\} \\
		116 & B_2 & \{2,2,4\} \\
		116 & A_2 & \{\} \\
	\end{array}&&
	\begin{array}{ccc}
		\text{SG} & \text{Irrep} & X_{\text{BS}} \\ \hline
		116 & B_1 & \{2,2,4\} \\
		116 & A_1 & \{\} \\
		117 & B_2 & \{2,2,4\} \\
		117 & A_2 & \{\} \\
		117 & B_1 & \{2,2,4\} \\
		117 & A_1 & \{\} \\
		118 & B_2 & \{2,2,4\} \\
		118 & A_2 & \{\} \\
		118 & B_1 & \{2,2,4\} \\
		118 & A_1 & \{\} \\
		119 & B_2 & \{2,2,4\} \\
		119 & A_2 & \{\} \\
		119 & B_1 & \{2,2,4\} \\
		119 & A_1 & \{\} \\
		120 & B_2 & \{2,2,4\} \\
		120 & A_2 & \{\} \\
		120 & B_1 & \{2,2,4\} \\
		120 & A_1 & \{\} \\
		121 & B_2 & \{2,2,4\} \\
		121 & A_2 & \{\} \\
		121 & B_1 & \{2,2,4\} \\
		121 & A_1 & \{\} \\
		122 & B_2 & \{2,2,4\} \\
		122 & A_2 & \{\} \\
		122 & B_1 & \{2,4\} \\
		122 & A_1 & \{\} \\
		123 & B_{2u} & \{2,4,4,8\} \\
		123 & B_{2g} & \{2,2,2\} \\
		123 & B_{1u} & \{2,4,4,8\} \\
		123 & B_{1g} & \{2,2,2\} \\
		123 & A_{2u} & \{2,2,2,2,4,8,16\} \\
		123 & A_{2g} & \{\} \\
		123 & A_{1u} & \{2,2,2,2,4,8,16\} \\
		123 & A_{1g} & \{\} \\
		124 & B_{2u} & \{2,4\} \\
		124 & B_{2g} & \{2,2,2\} \\
		124 & B_{1u} & \{2,4\} \\
		124 & B_{1g} & \{2,2,2\} \\
		124 & A_{2u} & \{2,2,2,2,8\} \\
		124 & A_{2g} & \{\} \\
		124 & A_{1u} & \{2,2,2,2,8\} \\
		124 & A_{1g} & \{\} \\
		125 & B_{2u} & \{2,2\} \\
		125 & B_{2g} & \{2,2\} \\
		125 & B_{1u} & \{2,2\} \\
		125 & B_{1g} & \{2,2\} \\
		125 & A_{2u} & \{2,2,4\} \\
		125 & A_{2g} & \{\} \\
		125 & A_{1u} & \{2,2,4\} \\
	\end{array}
\end{align*}
\begin{align*}
	\begin{array}{lll}
		\text{SG} & \text{Irrep} & X_{\text{BS}} \\ \hline
		125 & A_{1g} & \{\} \\
		126 & B_{2u} & \{2\} \\
		126 & B_{2g} & \{2,2\} \\
		126 & B_{1u} & \{2\} \\
		126 & B_{1g} & \{2,2\} \\
		126 & A_{2u} & \{2,4\} \\
		126 & A_{2g} & \{\} \\
		126 & A_{1u} & \{2,4\} \\
		126 & A_{1g} & \{\} \\
		127 & B_{2u} & \{4,4\} \\
		127 & B_{2g} & \{2,2,2\} \\
		127 & B_{1u} & \{4,4\} \\
		127 & B_{1g} & \{2,2,2\} \\
		127 & A_{2u} & \{2,2,2,4,8\} \\
		127 & A_{2g} & \{\} \\
		127 & A_{1u} & \{2,2,2,4,8\} \\
		127 & A_{1g} & \{\} \\
		128 & B_{2u} & \{4\} \\
		128 & B_{2g} & \{2,2,2\} \\
		128 & B_{1u} & \{4\} \\
		128 & B_{1g} & \{2,2,2\} \\
		128 & A_{2u} & \{2,2,2,8\} \\
		128 & A_{2g} & \{\} \\
		128 & A_{1u} & \{2,2,2,8\} \\
		128 & A_{1g} & \{\} \\
		129 & B_{2u} & \{2,2\} \\
		129 & B_{2g} & \{2,2\} \\
		129 & B_{1u} & \{2,2\} \\
		129 & B_{1g} & \{2,2\} \\
		129 & A_{2u} & \{2,2,4\} \\
		129 & A_{2g} & \{\} \\
		129 & A_{1u} & \{2,2,4\} \\
		129 & A_{1g} & \{\} \\
		130 & B_{2u} & \{2\} \\
		130 & B_{2g} & \{2,2\} \\
		130 & B_{1u} & \{2\} \\
		130 & B_{1g} & \{2,2\} \\
		130 & A_{2u} & \{2,4\} \\
		130 & A_{2g} & \{\} \\
		130 & A_{1u} & \{2,4\} \\
		130 & A_{1g} & \{\} \\
		131 & B_{2u} & \{2,4,4\} \\
		131 & B_{2g} & \{2\} \\
		131 & B_{1u} & \{2,4,4\} \\
		131 & B_{1g} & \{2\} \\
		131 & A_{2u} & \{2,2,2,4,8\} \\
		131 & A_{2g} & \{\} \\
		131 & A_{1u} & \{2,2,2,4,8\} \\
		131 & A_{1g} & \{\} \\
	\end{array}&&
	\begin{array}{lll}
		\text{SG} & \text{Irrep} & X_{\text{BS}} \\ \hline
		132 & B_{2u} & \{2,4\} \\
		132 & B_{2g} & \{2\} \\
		132 & B_{1u} & \{2,4\} \\
		132 & B_{1g} & \{2\} \\
		132 & A_{2u} & \{2,2,2,8\} \\
		132 & A_{2g} & \{\} \\
		132 & A_{1u} & \{2,2,2,8\} \\
		132 & A_{1g} & \{\} \\
		133 & B_{2u} & \{2\} \\
		133 & B_{2g} & \{2\} \\
		133 & B_{1u} & \{2\} \\
		133 & B_{1g} & \{2\} \\
		133 & A_{2u} & \{2,4\} \\
		133 & A_{2g} & \{\} \\
		133 & A_{1u} & \{2,4\} \\
		133 & A_{1g} & \{\} \\
		134 & B_{2u} & \{2\} \\
		134 & B_{2g} & \{2\} \\
		134 & B_{1u} & \{2\} \\
		134 & B_{1g} & \{2\} \\
		134 & A_{2u} & \{2,2,4\} \\
		134 & A_{2g} & \{\} \\
		134 & A_{1u} & \{2,2,4\} \\
		134 & A_{1g} & \{\} \\
		135 & B_{2u} & \{2\} \\
		135 & B_{2g} & \{2\} \\
		135 & B_{1u} & \{2\} \\
		135 & B_{1g} & \{2\} \\
		135 & A_{2u} & \{2,2,4\} \\
		135 & A_{2g} & \{\} \\
		135 & A_{1u} & \{2,2,4\} \\
		135 & A_{1g} & \{\} \\
		136 & B_{2u} & \{4\} \\
		136 & B_{2g} & \{2\} \\
		136 & B_{1u} & \{4\} \\
		136 & B_{1g} & \{2\} \\
		136 & A_{2u} & \{2,2,8\} \\
		136 & A_{2g} & \{\} \\
		136 & A_{1u} & \{2,2,8\} \\
		136 & A_{1g} & \{\} \\
		137 & B_{2u} & \{2\} \\
		137 & B_{2g} & \{2\} \\
		137 & B_{1u} & \{2\} \\
		137 & B_{1g} & \{2\} \\
		137 & A_{2u} & \{2,4\} \\
		137 & A_{2g} & \{\} \\
		137 & A_{1u} & \{2,4\} \\
		137 & A_{1g} & \{\} \\
		138 & B_{2u} & \{2\} \\
	\end{array}&&
	\begin{array}{lll}
		\text{SG} & \text{Irrep} & X_{\text{BS}} \\ \hline
		138 & B_{2g} & \{2\} \\
		138 & B_{1u} & \{2\} \\
		138 & B_{1g} & \{2\} \\
		138 & A_{2u} & \{2,2,4\} \\
		138 & A_{2g} & \{\} \\
		138 & A_{1u} & \{2,2,4\} \\
		138 & A_{1g} & \{\} \\
		139 & B_{2u} & \{2,2,4\} \\
		139 & B_{2g} & \{2,2\} \\
		139 & B_{1u} & \{2,2,4\} \\
		139 & B_{1g} & \{2,2\} \\
		139 & A_{2u} & \{2,2,2,2,16\} \\
		139 & A_{2g} & \{\} \\
		139 & A_{1u} & \{2,2,2,2,16\} \\
		139 & A_{1g} & \{\} \\
		140 & B_{2u} & \{2,4\} \\
		140 & B_{2g} & \{2,2\} \\
		140 & B_{1u} & \{2,4\} \\
		140 & B_{1g} & \{2,2\} \\
		140 & A_{2u} & \{2,2,2,8\} \\
		140 & A_{2g} & \{\} \\
		140 & A_{1u} & \{2,2,2,8\} \\
		140 & A_{1g} & \{\} \\
		141 & B_{2u} & \{2,2\} \\
		141 & B_{2g} & \{2\} \\
		141 & B_{1u} & \{2\} \\
		141 & B_{1g} & \{2\} \\
		141 & A_{2u} & \{2,2,2,4\} \\
		141 & A_{2g} & \{\} \\
		141 & A_{1u} & \{2,4\} \\
		141 & A_{1g} & \{\} \\
		142 & B_{2u} & \{2\} \\
		142 & B_{2g} & \{2\} \\
		142 & B_{1u} & \{2\} \\
		142 & B_{1g} & \{2\} \\
		142 & A_{2u} & \{2,4\} \\
		142 & A_{2g} & \{\} \\
		142 & A_{1u} & \{2,4\} \\
		142 & A_{1g} & \{\} \\
		143 & \text{A} & \{\} \\
		144 & \text{A} & \{\} \\
		145 & \text{A} & \{\} \\
		146 & \text{A} & \{\} \\
		147 & A_u & \{2,2,2,4,8\} \\
		147 & A_g & \{\} \\
		148 & A_u & \{2,2,2,4,8\} \\
		148 & A_g & \{\} \\
		149 & A_2 & \{2\} \\
		149 & A_1 & \{\} \\
	\end{array}&&
	\begin{array}{ccc}
		\text{SG} & \text{Irrep} & X_{\text{BS}} \\ \hline
		150 & A_2 & \{2\} \\
		150 & A_1 & \{\} \\
		151 & A_2 & \{2\} \\
		151 & A_1 & \{\} \\
		152 & A_2 & \{2\} \\
		152 & A_1 & \{\} \\
		153 & A_2 & \{2\} \\
		153 & A_1 & \{\} \\
		154 & A_2 & \{2\} \\
		154 & A_1 & \{\} \\
		155 & A_2 & \{2\} \\
		155 & A_1 & \{\} \\
		156 & A_2 & \{\} \\
		156 & A_1 & \{\} \\
		157 & A_2 & \{\} \\
		157 & A_1 & \{\} \\
		158 & A_2 & \{\} \\
		158 & A_1 & \{\} \\
		159 & A_2 & \{\} \\
		159 & A_1 & \{\} \\
		160 & A_2 & \{\} \\
		160 & A_1 & \{\} \\
		161 & A_2 & \{\} \\
		161 & A_1 & \{\} \\
		162 & A_{2u} & \{2,2,2,2,4\} \\
		162 & A_{2g} & \{\} \\
		162 & A_{1u} & \{2,2,4\} \\
		162 & A_{1g} & \{\} \\
		163 & A_{2u} & \{2,2,4\} \\
		163 & A_{2g} & \{\} \\
		163 & A_{1u} & \{2,4\} \\
		163 & A_{1g} & \{\} \\
		164 & A_{2u} & \{2,2,2,2,4\} \\
		164 & A_{2g} & \{\} \\
		164 & A_{1u} & \{2,2,4\} \\
		164 & A_{1g} & \{\} \\
		165 & A_{2u} & \{2,2,4\} \\
		165 & A_{2g} & \{\} \\
		165 & A_{1u} & \{2,4\} \\
		165 & A_{1g} & \{\} \\
		166 & A_{2u} & \{2,2,2,2,4\} \\
		166 & A_{2g} & \{\} \\
		166 & A_{1u} & \{2,2,4\} \\
		166 & A_{1g} & \{\} \\
		167 & A_{2u} & \{2,2,4\} \\
		167 & A_{2g} & \{\} \\
		167 & A_{1u} & \{2,4\} \\
		167 & A_{1g} & \{\} \\
		168 & \text{B} & \{4\} \\
	\end{array}
\end{align*}
\begin{align*}
	\begin{array}{lll}
		\text{SG} & \text{Irrep} & X_{\text{BS}} \\ \hline
		168 & \text{A} & \{\} \\
		169 & \text{B} & \{\} \\
		169 & \text{A} & \{\} \\
		170 & \text{B} & \{\} \\
		170 & \text{A} & \{\} \\
		171 & \text{B} & \{4\} \\
		171 & \text{A} & \{\} \\
		172 & \text{B} & \{4\} \\
		172 & \text{A} & \{\} \\
		173 & \text{B} & \{\} \\
		173 & \text{A} & \{\} \\
		174 & \text{A''} & \{2,2,2,2,2,6,6\} \\
		174 & \text{A'} & \{\} \\
		175 & B_u & \{4,4,4\} \\
		175 & B_g & \{2,6,6\} \\
		175 & A_u & \{2,2,2,2,2,2,2,2,6,12\} \\
		175 & A_g & \{\} \\
		176 & B_u & \{4\} \\
		176 & B_g & \{2,6\} \\
		176 & A_u & \{2,2,2,2,2,12\} \\
		176 & A_g & \{\} \\
		177 & B_1 & \{\} \\
		177 & B_2 & \{\} \\
		177 & A_2 & \{\} \\
		177 & A_1 & \{\} \\
		178 & B_1 & \{\} \\
		178 & B_2 & \{\} \\
		178 & A_2 & \{2\} \\
		178 & A_1 & \{\} \\
		179 & B_1 & \{\} \\
		179 & B_2 & \{\} \\
		179 & A_2 & \{2\} \\
		179 & A_1 & \{\} \\
		180 & B_1 & \{\} \\
		180 & B_2 & \{\} \\
		180 & A_2 & \{\} \\
		180 & A_1 & \{\} \\
		181 & B_1 & \{\} \\
		181 & B_2 & \{\} \\
		181 & A_2 & \{\} \\
		181 & A_1 & \{\} \\
		182 & B_1 & \{\} \\
		182 & B_2 & \{\} \\
		182 & A_2 & \{2\} \\
		182 & A_1 & \{\} \\
		183 & B_2 & \{\} \\
		183 & B_1 & \{\} \\
		183 & A_2 & \{\} \\
		183 & A_1 & \{\} \\
	\end{array}&&
	\begin{array}{lll}
		\text{SG} & \text{Irrep} & X_{\text{BS}} \\ \hline
		184 & B_2 & \{\} \\
		184 & B_1 & \{\} \\
		184 & A_2 & \{\} \\
		184 & A_1 & \{\} \\
		185 & B_2 & \{\} \\
		185 & B_1 & \{\} \\
		185 & A_2 & \{\} \\
		185 & A_1 & \{\} \\
		186 & B_2 & \{\} \\
		186 & B_1 & \{\} \\
		186 & A_2 & \{\} \\
		186 & A_1 & \{\} \\
		187 & A''_2 & \{2,6,6\} \\
		187 & A'_2 & \{\} \\
		187 & A''_1 & \{2,6,6\} \\
		187 & A'_1 & \{\} \\
		188 & A''_2 & \{2,2,2,6\} \\
		188 & A'_2 & \{\} \\
		188 & A''_1 & \{2,2,6\} \\
		188 & A'_1 & \{\} \\
		189 & A''_2 & \{2,6,6\} \\
		189 & A'_2 & \{\} \\
		189 & A''_1 & \{2,6,6\} \\
		189 & A'_1 & \{\} \\
		190 & A''_2 & \{2,2,2,6\} \\
		190 & A'_2 & \{\} \\
		190 & A''_1 & \{2,2,6\} \\
		190 & A'_1 & \{\} \\
		191 & B_{1u} & \{2,4,8\} \\
		191 & B_{1g} & \{3,6\} \\
		191 & B_{2u} & \{2,4,8\} \\
		191 & B_{2g} & \{3,6\} \\
		191 & A_{2u} & \{2,2,2,2,12,24\} \\
		191 & A_{2g} & \{\} \\
		191 & A_{1u} & \{2,2,2,2,12,24\} \\
		191 & A_{1g} & \{\} \\
		192 & B_{1u} & \{4\} \\
		192 & B_{1g} & \{6\} \\
		192 & B_{2u} & \{4\} \\
		192 & B_{2g} & \{6\} \\
		192 & A_{2u} & \{2,2,2,2,12\} \\
		192 & A_{2g} & \{\} \\
		192 & A_{1u} & \{2,2,2,2,12\} \\
		192 & A_{1g} & \{\} \\
		193 & B_{1u} & \{4\} \\
		193 & B_{1g} & \{6\} \\
		193 & B_{2u} & \{2,4\} \\
		193 & B_{2g} & \{6\} \\
		193 & A_{2u} & \{2,2,2,12\} \\
	\end{array}&&
	\begin{array}{lll}
		\text{SG} & \text{Irrep} & X_{\text{BS}} \\ \hline
		193 & A_{2g} & \{\} \\
		193 & A_{1u} & \{2,2,12\} \\
		193 & A_{1g} & \{\} \\
		194 & B_{1u} & \{2,4\} \\
		194 & B_{1g} & \{6\} \\
		194 & B_{2u} & \{4\} \\
		194 & B_{2g} & \{6\} \\
		194 & A_{2u} & \{2,2,2,12\} \\
		194 & A_{2g} & \{\} \\
		194 & A_{1u} & \{2,2,12\} \\
		194 & A_{1g} & \{\} \\
		195 & \text{A} & \{\} \\
		196 & \text{A} & \{\} \\
		197 & \text{A} & \{\} \\
		198 & \text{A} & \{\} \\
		199 & \text{A} & \{\} \\
		200 & A_u & \{2,2,2,4,8\} \\
		200 & A_g & \{\} \\
		201 & A_u & \{2,2,2,4\} \\
		201 & A_g & \{\} \\
		202 & A_u & \{2,2,2,8\} \\
		202 & A_g & \{\} \\
		203 & A_u & \{2,2,2,4\} \\
		203 & A_g & \{\} \\
		204 & A_u & \{2,2,2,8\} \\
		204 & A_g & \{\} \\
		205 & A_u & \{2,2,4\} \\
		205 & A_g & \{\} \\
		206 & A_u & \{2,2,2,4\} \\
		206 & A_g & \{\} \\
		207 & A_2 & \{2\} \\
		207 & A_1 & \{\} \\
		208 & A_2 & \{\} \\
		208 & A_1 & \{\} \\
		209 & A_2 & \{\} \\
		209 & A_1 & \{\} \\
		210 & A_2 & \{2\} \\
		210 & A_1 & \{\} \\
		211 & A_2 & \{\} \\
		211 & A_1 & \{\} \\
		212 & A_2 & \{\} \\
		212 & A_1 & \{\} \\
		213 & A_2 & \{\} \\
		213 & A_1 & \{\} \\
		214 & A_2 & \{\} \\
		214 & A_1 & \{\} \\
		215 & A_2 & \{2,2,4\} \\
		215 & A_1 & \{\} \\
		216 & A_2 & \{2,2,4\} \\
	\end{array}&&
	\begin{array}{ccc}
		\text{SG} & \text{Irrep} & X_{\text{BS}} \\ \hline
		216 & A_1 & \{\} \\
		217 & A_2 & \{2,2,4\} \\
		217 & A_1 & \{\} \\
		218 & A_2 & \{2,2,4\} \\
		218 & A_1 & \{\} \\
		219 & A_2 & \{2,2,4\} \\
		219 & A_1 & \{\} \\
		220 & A_2 & \{2,4\} \\
		220 & A_1 & \{\} \\
		221 & A_{2u} & \{2,4,8\} \\
		221 & A_{2g} & \{2,2,2\} \\
		221 & A_{1u} & \{2,2,2,2,8,16\} \\
		221 & A_{1g} & \{\} \\
		222 & A_{2u} & \{2,2\} \\
		222 & A_{2g} & \{2,2\} \\
		222 & A_{1u} & \{2,2,4\} \\
		222 & A_{1g} & \{\} \\
		223 & A_{2u} & \{2,4\} \\
		223 & A_{2g} & \{2\} \\
		223 & A_{1u} & \{2,2,2,8\} \\
		223 & A_{1g} & \{\} \\
		224 & A_{2u} & \{2,2\} \\
		224 & A_{2g} & \{2\} \\
		224 & A_{1u} & \{2,2,2,4\} \\
		224 & A_{1g} & \{\} \\
		225 & A_{2u} & \{2,2,4\} \\
		225 & A_{2g} & \{2,2\} \\
		225 & A_{1u} & \{2,2,2,2,16\} \\
		225 & A_{1g} & \{\} \\
		226 & A_{2u} & \{2,4\} \\
		226 & A_{2g} & \{2,2\} \\
		226 & A_{1u} & \{2,2,2,8\} \\
		226 & A_{1g} & \{\} \\
		227 & A_{2u} & \{2,2,2\} \\
		227 & A_{2g} & \{2\} \\
		227 & A_{1u} & \{2,2,4\} \\
		227 & A_{1g} & \{\} \\
		228 & A_{2u} & \{2,2\} \\
		228 & A_{2g} & \{2\} \\
		228 & A_{1u} & \{2,2,4\} \\
		228 & A_{1g} & \{\} \\
		229 & A_{2u} & \{2,2,4\} \\
		229 & A_{2g} & \{2,2\} \\
		229 & A_{1u} & \{2,2,2,2,16\} \\
		229 & A_{1g} & \{\} \\
		230 & A_{2u} & \{2,2\} \\
		230 & A_{2g} & \{2\} \\
		230 & A_{1u} & \{2,2,4\} \\
		230 & A_{1g} & \{\} \\
	\end{array}
\end{align*}

%% file: SG_Spinless.tex
\begin{align*}
	\begin{array}{lll}
		\text{SG} & \text{Irrep} & X_{\text{BS}} \\ \hline
		1 & \text{A} & \{\} \\
		2 & A_u & \{2,2,2,4,4,4,8\} \\
		2 & A_g & \{\} \\
		3 & \text{B} & \{2,2,4\} \\
		3 & \text{A} & \{\} \\
		4 & \text{B} & \{\} \\
		4 & \text{A} & \{\} \\
		5 & \text{B} & \{2\} \\
		5 & \text{A} & \{\} \\
		6 & \text{A''} & \{2\} \\
		6 & \text{A'} & \{\} \\
		7 & \text{A''} & \{\} \\
		7 & \text{A'} & \{\} \\
		8 & \text{A''} & \{\} \\
		8 & \text{A'} & \{\} \\
		9 & \text{A''} & \{\} \\
		9 & \text{A'} & \{\} \\
		10 & B_u & \{2,2,2,2,2,2,4,4,4\} \\
		10 & B_g & \{2\} \\
		10 & A_u & \{2,2,2,2,2,4\} \\
		10 & A_g & \{\} \\
		11 & B_u & \{2,2,4\} \\
		11 & B_g & \{\} \\
		11 & A_u & \{2,2,2,4\} \\
		11 & A_g & \{\} \\
		12 & B_u & \{2,2,2,2,4,4\} \\
		12 & B_g & \{\} \\
		12 & A_u & \{2,2,4,4\} \\
		12 & A_g & \{\} \\
		13 & B_u & \{2,2,2,4,4\} \\
		13 & B_g & \{2\} \\
		13 & A_u & \{2,2,4\} \\
		13 & A_g & \{\} \\
		14 & B_u & \{2,2,4\} \\
		14 & B_g & \{\} \\
		14 & A_u & \{2,2,4\} \\
		14 & A_g & \{\} \\
		15 & B_u & \{2,2,4,4\} \\
		15 & B_g & \{\} \\
		15 & A_u & \{2,4,4\} \\
		15 & A_g & \{\} \\
		16 & B_1 & \{2,2,2,2\} \\
		16 & B_2 & \{2,2,2,2\} \\
		16 & B_3 & \{2,2,2,2\} \\
		16 & A_1 & \{\} \\
		17 & B_1 & \{2,2,2\} \\
		17 & B_2 & \{2\} \\
		17 & B_3 & \{2\} \\
		17 & A_1 & \{\} \\
	\end{array}&&
	\begin{array}{lll}
		\text{SG} & \text{Irrep} & X_{\text{BS}} \\ \hline
		18 & B_1 & \{\} \\
		18 & B_2 & \{2\} \\
		18 & B_3 & \{2\} \\
		18 & A_1 & \{\} \\
		19 & B_1 & \{\} \\
		19 & B_2 & \{\} \\
		19 & B_3 & \{\} \\
		19 & A_1 & \{\} \\
		20 & B_1 & \{2\} \\
		20 & B_2 & \{\} \\
		20 & B_3 & \{\} \\
		20 & A_1 & \{\} \\
		21 & B_1 & \{2,2\} \\
		21 & B_2 & \{2,2\} \\
		21 & B_3 & \{2,2\} \\
		21 & A_1 & \{\} \\
		22 & B_1 & \{2\} \\
		22 & B_2 & \{2\} \\
		22 & B_3 & \{2\} \\
		22 & A_1 & \{\} \\
		23 & B_1 & \{2\} \\
		23 & B_2 & \{2\} \\
		23 & B_3 & \{2\} \\
		23 & A_1 & \{\} \\
		24 & B_1 & \{2\} \\
		24 & B_2 & \{2\} \\
		24 & B_3 & \{2\} \\
		24 & A_1 & \{\} \\
		25 & B_1 & \{2,2,2\} \\
		25 & B_2 & \{2,2,2\} \\
		25 & A_2 & \{\} \\
		25 & A_1 & \{\} \\
		26 & B_1 & \{2\} \\
		26 & B_2 & \{\} \\
		26 & A_2 & \{2\} \\
		26 & A_1 & \{\} \\
		27 & B_1 & \{2,2,4\} \\
		27 & B_2 & \{2,2,4\} \\
		27 & A_2 & \{\} \\
		27 & A_1 & \{\} \\
		28 & B_1 & \{2,2\} \\
		28 & B_2 & \{2\} \\
		28 & A_2 & \{\} \\
		28 & A_1 & \{\} \\
		29 & B_1 & \{\} \\
		29 & B_2 & \{\} \\
		29 & A_2 & \{\} \\
		29 & A_1 & \{\} \\
		30 & B_1 & \{2\} \\
	\end{array}&&
	\begin{array}{lll}
		\text{SG} & \text{Irrep} & X_{\text{BS}} \\ \hline
		30 & B_2 & \{2\} \\
		30 & A_2 & \{\} \\
		30 & A_1 & \{\} \\
		31 & B_1 & \{\} \\
		31 & B_2 & \{\} \\
		31 & A_2 & \{\} \\
		31 & A_1 & \{\} \\
		32 & B_1 & \{2\} \\
		32 & B_2 & \{2\} \\
		32 & A_2 & \{\} \\
		32 & A_1 & \{\} \\
		33 & B_1 & \{\} \\
		33 & B_2 & \{\} \\
		33 & A_2 & \{\} \\
		33 & A_1 & \{\} \\
		34 & B_1 & \{2\} \\
		34 & B_2 & \{2\} \\
		34 & A_2 & \{\} \\
		34 & A_1 & \{\} \\
		35 & B_1 & \{2,2\} \\
		35 & B_2 & \{2,2\} \\
		35 & A_2 & \{\} \\
		35 & A_1 & \{\} \\
		36 & B_1 & \{\} \\
		36 & B_2 & \{\} \\
		36 & A_2 & \{\} \\
		36 & A_1 & \{\} \\
		37 & B_1 & \{2,4\} \\
		37 & B_2 & \{2,4\} \\
		37 & A_2 & \{\} \\
		37 & A_1 & \{\} \\
		38 & B_1 & \{2,2\} \\
		38 & B_2 & \{\} \\
		38 & A_2 & \{\} \\
		38 & A_1 & \{\} \\
		39 & B_1 & \{2\} \\
		39 & B_2 & \{2,2\} \\
		39 & A_2 & \{\} \\
		39 & A_1 & \{\} \\
		40 & B_1 & \{2\} \\
		40 & B_2 & \{\} \\
		40 & A_2 & \{\} \\
		40 & A_1 & \{\} \\
		41 & B_1 & \{\} \\
		41 & B_2 & \{\} \\
		41 & A_2 & \{\} \\
		41 & A_1 & \{\} \\
		42 & B_1 & \{2\} \\
		42 & B_2 & \{2\} \\
	\end{array}&&
	\begin{array}{ccc}
		\text{SG} & \text{Irrep} & X_{\text{BS}} \\ \hline
		42 & A_2 & \{\} \\
		42 & A_1 & \{\} \\
		43 & B_1 & \{\} \\
		43 & B_2 & \{\} \\
		43 & A_2 & \{\} \\
		43 & A_1 & \{\} \\
		44 & B_1 & \{2\} \\
		44 & B_2 & \{2\} \\
		44 & A_2 & \{\} \\
		44 & A_1 & \{\} \\
		45 & B_1 & \{4\} \\
		45 & B_2 & \{4\} \\
		45 & A_2 & \{\} \\
		45 & A_1 & \{\} \\
		46 & B_1 & \{2\} \\
		46 & B_2 & \{\} \\
		46 & A_2 & \{\} \\
		46 & A_1 & \{\} \\
		47 & B_{1u} & \{2,2,2,2,2,2,2,2,2\} \\
		47 & B_{1g} & \{\} \\
		47 & B_{2u} & \{2,2,2,2,2,2,2,2,2\} \\
		47 & B_{2g} & \{\} \\
		47 & B_{3u} & \{2,2,2,2,2,2,2,2,2\} \\
		47 & B_{3g} & \{\} \\
		47 & A_u & \{2\} \\
		47 & A_g & \{\} \\
		48 & B_{1u} & \{2,2,4\} \\
		48 & B_{1g} & \{2\} \\
		48 & B_{2u} & \{2,2,4\} \\
		48 & B_{2g} & \{2\} \\
		48 & B_{3u} & \{2,2,4\} \\
		48 & B_{3g} & \{2\} \\
		48 & A_u & \{2\} \\
		48 & A_g & \{\} \\
		49 & B_{1u} & \{2,2,2,2,2\} \\
		49 & B_{1g} & \{\} \\
		49 & B_{2u} & \{2,2,2,2,2,4\} \\
		49 & B_{2g} & \{2,2\} \\
		49 & B_{3u} & \{2,2,2,2,2,4\} \\
		49 & B_{3g} & \{2,2\} \\
		49 & A_u & \{2\} \\
		49 & A_g & \{\} \\
		50 & B_{1u} & \{2,2,2,4\} \\
		50 & B_{1g} & \{2,2\} \\
		50 & B_{2u} & \{2,2,4\} \\
		50 & B_{2g} & \{2\} \\
		50 & B_{3u} & \{2,2,4\} \\
		50 & B_{3g} & \{2\} \\
		50 & A_u & \{2\} \\
	\end{array}
\end{align*}
\begin{align*}
	\begin{array}{lll}
		\text{SG} & \text{Irrep} & X_{\text{BS}} \\ \hline
		50 & A_g & \{\} \\
		51 & B_{1u} & \{2,2,2\} \\
		51 & B_{1g} & \{\} \\
		51 & B_{2u} & \{2,2,2,2\} \\
		51 & B_{2g} & \{\} \\
		51 & B_{3u} & \{2,2,2,2,2,2\} \\
		51 & B_{3g} & \{2\} \\
		51 & A_u & \{2,2\} \\
		51 & A_g & \{\} \\
		52 & B_{1u} & \{2,4\} \\
		52 & B_{1g} & \{\} \\
		52 & B_{2u} & \{2,2,4\} \\
		52 & B_{2g} & \{2\} \\
		52 & B_{3u} & \{2,4\} \\
		52 & B_{3g} & \{\} \\
		52 & A_u & \{4\} \\
		52 & A_g & \{\} \\
		53 & B_{1u} & \{2,2,2,2,4\} \\
		53 & B_{1g} & \{2\} \\
		53 & B_{2u} & \{2,2,2,4\} \\
		53 & B_{2g} & \{\} \\
		53 & B_{3u} & \{2,2,4\} \\
		53 & B_{3g} & \{\} \\
		53 & A_u & \{2,4\} \\
		53 & A_g & \{\} \\
		54 & B_{1u} & \{2,2\} \\
		54 & B_{1g} & \{\} \\
		54 & B_{2u} & \{2,2,4\} \\
		54 & B_{2g} & \{2\} \\
		54 & B_{3u} & \{2,2,2,4\} \\
		54 & B_{3g} & \{2,2\} \\
		54 & A_u & \{2\} \\
		54 & A_g & \{\} \\
		55 & B_{1u} & \{2,2,2\} \\
		55 & B_{1g} & \{\} \\
		55 & B_{2u} & \{2,2,2\} \\
		55 & B_{2g} & \{\} \\
		55 & B_{3u} & \{2,2,2\} \\
		55 & B_{3g} & \{\} \\
		55 & A_u & \{2,2,2\} \\
		55 & A_g & \{\} \\
		56 & B_{1u} & \{2\} \\
		56 & B_{1g} & \{\} \\
		56 & B_{2u} & \{2,2,4\} \\
		56 & B_{2g} & \{2\} \\
		56 & B_{3u} & \{2,2,4\} \\
		56 & B_{3g} & \{2\} \\
		56 & A_u & \{2\} \\
		56 & A_g & \{\} \\
	\end{array}&&
	\begin{array}{lll}
		\text{SG} & \text{Irrep} & X_{\text{BS}} \\ \hline
		57 & B_{1u} & \{2,2,2\} \\
		57 & B_{1g} & \{\} \\
		57 & B_{2u} & \{2,2\} \\
		57 & B_{2g} & \{2\} \\
		57 & B_{3u} & \{2\} \\
		57 & B_{3g} & \{\} \\
		57 & A_u & \{2,2\} \\
		57 & A_g & \{\} \\
		58 & B_{1u} & \{2,4\} \\
		58 & B_{1g} & \{\} \\
		58 & B_{2u} & \{2,2,2,4\} \\
		58 & B_{2g} & \{\} \\
		58 & B_{3u} & \{2,2,2,4\} \\
		58 & B_{3g} & \{\} \\
		58 & A_u & \{2,4\} \\
		58 & A_g & \{\} \\
		59 & B_{1u} & \{2\} \\
		59 & B_{1g} & \{\} \\
		59 & B_{2u} & \{2,2,2\} \\
		59 & B_{2g} & \{2\} \\
		59 & B_{3u} & \{2,2,2\} \\
		59 & B_{3g} & \{2\} \\
		59 & A_u & \{2,2\} \\
		59 & A_g & \{\} \\
		60 & B_{1u} & \{2,4\} \\
		60 & B_{1g} & \{\} \\
		60 & B_{2u} & \{4\} \\
		60 & B_{2g} & \{\} \\
		60 & B_{3u} & \{2,4\} \\
		60 & B_{3g} & \{\} \\
		60 & A_u & \{4\} \\
		60 & A_g & \{\} \\
		61 & B_{1u} & \{4\} \\
		61 & B_{1g} & \{\} \\
		61 & B_{2u} & \{4\} \\
		61 & B_{2g} & \{\} \\
		61 & B_{3u} & \{4\} \\
		61 & B_{3g} & \{\} \\
		61 & A_u & \{4\} \\
		61 & A_g & \{\} \\
		62 & B_{1u} & \{2\} \\
		62 & B_{1g} & \{\} \\
		62 & B_{2u} & \{2,2\} \\
		62 & B_{2g} & \{\} \\
		62 & B_{3u} & \{2\} \\
		62 & B_{3g} & \{\} \\
		62 & A_u & \{2,2\} \\
		62 & A_g & \{\} \\
		63 & B_{1u} & \{2,2,2,2\} \\
	\end{array}&&
	\begin{array}{lll}
		\text{SG} & \text{Irrep} & X_{\text{BS}} \\ \hline
		63 & B_{1g} & \{\} \\
		63 & B_{2u} & \{2,2\} \\
		63 & B_{2g} & \{\} \\
		63 & B_{3u} & \{2,2\} \\
		63 & B_{3g} & \{\} \\
		63 & A_u & \{2,2\} \\
		63 & A_g & \{\} \\
		64 & B_{1u} & \{2,2,2\} \\
		64 & B_{1g} & \{2\} \\
		64 & B_{2u} & \{2,2\} \\
		64 & B_{2g} & \{\} \\
		64 & B_{3u} & \{2,2,2\} \\
		64 & B_{3g} & \{\} \\
		64 & A_u & \{2,2\} \\
		64 & A_g & \{\} \\
		65 & B_{1u} & \{2,2,2,2,2,2\} \\
		65 & B_{1g} & \{\} \\
		65 & B_{2u} & \{2,2,2,2,2,2\} \\
		65 & B_{2g} & \{\} \\
		65 & B_{3u} & \{2,2,2,2,2,2\} \\
		65 & B_{3g} & \{\} \\
		65 & A_u & \{2,2\} \\
		65 & A_g & \{\} \\
		66 & B_{1u} & \{2,2,2,2\} \\
		66 & B_{1g} & \{\} \\
		66 & B_{2u} & \{2,2,2,2,4\} \\
		66 & B_{2g} & \{2\} \\
		66 & B_{3u} & \{2,2,2,2,4\} \\
		66 & B_{3g} & \{2\} \\
		66 & A_u & \{2,2\} \\
		66 & A_g & \{\} \\
		67 & B_{1u} & \{2,2,2,2\} \\
		67 & B_{1g} & \{2\} \\
		67 & B_{2u} & \{2,2,2,2,2\} \\
		67 & B_{2g} & \{\} \\
		67 & B_{3u} & \{2,2,2,2,2\} \\
		67 & B_{3g} & \{\} \\
		67 & A_u & \{2\} \\
		67 & A_g & \{\} \\
		68 & B_{1u} & \{2,2\} \\
		68 & B_{1g} & \{\} \\
		68 & B_{2u} & \{2,2,4\} \\
		68 & B_{2g} & \{2\} \\
		68 & B_{3u} & \{2,2,4\} \\
		68 & B_{3g} & \{2\} \\
		68 & A_u & \{2\} \\
		68 & A_g & \{\} \\
		69 & B_{1u} & \{2,2,2,2,2\} \\
		69 & B_{1g} & \{\} \\
	\end{array}&&
	\begin{array}{ccc}
		\text{SG} & \text{Irrep} & X_{\text{BS}} \\ \hline
		69 & B_{2u} & \{2,2,2,2,2\} \\
		69 & B_{2g} & \{\} \\
		69 & B_{3u} & \{2,2,2,2,2\} \\
		69 & B_{3g} & \{\} \\
		69 & A_u & \{2,2\} \\
		69 & A_g & \{\} \\
		70 & B_{1u} & \{2,4\} \\
		70 & B_{1g} & \{\} \\
		70 & B_{2u} & \{2,4\} \\
		70 & B_{2g} & \{\} \\
		70 & B_{3u} & \{2,4\} \\
		70 & B_{3g} & \{\} \\
		70 & A_u & \{4\} \\
		70 & A_g & \{\} \\
		71 & B_{1u} & \{2,2,2,2,2\} \\
		71 & B_{1g} & \{\} \\
		71 & B_{2u} & \{2,2,2,2,2\} \\
		71 & B_{2g} & \{\} \\
		71 & B_{3u} & \{2,2,2,2,2\} \\
		71 & B_{3g} & \{\} \\
		71 & A_u & \{2,2\} \\
		71 & A_g & \{\} \\
		72 & B_{1u} & \{2,2,2,2\} \\
		72 & B_{1g} & \{\} \\
		72 & B_{2u} & \{2,2,2\} \\
		72 & B_{2g} & \{2\} \\
		72 & B_{3u} & \{2,2,2\} \\
		72 & B_{3g} & \{2\} \\
		72 & A_u & \{2,2\} \\
		72 & A_g & \{\} \\
		73 & B_{1u} & \{2,2,2\} \\
		73 & B_{1g} & \{2\} \\
		73 & B_{2u} & \{2,2,2\} \\
		73 & B_{2g} & \{2\} \\
		73 & B_{3u} & \{2,2,2\} \\
		73 & B_{3g} & \{2\} \\
		73 & A_u & \{2\} \\
		73 & A_g & \{\} \\
		74 & B_{1u} & \{2,2,2\} \\
		74 & B_{1g} & \{\} \\
		74 & B_{2u} & \{2,2,2,2\} \\
		74 & B_{2g} & \{\} \\
		74 & B_{3u} & \{2,2,2,2\} \\
		74 & B_{3g} & \{\} \\
		74 & A_u & \{2,2\} \\
		74 & A_g & \{\} \\
		75 & \text{B} & \{2,2\} \\
		75 & \text{A} & \{\} \\
		76 & \text{B} & \{\} \\
	\end{array}
\end{align*}
\begin{align*}
	\begin{array}{lll}
		\text{SG} & \text{Irrep} & X_{\text{BS}} \\ \hline
		76 & \text{A} & \{\} \\
		77 & \text{B} & \{\} \\
		77 & \text{A} & \{\} \\
		78 & \text{B} & \{\} \\
		78 & \text{A} & \{\} \\
		79 & \text{B} & \{\} \\
		79 & \text{A} & \{\} \\
		80 & \text{B} & \{\} \\
		80 & \text{A} & \{\} \\
		81 & \text{B} & \{2,2,2,2,4,4\} \\
		81 & \text{A} & \{\} \\
		82 & \text{B} & \{2,2,4,4,4\} \\
		82 & \text{A} & \{\} \\
		83 & B_u & \{2,2,4\} \\
		83 & B_g & \{2,2,2,2,2,2\} \\
		83 & A_u & \{2,2,2,2,2,2,2,4,8\} \\
		83 & A_g & \{\} \\
		84 & B_u & \{2,2\} \\
		84 & B_g & \{2,2\} \\
		84 & A_u & \{2,2,2,2,2,4\} \\
		84 & A_g & \{\} \\
		85 & B_u & \{2,2\} \\
		85 & B_g & \{2,2,2,2\} \\
		85 & A_u & \{2,2,2,4\} \\
		85 & A_g & \{\} \\
		86 & B_u & \{2\} \\
		86 & B_g & \{2,2\} \\
		86 & A_u & \{2,2,2,4\} \\
		86 & A_g & \{\} \\
		87 & B_u & \{2,2,2\} \\
		87 & B_g & \{2,2,2,2\} \\
		87 & A_u & \{2,2,2,2,2,2,8\} \\
		87 & A_g & \{\} \\
		88 & B_u & \{4\} \\
		88 & B_g & \{2,4\} \\
		88 & A_u & \{2,2,4,4\} \\
		88 & A_g & \{\} \\
		89 & B_2 & \{2\} \\
		89 & B_1 & \{\} \\
		89 & A_2 & \{2,2,2\} \\
		89 & A_1 & \{\} \\
		90 & B_2 & \{\} \\
		90 & B_1 & \{2,2\} \\
		90 & A_2 & \{2\} \\
		90 & A_1 & \{\} \\
		91 & B_2 & \{2\} \\
		91 & B_1 & \{\} \\
		91 & A_2 & \{2,2\} \\
		91 & A_1 & \{\} \\
	\end{array}&&
	\begin{array}{lll}
		\text{SG} & \text{Irrep} & X_{\text{BS}} \\ \hline
		92 & B_2 & \{\} \\
		92 & B_1 & \{\} \\
		92 & A_2 & \{\} \\
		92 & A_1 & \{\} \\
		93 & B_2 & \{2\} \\
		93 & B_1 & \{\} \\
		93 & A_2 & \{2,2,2\} \\
		93 & A_1 & \{\} \\
		94 & B_2 & \{\} \\
		94 & B_1 & \{\} \\
		94 & A_2 & \{\} \\
		94 & A_1 & \{\} \\
		95 & B_2 & \{2\} \\
		95 & B_1 & \{\} \\
		95 & A_2 & \{2,2\} \\
		95 & A_1 & \{\} \\
		96 & B_2 & \{\} \\
		96 & B_1 & \{\} \\
		96 & A_2 & \{\} \\
		96 & A_1 & \{\} \\
		97 & B_2 & \{2\} \\
		97 & B_1 & \{2\} \\
		97 & A_2 & \{2,2\} \\
		97 & A_1 & \{\} \\
		98 & B_2 & \{\} \\
		98 & B_1 & \{\} \\
		98 & A_2 & \{2\} \\
		98 & A_1 & \{\} \\
		99 & B_2 & \{\} \\
		99 & B_1 & \{\} \\
		99 & A_2 & \{\} \\
		99 & A_1 & \{\} \\
		100 & B_2 & \{\} \\
		100 & B_1 & \{2\} \\
		100 & A_2 & \{\} \\
		100 & A_1 & \{\} \\
		101 & B_2 & \{\} \\
		101 & B_1 & \{\} \\
		101 & A_2 & \{\} \\
		101 & A_1 & \{\} \\
		102 & B_2 & \{\} \\
		102 & B_1 & \{\} \\
		102 & A_2 & \{\} \\
		102 & A_1 & \{\} \\
		103 & B_2 & \{2\} \\
		103 & B_1 & \{2\} \\
		103 & A_2 & \{\} \\
		103 & A_1 & \{\} \\
		104 & B_2 & \{\} \\
	\end{array}&&
	\begin{array}{lll}
		\text{SG} & \text{Irrep} & X_{\text{BS}} \\ \hline
		104 & B_1 & \{\} \\
		104 & A_2 & \{\} \\
		104 & A_1 & \{\} \\
		105 & B_2 & \{\} \\
		105 & B_1 & \{\} \\
		105 & A_2 & \{\} \\
		105 & A_1 & \{\} \\
		106 & B_2 & \{\} \\
		106 & B_1 & \{\} \\
		106 & A_2 & \{\} \\
		106 & A_1 & \{\} \\
		107 & B_2 & \{\} \\
		107 & B_1 & \{\} \\
		107 & A_2 & \{\} \\
		107 & A_1 & \{\} \\
		108 & B_2 & \{\} \\
		108 & B_1 & \{2\} \\
		108 & A_2 & \{\} \\
		108 & A_1 & \{\} \\
		109 & B_2 & \{\} \\
		109 & B_1 & \{\} \\
		109 & A_2 & \{\} \\
		109 & A_1 & \{\} \\
		110 & B_2 & \{\} \\
		110 & B_1 & \{\} \\
		110 & A_2 & \{\} \\
		110 & A_1 & \{\} \\
		111 & B_2 & \{2,2,2,2\} \\
		111 & A_2 & \{2\} \\
		111 & B_1 & \{2\} \\
		111 & A_1 & \{\} \\
		112 & B_2 & \{2,2,2,2,2\} \\
		112 & A_2 & \{2\} \\
		112 & B_1 & \{2,2\} \\
		112 & A_1 & \{\} \\
		113 & B_2 & \{2,2\} \\
		113 & A_2 & \{\} \\
		113 & B_1 & \{2,2,2\} \\
		113 & A_1 & \{\} \\
		114 & B_2 & \{2,2\} \\
		114 & A_2 & \{\} \\
		114 & B_1 & \{2,2\} \\
		114 & A_1 & \{\} \\
		115 & B_2 & \{2,2,2\} \\
		115 & A_2 & \{\} \\
		115 & B_1 & \{2\} \\
		115 & A_1 & \{\} \\
		116 & B_2 & \{2,2,2,2\} \\
		116 & A_2 & \{\} \\
	\end{array}&&
	\begin{array}{ccc}
		\text{SG} & \text{Irrep} & X_{\text{BS}} \\ \hline
		116 & B_1 & \{2,2\} \\
		116 & A_1 & \{\} \\
		117 & B_2 & \{2,2,2,4\} \\
		117 & A_2 & \{2\} \\
		117 & B_1 & \{2,2\} \\
		117 & A_1 & \{\} \\
		118 & B_2 & \{2,2,4\} \\
		118 & A_2 & \{\} \\
		118 & B_1 & \{2,2\} \\
		118 & A_1 & \{\} \\
		119 & B_2 & \{2,2,2\} \\
		119 & A_2 & \{\} \\
		119 & B_1 & \{2,2\} \\
		119 & A_1 & \{\} \\
		120 & B_2 & \{2,2,2,2,2\} \\
		120 & A_2 & \{2\} \\
		120 & B_1 & \{2,2,2\} \\
		120 & A_1 & \{\} \\
		121 & B_2 & \{2,2,2\} \\
		121 & A_2 & \{\} \\
		121 & B_1 & \{2,2\} \\
		121 & A_1 & \{\} \\
		122 & B_2 & \{2,2,4\} \\
		122 & A_2 & \{\} \\
		122 & B_1 & \{2,4\} \\
		122 & A_1 & \{\} \\
		123 & B_{2u} & \{2,2,2\} \\
		123 & B_{2g} & \{\} \\
		123 & B_{1u} & \{2\} \\
		123 & B_{1g} & \{\} \\
		123 & A_{2u} & \{2,2,2,2,2,2,2,2,2\} \\
		123 & A_{2g} & \{\} \\
		123 & A_{1u} & \{2,2,2\} \\
		123 & A_{1g} & \{\} \\
		124 & B_{2u} & \{2,2\} \\
		124 & B_{2g} & \{2,2\} \\
		124 & B_{1u} & \{2\} \\
		124 & B_{1g} & \{2,2\} \\
		124 & A_{2u} & \{2,2,2,2,2,4\} \\
		124 & A_{2g} & \{\} \\
		124 & A_{1u} & \{2,2,4\} \\
		124 & A_{1g} & \{\} \\
		125 & B_{2u} & \{2\} \\
		125 & B_{2g} & \{2\} \\
		125 & B_{1u} & \{2\} \\
		125 & B_{1g} & \{2\} \\
		125 & A_{2u} & \{2,2,2,2\} \\
		125 & A_{2g} & \{2\} \\
		125 & A_{1u} & \{2\} \\
	\end{array}
\end{align*}
\begin{align*}
	\begin{array}{lll}
		\text{SG} & \text{Irrep} & X_{\text{BS}} \\ \hline
		125 & A_{1g} & \{\} \\
		126 & B_{2u} & \{2,2\} \\
		126 & B_{2g} & \{2,2,2\} \\
		126 & B_{1u} & \{2\} \\
		126 & B_{1g} & \{2,2\} \\
		126 & A_{2u} & \{2,2,2,2\} \\
		126 & A_{2g} & \{2\} \\
		126 & A_{1u} & \{2,2\} \\
		126 & A_{1g} & \{\} \\
		127 & B_{2u} & \{2\} \\
		127 & B_{2g} & \{\} \\
		127 & B_{1u} & \{2,2\} \\
		127 & B_{1g} & \{2,2,2\} \\
		127 & A_{2u} & \{2,2,2,2,2,2\} \\
		127 & A_{2g} & \{\} \\
		127 & A_{1u} & \{2,2,2,2\} \\
		127 & A_{1g} & \{\} \\
		128 & B_{2u} & \{2,2\} \\
		128 & B_{2g} & \{2,2\} \\
		128 & B_{1u} & \{2,2,2\} \\
		128 & B_{1g} & \{2,2,2\} \\
		128 & A_{2u} & \{2,2,2,2,4\} \\
		128 & A_{2g} & \{\} \\
		128 & A_{1u} & \{2,2,2,4\} \\
		128 & A_{1g} & \{\} \\
		129 & B_{2u} & \{\} \\
		129 & B_{2g} & \{\} \\
		129 & B_{1u} & \{2\} \\
		129 & B_{1g} & \{2\} \\
		129 & A_{2u} & \{2,2,2\} \\
		129 & A_{2g} & \{\} \\
		129 & A_{1u} & \{2\} \\
		129 & A_{1g} & \{\} \\
		130 & B_{2u} & \{2\} \\
		130 & B_{2g} & \{2,2\} \\
		130 & B_{1u} & \{2,2\} \\
		130 & B_{1g} & \{2,2,2\} \\
		130 & A_{2u} & \{2,2,2\} \\
		130 & A_{2g} & \{\} \\
		130 & A_{1u} & \{2,2\} \\
		130 & A_{1g} & \{\} \\
		131 & B_{2u} & \{2,2,2\} \\
		131 & B_{2g} & \{\} \\
		131 & B_{1u} & \{\} \\
		131 & B_{1g} & \{\} \\
		131 & A_{2u} & \{2,2,2,2,2,2\} \\
		131 & A_{2g} & \{\} \\
		131 & A_{1u} & \{2\} \\
		131 & A_{1g} & \{\} \\
	\end{array}&&
	\begin{array}{lll}
		\text{SG} & \text{Irrep} & X_{\text{BS}} \\ \hline
		132 & B_{2u} & \{2\} \\
		132 & B_{2g} & \{\} \\
		132 & B_{1u} & \{2\} \\
		132 & B_{1g} & \{\} \\
		132 & A_{2u} & \{2,2,2,2,2\} \\
		132 & A_{2g} & \{\} \\
		132 & A_{1u} & \{2\} \\
		132 & A_{1g} & \{\} \\
		133 & B_{2u} & \{2,2\} \\
		133 & B_{2g} & \{2,2,2\} \\
		133 & B_{1u} & \{2\} \\
		133 & B_{1g} & \{2,2\} \\
		133 & A_{2u} & \{2,2,2,2,2\} \\
		133 & A_{2g} & \{2,2\} \\
		133 & A_{1u} & \{2,2\} \\
		133 & A_{1g} & \{\} \\
		134 & B_{2u} & \{2\} \\
		134 & B_{2g} & \{2\} \\
		134 & B_{1u} & \{2\} \\
		134 & B_{1g} & \{2\} \\
		134 & A_{2u} & \{2,2,2,2\} \\
		134 & A_{2g} & \{2\} \\
		134 & A_{1u} & \{2\} \\
		134 & A_{1g} & \{\} \\
		135 & B_{2u} & \{2\} \\
		135 & B_{2g} & \{\} \\
		135 & B_{1u} & \{2,2\} \\
		135 & B_{1g} & \{2\} \\
		135 & A_{2u} & \{2,2,2,2\} \\
		135 & A_{2g} & \{\} \\
		135 & A_{1u} & \{2,2,2\} \\
		135 & A_{1g} & \{\} \\
		136 & B_{2u} & \{2\} \\
		136 & B_{2g} & \{\} \\
		136 & B_{1u} & \{2,2\} \\
		136 & B_{1g} & \{2\} \\
		136 & A_{2u} & \{2,2,2,2\} \\
		136 & A_{2g} & \{\} \\
		136 & A_{1u} & \{2,2,2\} \\
		136 & A_{1g} & \{\} \\
		137 & B_{2u} & \{\} \\
		137 & B_{2g} & \{2\} \\
		137 & B_{1u} & \{2\} \\
		137 & B_{1g} & \{2\} \\
		137 & A_{2u} & \{2,2\} \\
		137 & A_{2g} & \{\} \\
		137 & A_{1u} & \{2,2\} \\
		137 & A_{1g} & \{\} \\
		138 & B_{2u} & \{\} \\
	\end{array}&&
	\begin{array}{lll}
		\text{SG} & \text{Irrep} & X_{\text{BS}} \\ \hline
		138 & B_{2g} & \{\} \\
		138 & B_{1u} & \{2\} \\
		138 & B_{1g} & \{2,2\} \\
		138 & A_{2u} & \{2,2,2\} \\
		138 & A_{2g} & \{\} \\
		138 & A_{1u} & \{2,2\} \\
		138 & A_{1g} & \{\} \\
		139 & B_{2u} & \{2\} \\
		139 & B_{2g} & \{\} \\
		139 & B_{1u} & \{2\} \\
		139 & B_{1g} & \{\} \\
		139 & A_{2u} & \{2,2,2,2,2,2\} \\
		139 & A_{2g} & \{\} \\
		139 & A_{1u} & \{2,2\} \\
		139 & A_{1g} & \{\} \\
		140 & B_{2u} & \{\} \\
		140 & B_{2g} & \{\} \\
		140 & B_{1u} & \{2\} \\
		140 & B_{1g} & \{2\} \\
		140 & A_{2u} & \{2,2,2,2,2\} \\
		140 & A_{2g} & \{\} \\
		140 & A_{1u} & \{2,2\} \\
		140 & A_{1g} & \{\} \\
		141 & B_{2u} & \{2\} \\
		141 & B_{2g} & \{2\} \\
		141 & B_{1u} & \{2\} \\
		141 & B_{1g} & \{2\} \\
		141 & A_{2u} & \{2,2,2\} \\
		141 & A_{2g} & \{\} \\
		141 & A_{1u} & \{2,2\} \\
		141 & A_{1g} & \{\} \\
		142 & B_{2u} & \{2\} \\
		142 & B_{2g} & \{2,2\} \\
		142 & B_{1u} & \{2\} \\
		142 & B_{1g} & \{2,2\} \\
		142 & A_{2u} & \{2,2,2,2\} \\
		142 & A_{2g} & \{2\} \\
		142 & A_{1u} & \{2,2\} \\
		142 & A_{1g} & \{\} \\
		143 & \text{A} & \{\} \\
		144 & \text{A} & \{\} \\
		145 & \text{A} & \{\} \\
		146 & \text{A} & \{\} \\
		147 & A_u & \{2,2,2,4,8\} \\
		147 & A_g & \{\} \\
		148 & A_u & \{2,2,2,4,8\} \\
		148 & A_g & \{\} \\
		149 & A_2 & \{2\} \\
		149 & A_1 & \{\} \\
	\end{array}&&
	\begin{array}{ccc}
		\text{SG} & \text{Irrep} & X_{\text{BS}} \\ \hline
		150 & A_2 & \{2\} \\
		150 & A_1 & \{\} \\
		151 & A_2 & \{2\} \\
		151 & A_1 & \{\} \\
		152 & A_2 & \{2\} \\
		152 & A_1 & \{\} \\
		153 & A_2 & \{2\} \\
		153 & A_1 & \{\} \\
		154 & A_2 & \{2\} \\
		154 & A_1 & \{\} \\
		155 & A_2 & \{2\} \\
		155 & A_1 & \{\} \\
		156 & A_2 & \{\} \\
		156 & A_1 & \{\} \\
		157 & A_2 & \{\} \\
		157 & A_1 & \{\} \\
		158 & A_2 & \{\} \\
		158 & A_1 & \{\} \\
		159 & A_2 & \{\} \\
		159 & A_1 & \{\} \\
		160 & A_2 & \{\} \\
		160 & A_1 & \{\} \\
		161 & A_2 & \{\} \\
		161 & A_1 & \{\} \\
		162 & A_{2u} & \{2,2,2,2,4\} \\
		162 & A_{2g} & \{\} \\
		162 & A_{1u} & \{2,2,4\} \\
		162 & A_{1g} & \{\} \\
		163 & A_{2u} & \{2,2,4\} \\
		163 & A_{2g} & \{\} \\
		163 & A_{1u} & \{2,4\} \\
		163 & A_{1g} & \{\} \\
		164 & A_{2u} & \{2,2,2,2,4\} \\
		164 & A_{2g} & \{\} \\
		164 & A_{1u} & \{2,2,4\} \\
		164 & A_{1g} & \{\} \\
		165 & A_{2u} & \{2,2,4\} \\
		165 & A_{2g} & \{\} \\
		165 & A_{1u} & \{2,4\} \\
		165 & A_{1g} & \{\} \\
		166 & A_{2u} & \{2,2,2,2,4\} \\
		166 & A_{2g} & \{\} \\
		166 & A_{1u} & \{2,2,4\} \\
		166 & A_{1g} & \{\} \\
		167 & A_{2u} & \{2,2,4\} \\
		167 & A_{2g} & \{\} \\
		167 & A_{1u} & \{2,4\} \\
		167 & A_{1g} & \{\} \\
		168 & \text{B} & \{4\} \\
	\end{array}
\end{align*}
\begin{align*}
	\begin{array}{lll}
		\text{SG} & \text{Irrep} & X_{\text{BS}} \\ \hline
		168 & \text{A} & \{\} \\
		169 & \text{B} & \{\} \\
		169 & \text{A} & \{\} \\
		170 & \text{B} & \{\} \\
		170 & \text{A} & \{\} \\
		171 & \text{B} & \{4\} \\
		171 & \text{A} & \{\} \\
		172 & \text{B} & \{4\} \\
		172 & \text{A} & \{\} \\
		173 & \text{B} & \{\} \\
		173 & \text{A} & \{\} \\
		174 & \text{A''} & \{2,2,2,2,2,6,6\} \\
		174 & \text{A'} & \{\} \\
		175 & B_u & \{4,4,4\} \\
		175 & B_g & \{2,6,6\} \\
		175 & A_u & \{2,2,2,2,2,2,2,2,6,12\} \\
		175 & A_g & \{\} \\
		176 & B_u & \{4\} \\
		176 & B_g & \{2,6\} \\
		176 & A_u & \{2,2,2,2,2,12\} \\
		176 & A_g & \{\} \\
		177 & B_1 & \{2\} \\
		177 & B_2 & \{2\} \\
		177 & A_2 & \{2,2\} \\
		177 & A_1 & \{\} \\
		178 & B_1 & \{\} \\
		178 & B_2 & \{\} \\
		178 & A_2 & \{2\} \\
		178 & A_1 & \{\} \\
		179 & B_1 & \{\} \\
		179 & B_2 & \{\} \\
		179 & A_2 & \{2\} \\
		179 & A_1 & \{\} \\
		180 & B_1 & \{2\} \\
		180 & B_2 & \{2\} \\
		180 & A_2 & \{2,2\} \\
		180 & A_1 & \{\} \\
		181 & B_1 & \{2\} \\
		181 & B_2 & \{2\} \\
		181 & A_2 & \{2,2\} \\
		181 & A_1 & \{\} \\
		182 & B_1 & \{\} \\
		182 & B_2 & \{\} \\
		182 & A_2 & \{2\} \\
		182 & A_1 & \{\} \\
		183 & B_2 & \{2\} \\
		183 & B_1 & \{2\} \\
		183 & A_2 & \{\} \\
		183 & A_1 & \{\} \\
	\end{array}%
\end{align*}

%% file: LG_Spinful.tex
\begin{align*}
	\begin{array}{lll}
		\text{LG} & \text{Irrep} & X_{\text{BS}} \\ \hline
		1 & \text{A} & \{\} \\
		2 & A_g & \{\} \\
		2 & A_u & \{2,2,4\} \\
		3 & \text{A} & \{\} \\
		3 & \text{B} & \{2,2,4\} \\
		4 & \text{A'} & \{\} \\
		4 & \text{A''} & \{\} \\
		5 & \text{A'} & \{\} \\
		5 & \text{A''} & \{\} \\
		6 & A_g & \{\} \\
		6 & A_u & \{2\} \\
		6 & B_g & \{2\} \\
		6 & B_u & \{2,2,2,2,4,4\} \\
		7 & A_g & \{\} \\
		7 & A_u & \{2\} \\
		7 & B_g & \{2\} \\
		7 & B_u & \{2,2,4\} \\
		8 & \text{A} & \{\} \\
		8 & \text{B} & \{2\} \\
		9 & \text{A} & \{\} \\
		9 & \text{B} & \{\} \\
		10 & \text{A} & \{\} \\
		10 & \text{B} & \{\} \\
		11 & \text{A'} & \{\} \\
		11 & \text{A''} & \{2\} \\
		12 & \text{A'} & \{\} \\
		12 & \text{A''} & \{\} \\
		13 & \text{A'} & \{\} \\
		13 & \text{A''} & \{\} \\
		14 & A_g & \{\} \\
		14 & A_u & \{2,2,2\} \\
		14 & B_g & \{\} \\
		14 & B_u & \{2,2,2\} \\
		15 & A_g & \{\} \\
		15 & A_u & \{2,2\} \\
		15 & B_g & \{\} \\
		15 & B_u & \{2\} \\
		16 & A_g & \{\} \\
		16 & A_u & \{2\} \\
		16 & B_g & \{\} \\
		16 & B_u & \{2,2\} \\
		17 & A_g & \{\} \\
		17 & A_u & \{2\} \\
	\end{array}%
\end{align*}

%% file: LG_Spinless.tex
\begin{align*}
	%
\end{align*}